\documentclass[%
rmp,
 amsmath,amssymb,
 aps,
]{revtex4-2}
\usepackage{natbib}
\usepackage[section]{placeins}
\usepackage{booktabs,threeparttable}

\usepackage{hyperref}
\usepackage{graphicx}
\usepackage{dcolumn}
\usepackage{bm}
\usepackage{float}
\usepackage{caption}
\usepackage{calrsfs}
\usepackage{bbold}
\DeclareMathAlphabet{\pazocal}{OMS}{zplm}{m}{n}
\newcommand{\Aa}{\mathcal{A}}
\usepackage{amsthm}

\theoremstyle{definition}
\newtheorem{definition}{Definition}

\newtheorem{proposition}{Proposition}

\usepackage{enumitem}
\usepackage{eurosym}
\usepackage{xfrac}

\usepackage[
scale=0.8, 
]{geometry}

\usepackage{color}

\begin{document}

\title{Price impact in equity auctions: zero, then linear}

\author{Mohammed Salek}%
 \email{mohammed.salek@centralesupelec.fr}
\author{Damien Challet}%
 \email{damien.challet@centralesupelec.fr}
\author{Ioane Muni Toke}%
 \email{ioane.muni-toke@centralesupelec.fr}

\affiliation{Université Paris-Saclay, CentraleSupélec,  Laboratoire de Mathématiques et Informatique pour la Complexité et les Systèmes, 
  91192 Gif-sur-Yvette, France}%

\date{\today}

\begin{abstract}

Using high-quality data, we report several statistical regularities of equity auctions in the Paris stock exchange. 
First, the average order book density is linear around the auction price at the time of auction clearing and has a large peak at the auction price. While the peak is due to slow traders, the order density shape is the result of subtle dynamics. The impact of a new market order or cancellation at the auction time can be decomposed into three parts as a function of the size of the additional order: (1) zero impact, caused by the discrete nature of prices, sometimes up to a surprisingly large additional volume relative to the auction volume (2) linear impact for additional orders up to a large fraction of the auction volume (3) for even larger orders price impact is non-linear, frequently super-linear.
\end{abstract}

\maketitle

\tableofcontents

\parskip 0.1cm

\section{Introduction}


Most electronic markets rely on auctions to start and end trading days in an orderly way. Because the volume involved during auctions is larger than the liquidity available at a given time in a typical open-market limit order book, auctions reduce price impact and fluctuations. 
The share of the closing auction in the total exchanged volume has significantly increased over the years \cite{Blackrock2020}, especially in European markets \cite{raillon2020growing}. This increase highlights the importance of the auction mechanism in the price formation process.

In contrast to the abundant literature about open-market dynamics, work on auctions is scarce. On the theoretical side, \citet{toke2015exact} derives the distribution of the exchanged volume and the auction price using a stochastic order flow model during a standard call auction. In the same vein, \citet{derksen2020clearing} propose a stochastic model for call auctions which produces a concave price impact function of market orders; in addition, \citet{derksen2022heavy} build on the previous model to demonstrate the heavy-tailed nature of price and volume in closing auctions. Besides, \citet{donier2016walras} show that under sufficient regularity conditions (continuous price and time) and using a first-order Taylor expansion of supply and demand curves, price impact in Walrasian auctions is linear in the vicinity of the auction price.

Empirically, \citet{pagano2003closing} find that introducing opening and closing call auctions improves market quality and lowers execution costs in the Paris stock exchange. \citet{boussetta2017role} add that although opening volumes are decreasing and the market is fragmenting, the opening auction still improves market quality on Euronext Paris. They also report that slow brokers submit orders early, whereas high-frequency traders tend to act moments before the clearing. \citet{challet2018dynamical} analyze US equities data and compute the auction price response functions conditional on the addition, and cancellation of an order.
In addition, \citet{challet2019strategic} demonstrates that a strategic behavior of agents is needed to explain the antagonistic effects of activity acceleration and indicative price volatility decrease as the auction end approaches.

More recently, \citet{jegadeesh2022closing} assess the robustness of closing auctions by comparing the price impact between NASDAQ and NYSE exchanges and find that the cost of trading during closing auctions is generally smaller than during trading hours. They also find that closing auctions mainly attract uninformed and passive investors, while informed traders prefer to act during continuous market hours. In the same spirit, \citet{euronext2021} analyze the closing auction in European markets and use a linear function to fit the impact of market orders; they report a smaller instantaneous impact for later submissions, and an overall cost of trading on close two to three times smaller than during trading hours.

Here, we characterize in detail the empirical properties of liquidity and price impact in equity auctions. At auction time, price impact is fully determined by the state of the order book, and we focus on the instantaneous impact caused by an order if sent just before the clearing. We do not find a straightforward linear impact: while adding or canceling a market (or marketable) order at the auction time has a linear component, the discreteness of the limit order book mechanically leads to zero price impact for small enough orders. These free-of-cost volumes can represent a fairly large fraction of the total matched volume. Before auction time, the order book shape yields a virtual/instantaneous price impact that can differ from that of actual submissions/cancellations. However, we find that the average impact of actual orders is of the same nature, i.e., linear. We argue that the opacity of limit order books during auctions causes the absence of selective liquidity taking, which results in turn in a linear impact.

This paper is organized as follows: first, we introduce a discrete-price auction mathematical framework (Section \ref{framework}) suitable to derive the conditions under which price impact is zero or linear. Next, we present the high-quality data used in this work: a large dataset from the European high-frequency financial (BEDOFIH) database (Section \ref{sec:data}). The main part consists in a detailed study of several statistical regularities of auctions, focusing on limit order book shapes and price impact during the auctions (Sections \ref{avshape} and \ref{Price impact}). Our main results are as follows:
\begin{enumerate}
     \item the average limit order book of buy (sell) orders has a skewed bell shape whose maximum is below (above) the auction price. Both distributions roughly mirror each other and can  be considered linear in the vicinity of the auction price;
     \item there is an often large peak of volume at the auction price that builds up towards the end of the auction;
    \item breaking down  the average limit order book densities by the agent latency (HFT, MIXED, NON) and their account type (own account, client account, market maker, parent company, retail market organization \dots) makes it clear that each category has a different behavior; the peak is not due to HFTs but to slower traders, and some traders post buy and sell orders asymmetrically;
    \item at any time during the auction, instantaneous price impact is zero for small enough volumes for both buy and sell orders simultaneously because of the discreteness of prices. The presence of a peak for both buy and sell limit order densities increases the importance of zero impact in auctions;
    \item for large enough volumes, instantaneous price impact is linear for most of the days and not only on average. This holds when the sum of the buy and sell order densities is constant as a function of the price around the indicative/auction price, which happens on most days. Using a change point detection algorithm, we characterize the linear impact price region day by day and asset by asset at the auction time;
    \item the average price impact of actual submissions/cancellations during the accumulation period is linear as well. This contrasts with open markets where the dependence of the average impact on the order size is much weaker. Since limit order books are not disseminated during auctions,  selective liquidity taking is not possible;
    \item price impact at auction time is smaller during option expiry dates.

\end{enumerate}

\section{A mathematical framework for auctions}\label{framework}

In Euronext markets, equity auctions start with an accumulation period and end with a clearing process. During the accumulation period, participants send their orders (quantity, price, side, order type, \dots) to the exchange. Types of orders include market orders, limit orders, activated stop orders, and valid for auction orders. Modifications and cancellations are allowed, but transactions cannot occur. At any time during the accumulation process and at the end of the auction, the price that maximizes the matched volume and minimizes the imbalance is computed. At the auction time, buy (resp. sell) orders whose prices are larger (resp. smaller) than the auction price are executed, while limit orders whose price equals the auction price may be matched or remain in the order book after the auction.
 
\begin{definition}[\textbf{Supply and demand}]
For an auction $\Aa = (a,d)$, where $a$ is the auction type (open, close, $\dots$) at date $d$,  we define the available supply $S(p,t)$ and demand $D(p,t)$ at a price $p$ and time $t$ as, dropping the $(a,d)$ for the sake of  clarity,
\begin{equation}
\begin{aligned}
S(p,t) = \sum_{p' \leq p} V_S(p',t), \\
D(p,t) = \sum_{p' \geq p} V_B(p',t),
\end{aligned}
\end{equation}
where $V_S(p',t)$ (resp.  $V_B(p',t)$) is the available sell volume (resp. buy volume) at a price $p'$ and time $t$.
\end{definition}

Limit orders can only be submitted on a discrete price grid. Therefore, at any time $t$, $p\mapsto S(p,t)$ is a non-decreasing right-continuous step function, and $p\mapsto D(p,t)$ is a non-increasing left-continuous step function.

\begin{definition}[\textbf{Auction price and volume}]
For an auction $\Aa = (a,d)$, the auction volume $Q_a^d$ noted $Q_a$ is the one maximizing the exchanged quantity between buyers and sellers at the time of the clearing $T_a^d$ noted $T_a$. For a given price $p$ at time $t$, buyers and sellers can exchange a volume equal to $\min\{S(p,t),D(p,t)\}$ at most. Thus 
\begin{equation*}
    Q_a = \underset{p}{\max}\min\left\{S(p,T_a),D(p,T_a)\right\}.
\end{equation*}
The auction price $p_a^d$ noted $p_a$ is the price that maximizes the exchanged quantity. As it may not be unique, we have 
\begin{equation*}
    p_a \in \left\{p \mid Q_a = \min\left\{S(p,T_a),D(p,T_a)\right\} \right\}.
\end{equation*}
\end{definition}
In this work we will always assume that supply $S(p,T_a)$ and demand $D(p,T_a)$ intersect, so that $Q_a$ always exists and is unique. Note however that $p_a$ is often not uniquely defined by the maximization of the exchanged volume alone; this is why exchanges implement a complementary set of rules such that $p_a$ is always well defined. In the case of the Euronext markets used in this work, when multiple prices maximize the exchanged volume, the chosen $p_a$ is the one with the smallest imbalance. Then, if multiple prices with the highest executable volume and the smallest imbalance coexist, the auction price is the one closest to the reference price (last traded price).

\begin{definition}[\textbf{Indicative price and volume}]
For an auction $\Aa$, the indicative price $p^\text{ind}_t$ and the indicative volume $Q^\text{ind}_t$ at time $t\leq T_a$ are the hypothetical auction price and the total matched volume if the clearing took place at time $t$.
\end{definition}

Obviously, we have $p_a = p^\text{ind}_{T_a} $ and $Q_a = Q^\text{ind}_{T_a}$. From now on, the time notation will be omitted when we work at time $t=T_a$ (e.g., $S(p)$ stands for $S(p, T_a)$). Note however that subsequent definitions and results can be stated for any time $t\leq T_a$ using time-dependent notations and substituting $p_a$ with $p^\text{ind}_{t}$ and $Q_a$ with $Q^\text{ind}_{t}$. 

\begin{definition}[\textbf{Buy and sell densities}]
\label{def:bsdensities}
For an auction $\Aa = (a,d)$, we define the buy (resp. sell) density $\rho_{B}^d$ (resp. $\rho_{S}^d$) at a price $p$ as
\begin{equation}
    \rho_{\bullet}^d(p) = \frac{V_{\bullet}(p)}{\delta p}, \quad \bullet \in \{B,S\},
\end{equation}
where $\delta p$ is the difference between the price $p$ and the next non-empty tick price when $\bullet = B$, and $\delta p$  is the difference between $p$ and the previous non-empty tick price when $\bullet = S$.
\end{definition}
To define a meaningful average density over a large number of days, volumes can be scaled by the auction volume $Q_a^d$ at day $d$, and prices can be substituted with log-price differences from the auction price $p \leftarrow \log(p/p_a)$.
\begin{definition}[\textbf{Scaled buy and sell densities}]
\label{def:scaledbsdensities}
For an auction $\Aa = (a,d)$, we define the scaled buy and sell densities as
\begin{equation}
\tilde{\rho}^d_{\bullet}(x) = \frac{ \rho^d_{\bullet}(p_a e^x) }{Q_a^d}, \quad \bullet \in \{B,S\},
\end{equation}
where $x = \log\left(\frac{p}{p_a}\right)$. Furthermore, if we substitute $\delta p$ by a constant $\delta x$, we can compute for a given stock the average scaled density as
\begin{equation}
\left<\tilde{\rho}_{\bullet}(x)\right>= \left<\frac{ V_{\bullet}(p_a e^x) }{Q_a^d \ \times \delta x}\right>, \quad \bullet \in \{B,S\},
\end{equation}
where $\left<\cdot\right>_d$ denotes the average across days of the computed quantity at time $t=T_a$. 
\end{definition} 
Observe that this quantity is a discrete version of the continuous marginal supply and demand curves defined in \citet{donier2016walras}, where $\rho_B(p) = -\partial_p D$ and $\rho_S(p) = \partial_p S$.

\begin{definition}[\textbf{Matched and remaining volumes}]
For an auction $\Aa$, we define $V_{\bullet}^M(p)$ as the matched (executed) volume at a price $p$ and side $\bullet \in \{B,S\}$, and $V_{\bullet}^R(p)$ as the remaining (non-executed) volume at a price $p$ and side $\bullet$. Hence, any limit volume $V_{\bullet}(p)$ at price $p$ is the sum of the matched and remaining volumes 
\begin{equation}
    V_{\bullet}(p) = V_{\bullet}^M(p)+V_{\bullet}^R(p), \quad \bullet \in \{B,S\}.
\end{equation}
\end{definition}

Obviously, for any price $p > p_a$, all the buy volume is matched and all the sell volume remains. Thus $V_B^M(p)=V_B(p)$, $V_S^M(p)=0$, $V_B^R(p)=0$, and $V_S^R(p)=V_S(p)$. Symmetrically, for any, price $p < p_a$, we have $V_B^M(p)=0$, $V_B^R(p) = V_B(p)$, $V_S^M(p) = V_S(p)$, and $V_S^R(p)=0$. Consequently, $V_{\bullet}^M(p) \times V_{\bullet}^R(p)$ can be non-zero only if $p=p_a$.

\begin{proposition}
Let $\Aa$ be an auction with an auction price $p_a$ and an auction volume $Q_a$. The following equalities stand:
\begin{enumerate}[label=(\alph*)]
    \item $Q_a = S(p_a) - V_S^R(p_a) = D(p_a) - V_B^R(p_a)$ ;
    \item $V_S^R(p_a) \times V_B^R(p_a) = 0$.
\end{enumerate}
\label{prop:2}
\end{proposition}
\begin{proof}
(a): as the auction volume $Q_a$ is the sum of all matched volumes, we have
\begin{equation*}
\begin{aligned}
    Q_a &= \sum_p V_B^M(p) = \sum_p V_S^M(p), \\
        &= V_B^M(p_a) + \sum_{p>p_a} V_B^M(p) = V_S^M(p_a) + \sum_{p<p_a} V_S^M(p_a),\\
        &= V_B(p_a) - V_B^R(p_a) + \sum_{p>p_a} V_B(p) = V_S(p_a) - V_S^M(p_a) + \sum_{p<p_a} V_S(p_a),\\
        &= D(p_a) - V_B^R(p_a) = S(p_a) - V_S^R(p_a).
\end{aligned}
\end{equation*}
(b) is proved by contradiction: if $ V_S^R(p_a) \times V_B^R(p_a) \neq 0$ , then $\left(V_S^R(p_a) , V_B^R(p_a)\right) \neq (0,0)$. This implies that a residual volume $\delta V = \min\left(V_S^R(p_a), V_B^R(p_a)\right) > 0$ can be matched between buyers and sellers at the auction price and thus contradicts the fact that $Q_a$ is maximizing the exchanged volume during the auction.
\end{proof}

Let us now introduce volumes scaled by the auction volume: given an integer volume of shares $q \in \mathbb{N}$, we define the scaled volume $\omega=q/Q_a$.

\begin{definition}[\textbf{Price impact}]
For an auction $\Aa$, for any $\omega>0$, we define the price impact before the auction clearing of a buy (resp. sell) market order $I_B(\omega)$ (resp. $I_S(\omega)$) as the absolute change in the auction log-price immediately after submitting a buy (resp. sell) market order of size $q = \omega \times Q_a$
\begin{equation}
    I_{\bullet}(\omega) = \left| \log \left(\frac{p_{\omega}}{p_a}\right)\right|, \quad \bullet \in \{B,S\},
\end{equation}
where $p_{\omega}$ is the new auction price after injecting the market order.
\end{definition}

Note that $I_{\bullet}$ refers to the instantaneous impact of an order submission at auction time $t=T_a$, i.e., assuming a market order is sent just before the clearing. In this case, the market can not react to this submission as the clearing happens right away, and no relaxation can occur. However, if a submission/cancellation is sent to the exchange way before the clearing, the corresponding price impact $I_{\bullet}$ at $t<T_a$ with $p_a \leftarrow p_t^{\text{ind}}$ and $Q_a \leftarrow Q_t^{\text{ind}}$ refers to a virtual/instantaneous price impact that may differ from the price impact of an actual submission/cancellation since the market can still react to it.

\begin{proposition}
\label{prop:T}
Let $\Aa$ be an auction with an auction price $p_a$ and an auction volume $Q_a$. We inject a market order of size $q = \omega Q_a$ before the auction clearing. The new auction price is $p_{\omega}$. We have:

\begin{enumerate}[label=(\alph*)]
    \item The function $I_{\bullet}: \omega \mapsto \left| \log\left(\frac{p_{\omega}}{p_a}\right) \right|$, for $\bullet \in \{B,S\}$ and $\omega > 0$, is a non-decreasing and right-continuous step function.
    
    \item Let $(\omega_B^{(i)})_{i\geq0}$ be the ordered points of discontinuity of $I_B$. Then
    \begin{equation}
        \begin{aligned}
    \omega^{(0)}_B  & =  \frac{V_S^R(p_a) + V^M_B(p_a)}{Q_a},\\
    \omega^{(i)}_B  & = \omega^{(i-1)}_B +  \frac{V_S(p_B^{(i)})+V_B(p_B^{(i)})}{Q_a} \quad , \quad i\geq 1,
        \end{aligned}
    \end{equation}
    where $p_B^{(i)}>p_a$ is the $i^{\text{th}}$ non-empty price tick strictly greater than the auction price.
    \item Let $(\omega_S^{(i)})_{i\geq0}$ be the ordered points of discontinuity of $I_S$. Then
    \begin{equation}
        \begin{aligned}
    \omega^{(0)}_S  & = \frac{V_S^M(p_a) + V_B^R(p_a)}{Q_a}, \\
    \omega^{(i)}_S  & = \omega^{(i-1)}_S + \frac{V_S(p_S^{(i)})+V_B(p_S^{(i)})}{Q_a} \quad , \quad i\geq 1,
        \end{aligned}
    \end{equation}
    where $p_S^{(i)}<p_a$ is the $i^{\text{th}}$ non-empty price tick strictly lower than the auction price.
\end{enumerate}
\end{proposition}

\begin{figure}
    \begin{tabular}{cc}
         \includegraphics[scale=0.32]{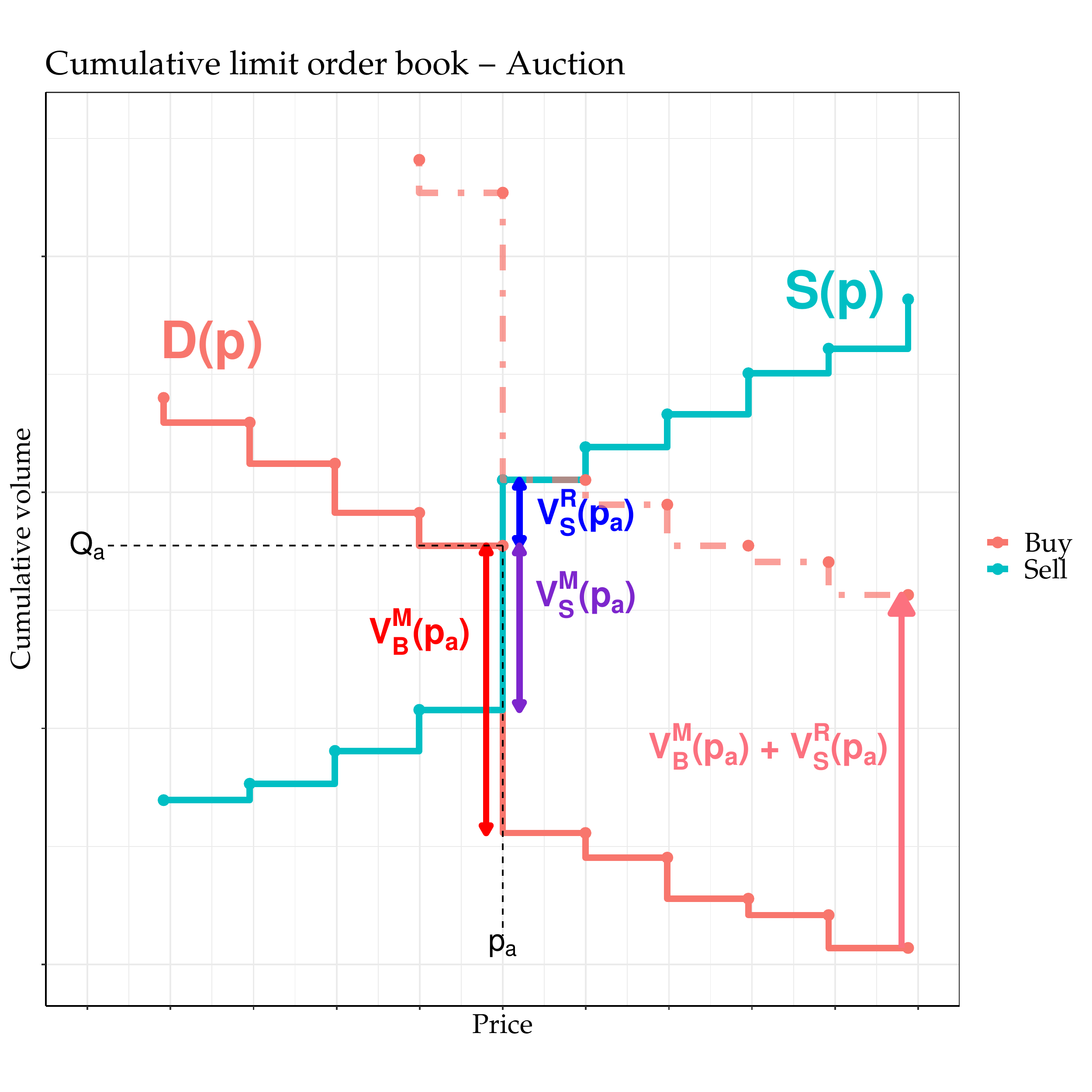}
         &  
         \includegraphics[scale=0.32]{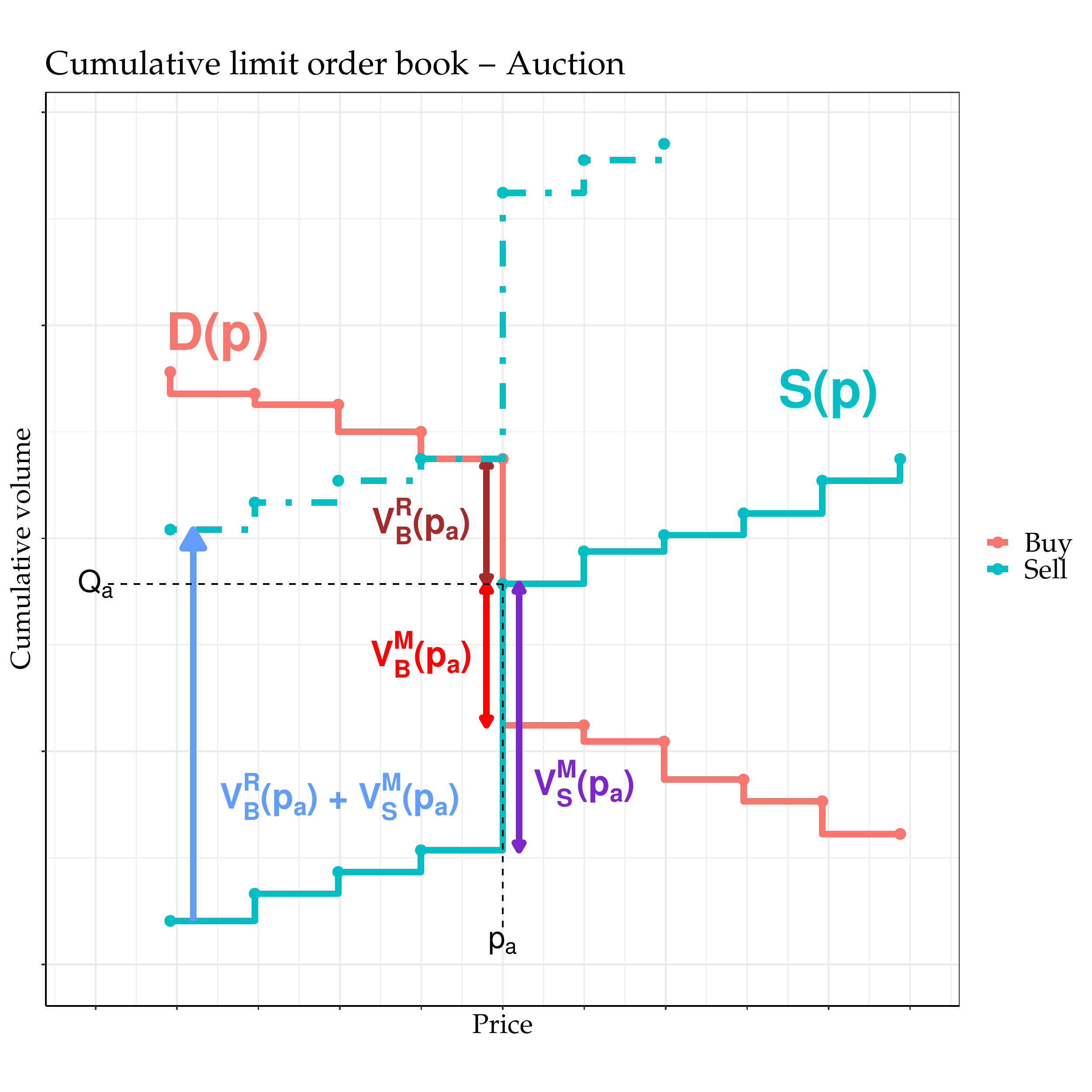}
    \end{tabular}
    \caption{Cumulative buy (red curves) and sell curves (blue curves) during hypothetical auctions. Left panel: the buy volume is totally matched at the auction price; right panel: the sell volume at the auction price is totally matched at the auction price. Dash-dotted lines: effect of an addition buy market order (left plot) and sell market order (right plot): the auction price can change only when the market order is larger than the matched volume plus the imbalance, which explains why zero impact is prevalent.}
    \label{fig:remV} 
\end{figure}

Obviously $I_{\bullet}(\omega^{(i)}_{\bullet}) = \left|\log(p_{\bullet}^{(i+1)}/p_a)\right|$. Also, remark that if all price ticks contain non null volume ($V_B + V_S > 0$), then $p_{\bullet}^{(i)} = p_a \pm i \theta $, where $\theta$ is the tick size. The proof of Proposition \ref{prop:T} is given in Appendix \ref{sec:proofs}. Proposition \ref{prop:T} allows us to compute the impact function at any time of a given auction, including during the accumulation period. 
In addition, the price impact of a new order is zero if its size is smaller than $\omega_{\bullet}^{(0)} Q_a$.
Figure \ref{fig:remV} provides a graphical explanation of $\omega_{\bullet}^{(0)}$ formulas.
On the left panel for example, the buy volume at the auction price is totally matched ($V_B(p_a) = V_B^M(p_a)$ and $V_B^R(p_a)=0$). In this case, in order to shift the price, a buyer would need to execute a market buy order of minimal volume $V^R_S(p_a) + V_B(p_a)$. Alternatively, a seller would need to execute a market sell order of minimal volume $V_S^M(p_a)$. The right panel of Figure \ref{fig:remV} illustrates the symmetric case in which the sell volume at the auction price is totally matched ($V_S(p_a) = V_S^M(p_a)$ and $V_S^R(p_a)=0$). Moreover, observe that if a trader sends a market order of exact size $q=\omega_{\bullet}^{(0)} \times Q_a \in \mathbb{N}$, then both $p_a$ and $p_{\bullet}^{(1)}$ maximize the auction volume. As explained above, the new auction price would be the one with the smallest imbalance, i.e. equal remaining volumes. If $p_a$ and $p_{\bullet}^{(1)}$ have equal imbalances, then the new auction price is the closest to the reference price. Here, we assumed that whenever $q=\omega_{\bullet}^{(0)} \times Q_a \in \mathbb{N}$, the price automatically shifts to $p_{\bullet}^{(1)}$.

Also, by Proposition \ref{prop:T}, $V_S(p_{\bullet}^{(i)})+V_B(p_{\bullet}^{(i)}) = Q_a \times ( \omega^{(i)}_{\bullet}  - \omega^{(i-1)}_{\bullet})$ for $i \geq 1$ is the necessary volume to take the price from $p_{\bullet}^{(i)}$ to $p_{\bullet}^{(i+1)}$. We therefore define $\delta \omega^{(i)}_{\bullet} =  \omega^{(i)}_{\bullet}  - \omega^{(i-1)}_{\bullet}$ for $i \geq 1$ to denote this scaled incremental volume, with the convention that $\delta \omega^{(0)}_{\bullet} = \omega^{(0)}_{\bullet}$. Finally, notice that a cancellation of a buy market order of size $q$ affects the price in the same way as submitting a sell market order of the same size: in both cases the new price $p_{\omega}$ is a solution of $S(p_{\omega})+q = D(p_{\omega})$. Similarly, cancelling a sell market order has the same effect as submitting a buy market order. Consequently, we only focus on the price impact of market order submissions in the following.

\section{Data}\label{sec:data}

The dataset used in this work is part of the BEDOFIH database (Base Européenne de Données Financières à Haute-fréquence) built by the European Financial Data Institute (EUROFIDAI). The dataset provides detailed order data for all stocks traded on Euronext Paris between 2013 and 2017. For each stock and each trading day, information is provided in four files: 
\begin{itemize}[noitemsep,topsep=0pt]
    \item a history orders file that contains all the orders that remained in the central limit order book from the previous trading day ;
    \item a current orders file that contains all submissions, modifications, and cancellations for the current trading day ;
    \item a trades file that lists all the transactions that took place during the current trading day ;
    \item an events file that lists special market events, if any, such as a delayed opening, a halt in trading, etc.
\end{itemize}
In addition to standard information such as time with microsecond precision, price, side (buy/sell), quantity, and price threshold for stop orders, we have access to additional order details in these files, some of which are computed \textit{ex-post}. These include the order type and its temporal validity (market, limit, valid-for-auction, valid-for-closing, etc.), the high-frequency status of the market participant (HFT, NON-HFT, or MIXED), and the account type (own account, client account, market maker, parent company, retail liquidity provider, retail market organization).

In order to reconstruct the exact state of the limit order book (LOB) at any point during the auction, we combine the information from the four different files for each stock and each trading day to create a snapshot. We select the 34 most traded stocks on Euronext Paris between 2013 and 2017 and analyze 2 to 5 years worth of data for each stock, totaling $N = 34,977$ stock-days. A small number of these stock-days result in errors or mismatches (e.g., dataset errors, non-crossing supply and demand for the opening auction, or half-day trading/halted trading before 17:30 for the closing auction). After removing these invalid snapshots, we are left with $N_o = 34,971$ valid snapshots at the opening auction time and $N_c = 34,820$ valid snapshots at the closing auction time.

Using these reconstructed snapshots just before the auction time, we compute reconstructed prices and volumes as per Euronext rules, i.e., by maximizing the exchanged volume and minimizing the imbalance. This boils down to finding the intersection of the reconstructed supply and demand curves. Table \ref{tab:check} reports the percentage of snapshots for which the reconstructed price (resp. volume) matches the actual auction price (resp. volume) among valid snapshots.
\begin{table}
\caption{Percentages of auction snapshots with accurate reconstruction.}
\begin{ruledtabular}
\begin{tabular}{ccc}
 & Opening auction & Closing auction \\
\hline
Number of valid snapshots & 34,971 & 34,820 \\
\% snapshots matching the auction price & 99.6\% & 99.9\% \\
\% snapshots matching the auction volume & 99.0\% & 99.7\% \\
\% snapshots matching both & 98.9\% & 99.6\% \\
\end{tabular}
\end{ruledtabular}
\label{tab:check}
\end{table}
The remaining discrepancies may be a result of using simplified rules to account for stop orders and occasional contradictions between recorded data in the orders file and the trades file. For these few unmatched snapshots, we note that the discrepancies between computed and actual quantities are small: less than 1 basis point on the absolute average difference from the auction price and 0.2\% on the absolute average distance from the auction volume. These few unmatched auctions are discarded from the sample in the subsequent analysis, though they would not alter the outcome of our experiments.

\section{Average shape of the auction limit order book}\label{avshape}

This section investigates the typical shape of the limit order book at auction time $T_a$, what it implies for post-clearing price impact, and how the average LOB shape can be broken down by latency and account type of market participants.

\subsection{Pre-clearing vs. post-clearing LOB shape}

\begin{figure}
    \centering
    \includegraphics[scale=0.33]{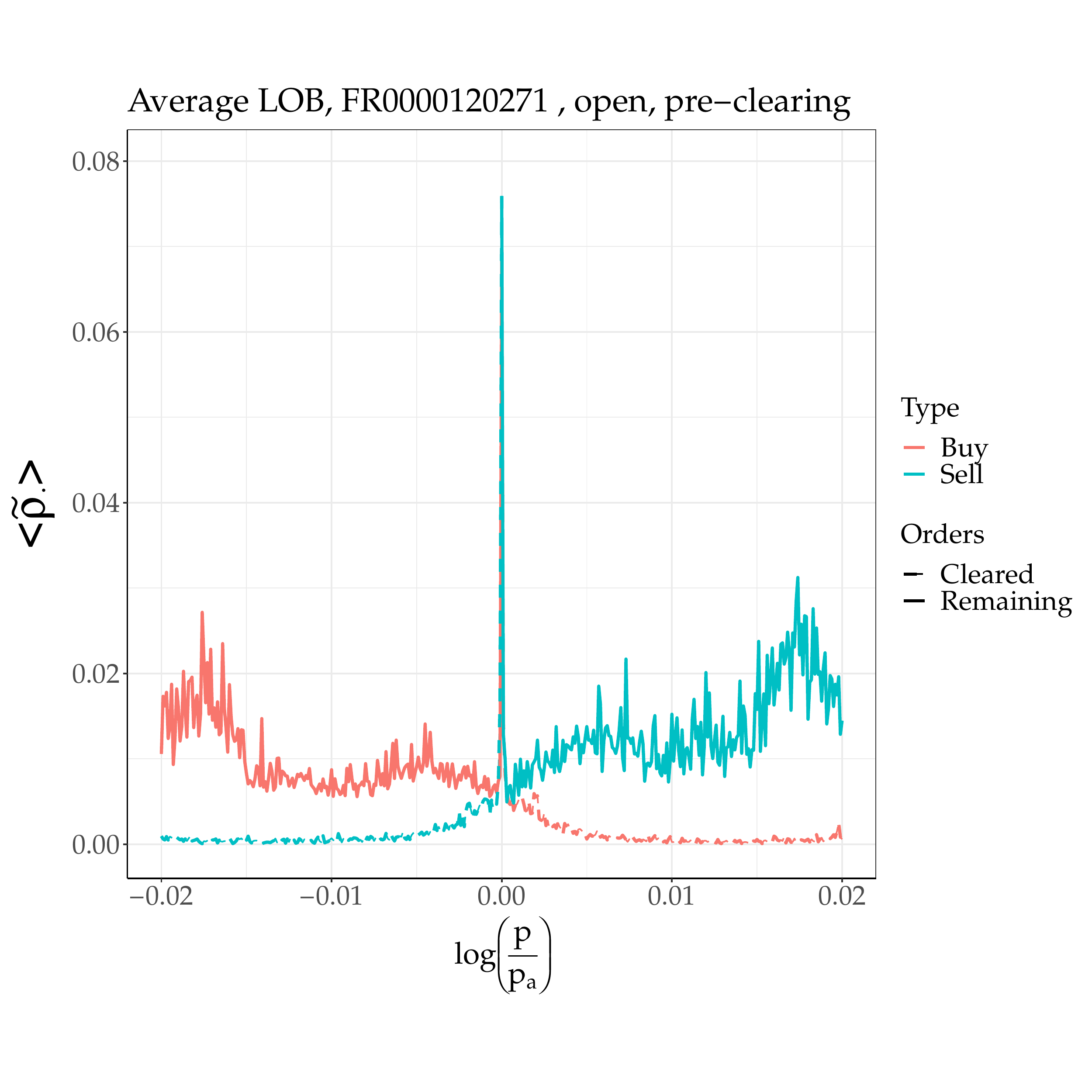}
    \includegraphics[scale=0.33]{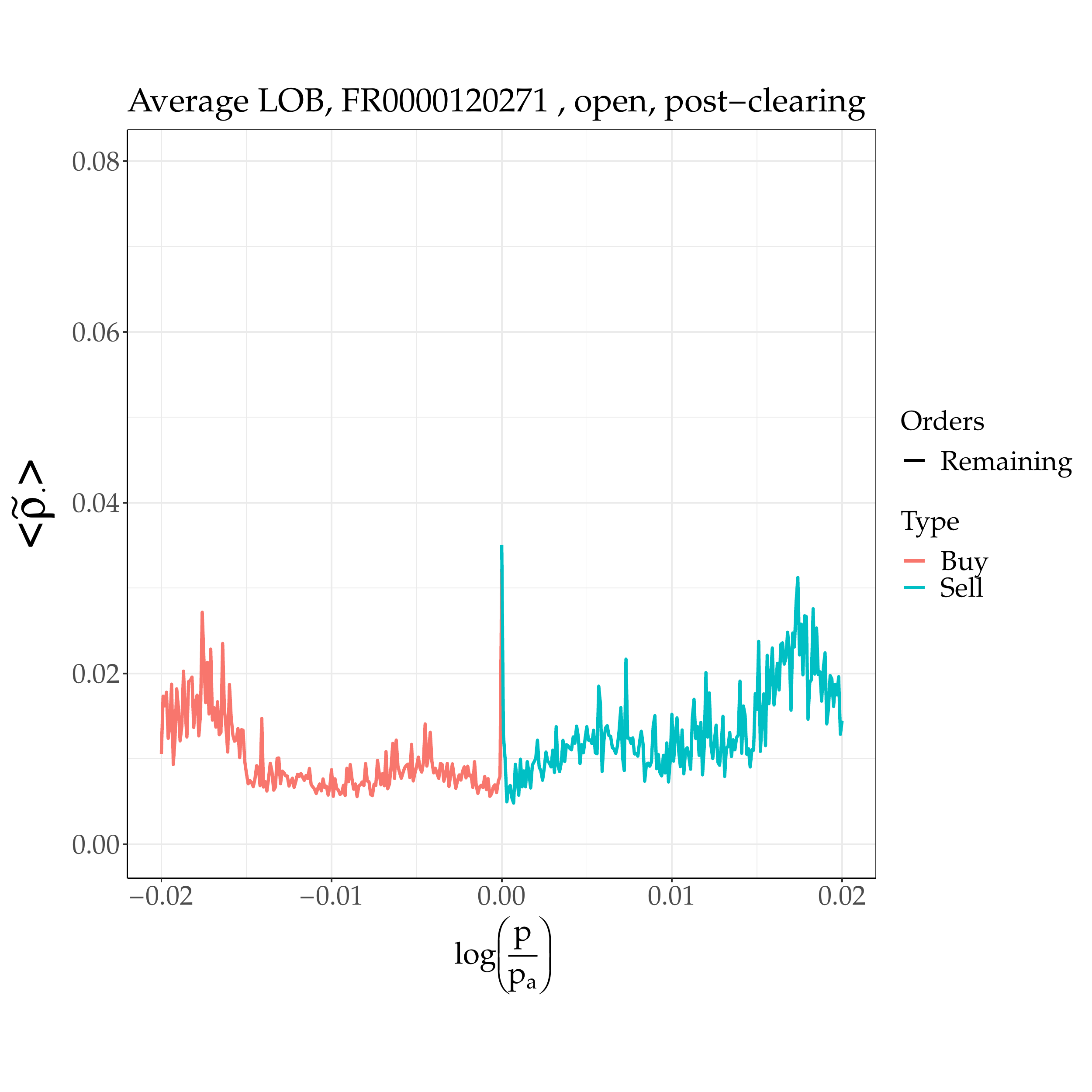}
        \includegraphics[scale=0.33]{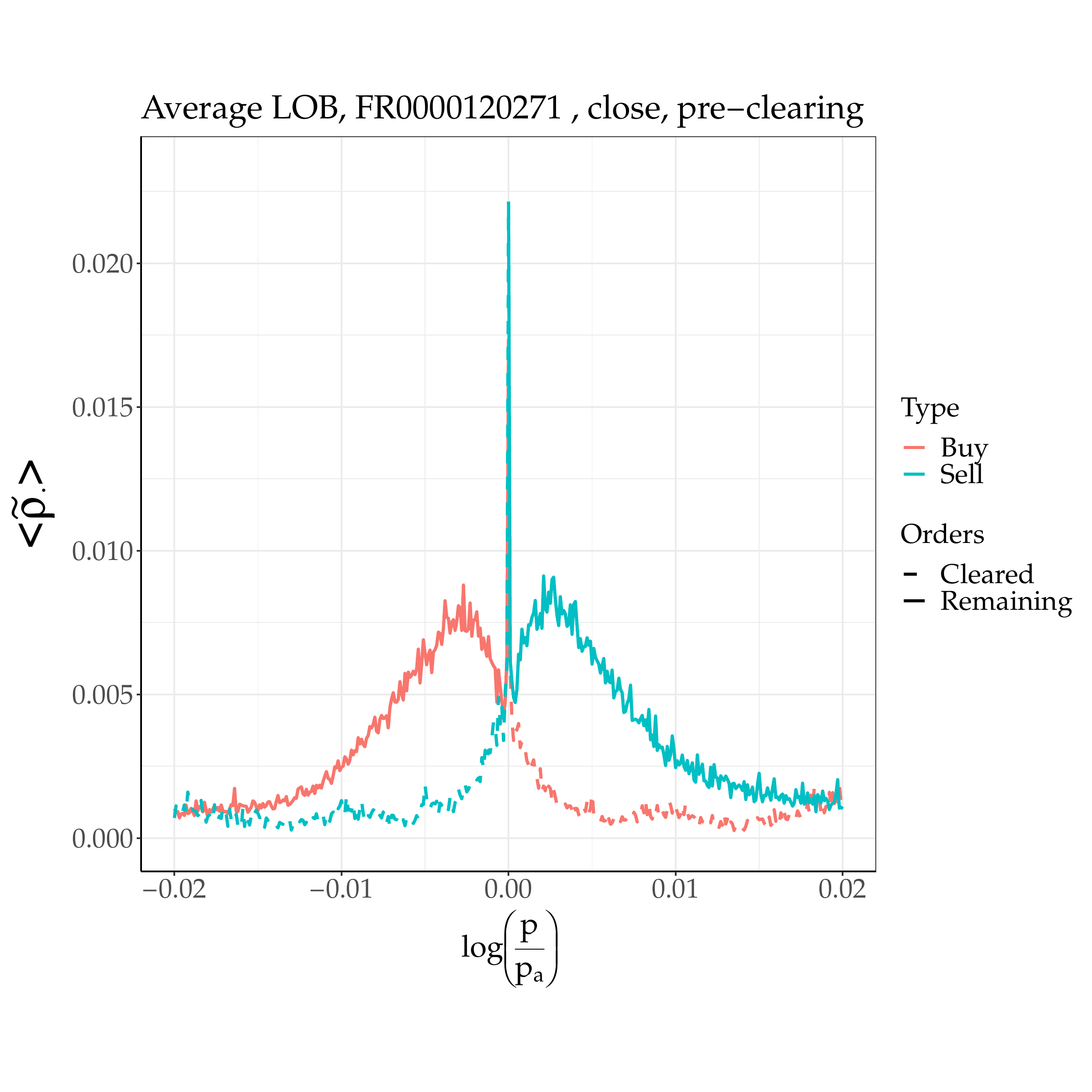}
    \includegraphics[scale=0.33]{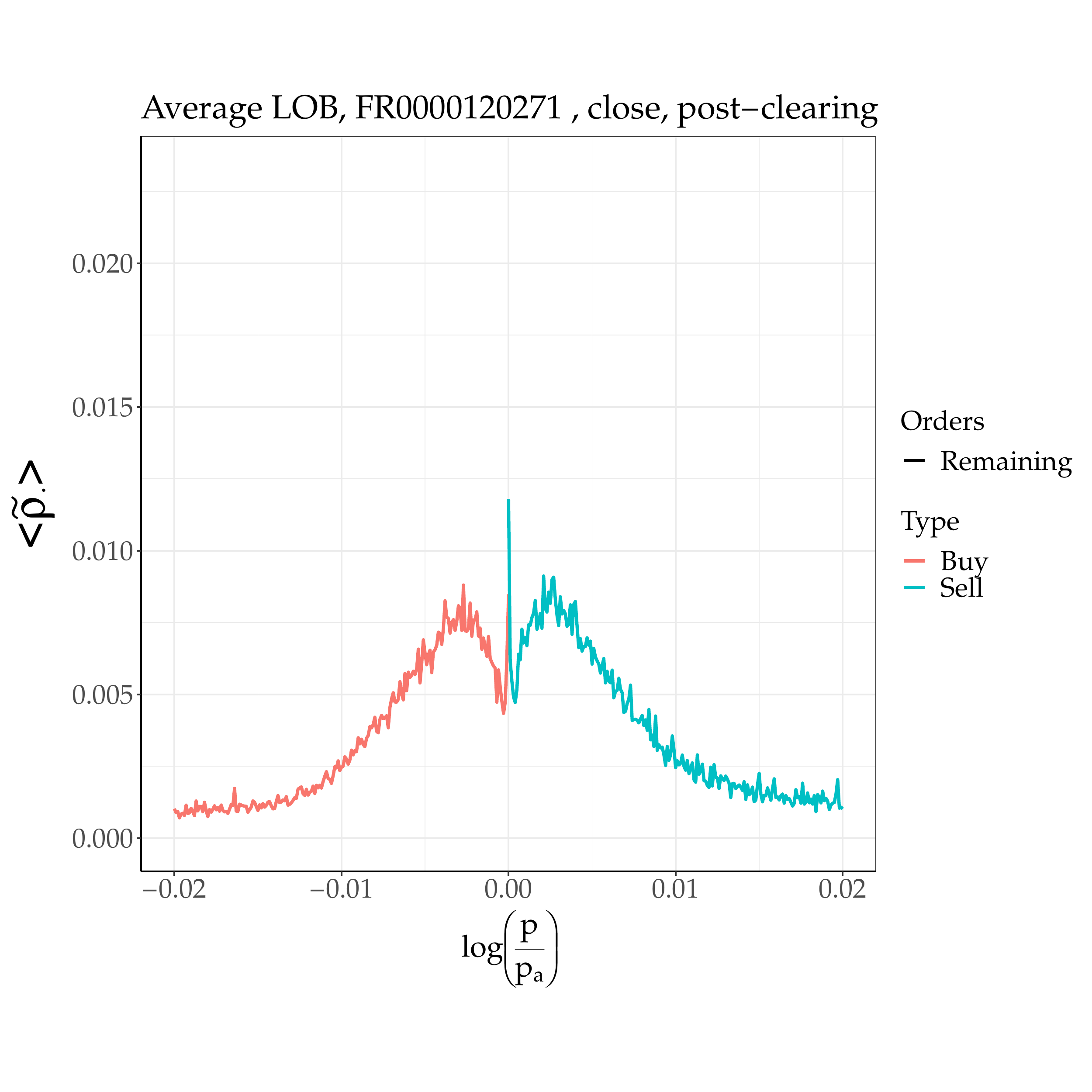}
    \caption{Average density of the limit order book $\left<\tilde \rho_{\bullet}\right>$ as a function of the log difference from the auction price $p_a$ at the opening auction (top) and at the closing auction (bottom); left plots:  pre-clearing, right plots: post-clearing (right). TTE.PA  (TotalEnergies) between 2013 and 2017. In all panels, the mean density is computed on price intervals of size $\delta x = 1 \text{bp}$ over $N=1266$ days.}
    \label{fig:average rhos}
\end{figure}

For each stock of the dataset, we compute the buy and sell average empirical densities $\left<\tilde{\rho}_{\bullet}\right>$ (see Definition \ref{def:scaledbsdensities}) as a function of the log-price difference $x= \log\left(\frac{p}{p_a}\right)$. Figure \ref{fig:average rhos} shows the average LOB density for the most traded stock in our dataset (ISIN FR0000120271, TTE.PA, TotalEnergies). We distinguish the orders that are cleared by the auction process (dotted lines) from the ones that remain in the LOB after the end of the auction (full lines). Average LOB densities are very similar across all the studied stocks.

Figure \ref{fig:average rhos} shows the average LOB densities at the closing auction:  the buy and sell densities have a skewed bell-shaped curve around the auction price. Opening and closing auctions have clearly different LOB densities. As expected, the average LOB density is noisier at the opening auction than at the closing auction which reflects the typical liquidity available at either auction \cite{challet2019strategic}. 
However, the following remarks hold for both auctions:
\begin{itemize}[noitemsep,topsep=0pt]
    \item there is a peak at the auction price, i.e. $\left<\tilde{\rho}_{\bullet}\right>(0)$ is larger than typical values taken near 0. This translates an accumulation of orders on $p=p_a$ on average at the time of the clearing ;
    \item $\left<\tilde{\rho}_{\bullet}\right>$ is linear around $x=0$, i.e. $p=p_a$.
\end{itemize}

As shown by Fig.\ \ref{fig:average rhos}, all buy orders with $p>p_a$ are cleared, and all buy orders with $p<p_a$ remain in the LOB after the auction as long as their temporal validity extends beyond the clearing; similarly, all sell orders with $p<p_a$ are cleared and all sell orders with $p>p_a$ remain in the LOB after the auction. For $p = p_a$, some orders are matched, some are not. This explains why the peaks of buy and sell volumes at $p_a$ are reduced after the clearing. Finally, the auction-only orders are removed from the LOB after the clearing if they are not executed.

\subsection{Post-clearing instantaneous price impact}
Let us briefly discuss the instantaneous post-clearing price impact during a continuous trading phase just after an auction. The following remarks are valid whenever there is a continuous trading phase right after the auction clearing (that is, after the open auction here).
Consider the case of a trader sending a buy market order during the continuous trading phase just after auction clearing. This trader can expect to match up to all the remaining sell orders at $p_a$ without impacting the price. Once the liquidity at $p_a$ is consumed, sending an additional buy volume $q>0$ will result in a sub-linear price impact.
Indeed, since $\left<\tilde{\rho}_S\right>$ has been observed to be linear around $0$ (peak excluded), we may write $\left<\tilde{\rho}_S\right>(x) = a_1 + b_1 x$ on this neighborhood so that we have on average
\begin{equation}
    \int_0^{x}\left<\tilde{\rho}_S\right>(u)\mathrm{d}u = q,
\end{equation}
which implies
\begin{equation}
    \frac{b_1}{2}x^2 + a_1 x - q = 0.
\end{equation}
Hence, the post-clearing instantaneous price impact $x$ is sub-linear and ranges between a square root limit when $q \gg \frac{a_1^2}{2b_1}$ and a linear impact limit $q \ll \frac{a_1^2}{2b_1}$. This reproduces in a stylized way the crossover between linear and square-root market impact observed in continuous double auctions \cite{bucci2019crossover}. The latter can be explained for example by assuming the existence of a hidden, latent LOB  \citet{toth2011anomalous}, which is only partially revealed but whose shape largely determines that of market impact. At auction times instead, market participants are forced to reveal their intentions at least in the vicinity of $p_a$, and one can relate the auction LOB with the latent LOB.

\subsection{Breakdown by market participant latency}

\begin{figure}
    \centering
    \includegraphics[scale=0.33]{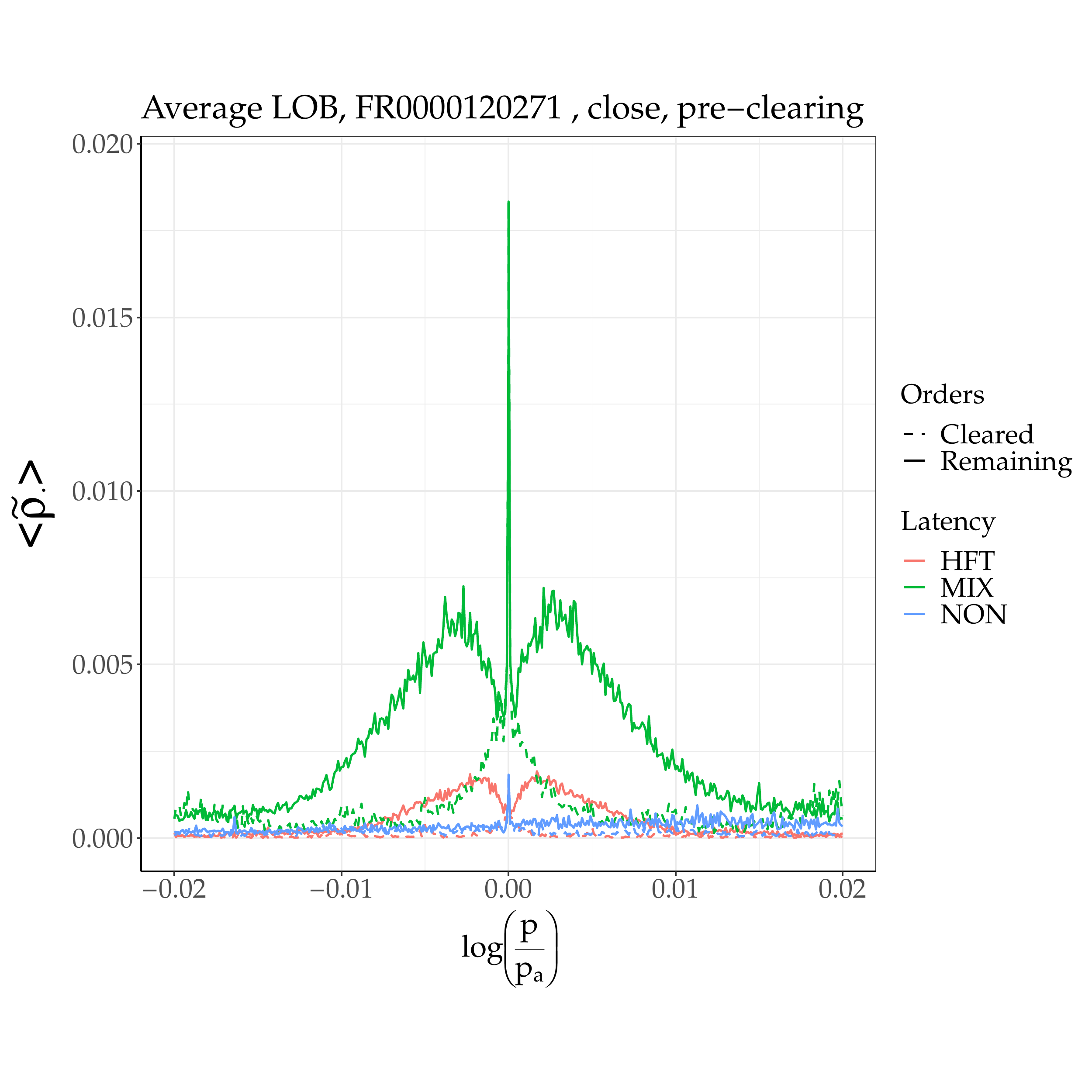}
    \includegraphics[scale=0.33]{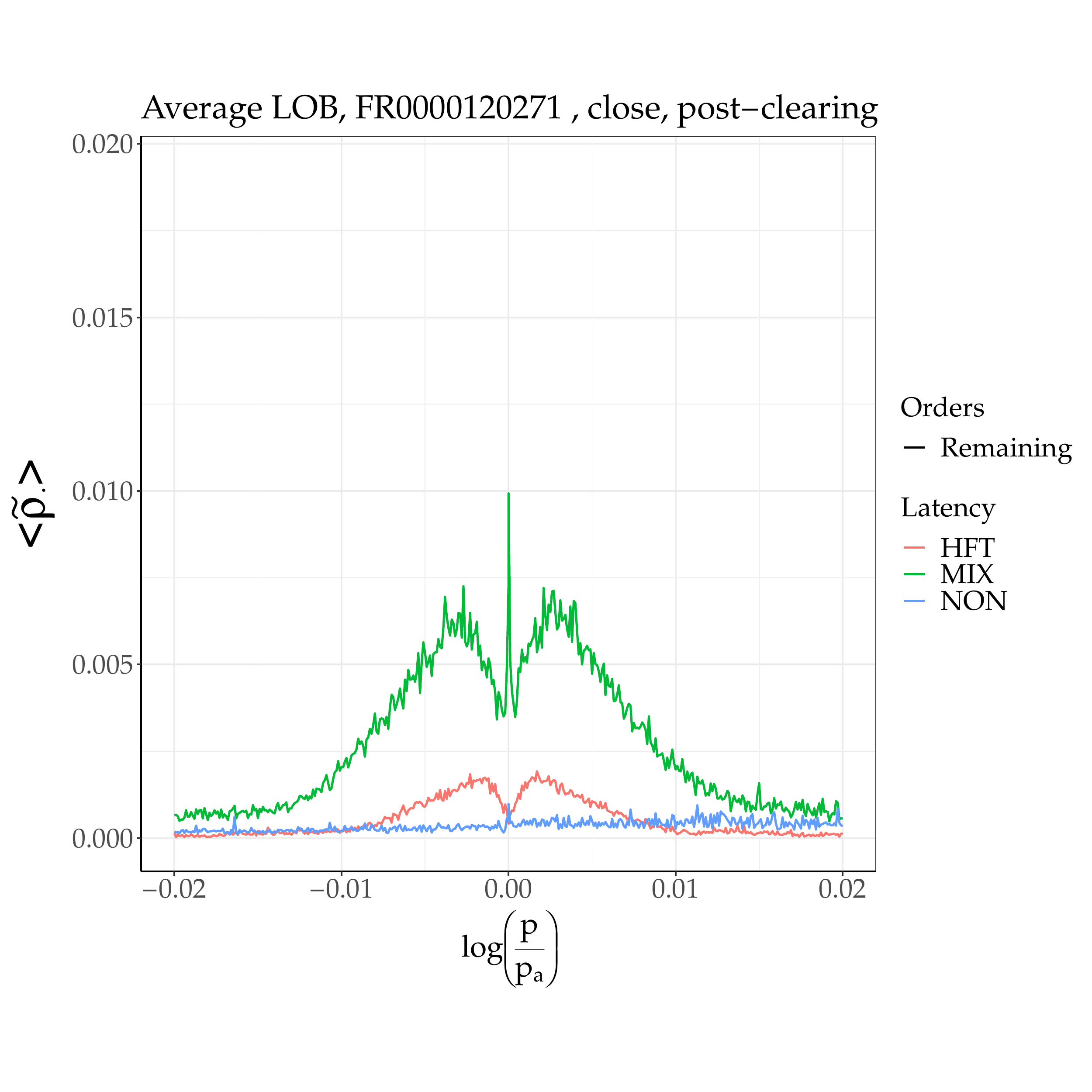}
    \includegraphics[scale=0.35]{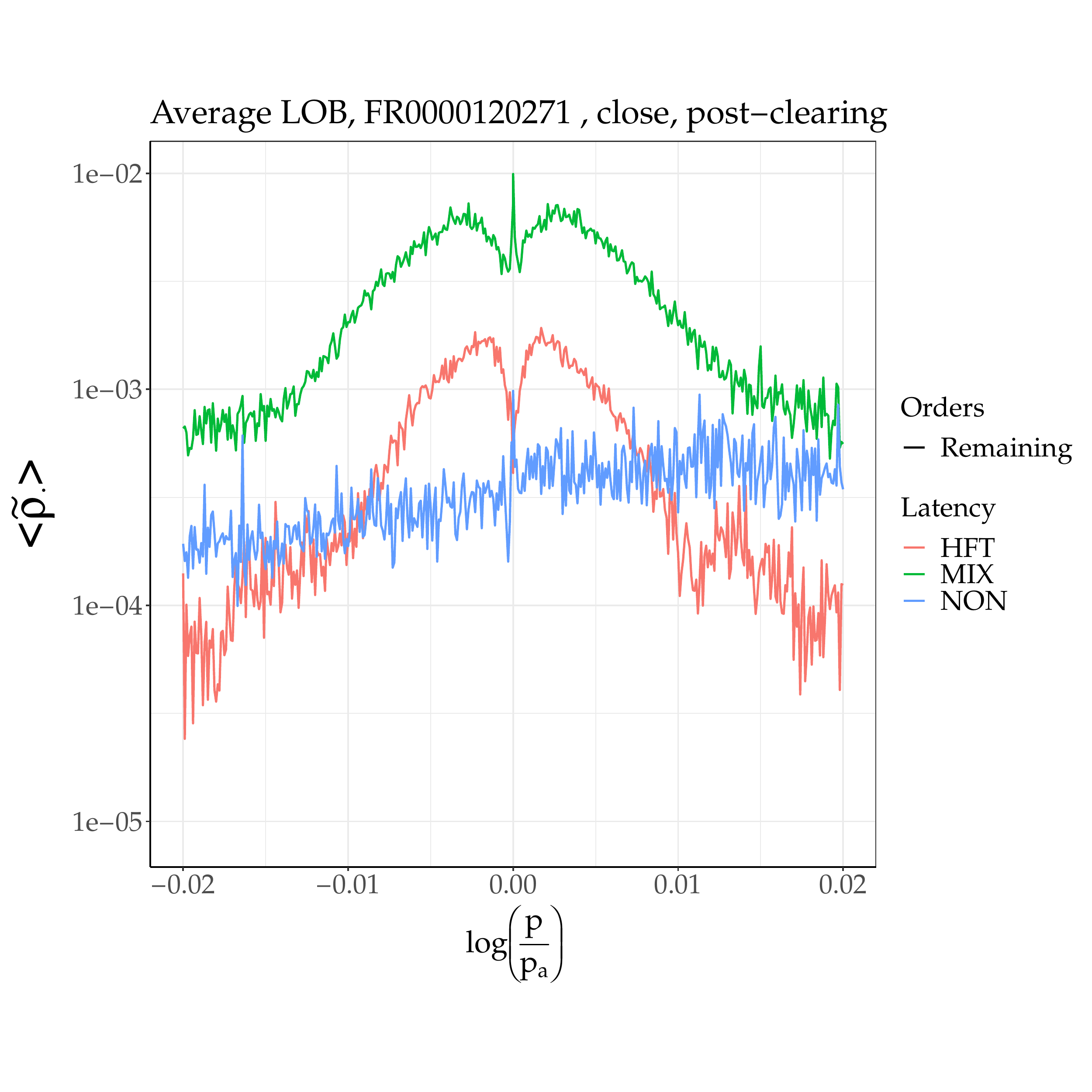}
    \caption{Average density of the limit order book $\left<\tilde \rho_{\bullet}\right>$ as a function of the log difference from the auction price $p_a$: breakdown by user latency of the average LOB during the closing auction just before the clearing (left), right after the clearing (right and bottom), with Y-axis in a log-scale (bottom) for TTE.PA between 2013 and 2017. The HFT flag denotes pure high frequency traders, MIX denotes investment banks with high frequency trading activities, and NON denotes traders without HFT activities.}
    \label{fig:bylatency}
\end{figure}

Figure \ref{fig:bylatency} displays a breakdown of the average empirical densities $\left<\tilde{\rho}_{\bullet}\right>$ at the closing auctions by the speed of market participants. We used the latency flag in our data which specifies the HFT category of the order sender as per the AMF definition\footnote{A participant is considered a high-frequency trader (HFT) if he meets one of the two following conditions:
\begin{itemize}[noitemsep,topsep=0pt]
    \item The average lifetime of its canceled orders is less than the average lifetime of all orders in the book, and it has canceled at least 100,000 orders during the year.
    \item The participant must have canceled at least 500,000 orders with a lifetime of fewer than 0.1 seconds, and the top percentile of the lifetime of its canceled orders must be less than 500 microseconds.
\end{itemize}
An investment bank meeting one of these conditions is described as mixed-HFT (MIX). If a participant does not meet any of the above conditions, it is a non-HFT (NON).}.
Let us make three remarks regarding Figure \ref{fig:bylatency}. First, we notice that the MIX LOB has the same order of magnitude and shape as the total LOB (Fig. \ref{fig:average rhos}, bottom). This indicates that the contribution of traders flagged as fast (HFT) and slow (NON) to the liquidity provision of the closing auction (limit orders in the neighbourhood of the auction price) is smaller than the contribution of investment banks (flagged MIX). Second, the HFT LOB does not display an outstanding peak of volumes at the auction price. This suggests that this peak is actually caused by slow traders and may result in  auction price pinning. Third, Figure \ref{fig:bylatency} deals with the most liquid stock of the sample, but some stocks have a very small HFT-flagged LOB with the same order of magnitude as the low frequency LOB: HFT-flagged traders do not place sizeable limit orders in the closing auction of all stocks.

As stated in \citet{AMF2017,Benzaquen2018}, open markets are dominated by fast trading algorithms, which suggests considering the HFT LOB only (up to a multiplicative constant) when relating the auction LOB with the latent continuous-auction LOB. In this setting, the post-clearing price impact is much closer to a square root because of the sharp linear shape of the HFT LOB that vanishes around the current price.

\subsection{Breakdown by account type}

\begin{figure}
    \centering
    \includegraphics[scale=0.33]{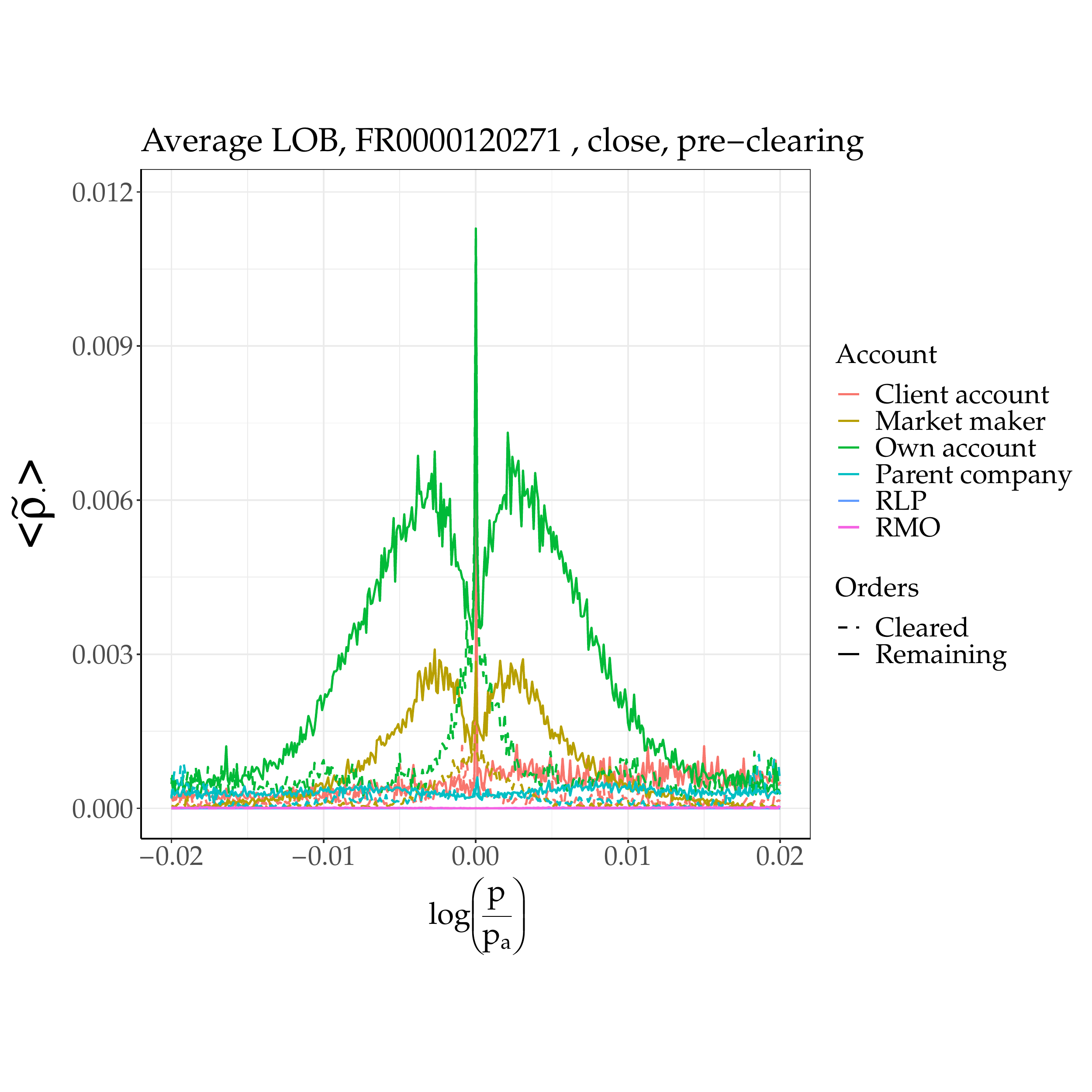}
    \includegraphics[scale=0.33]{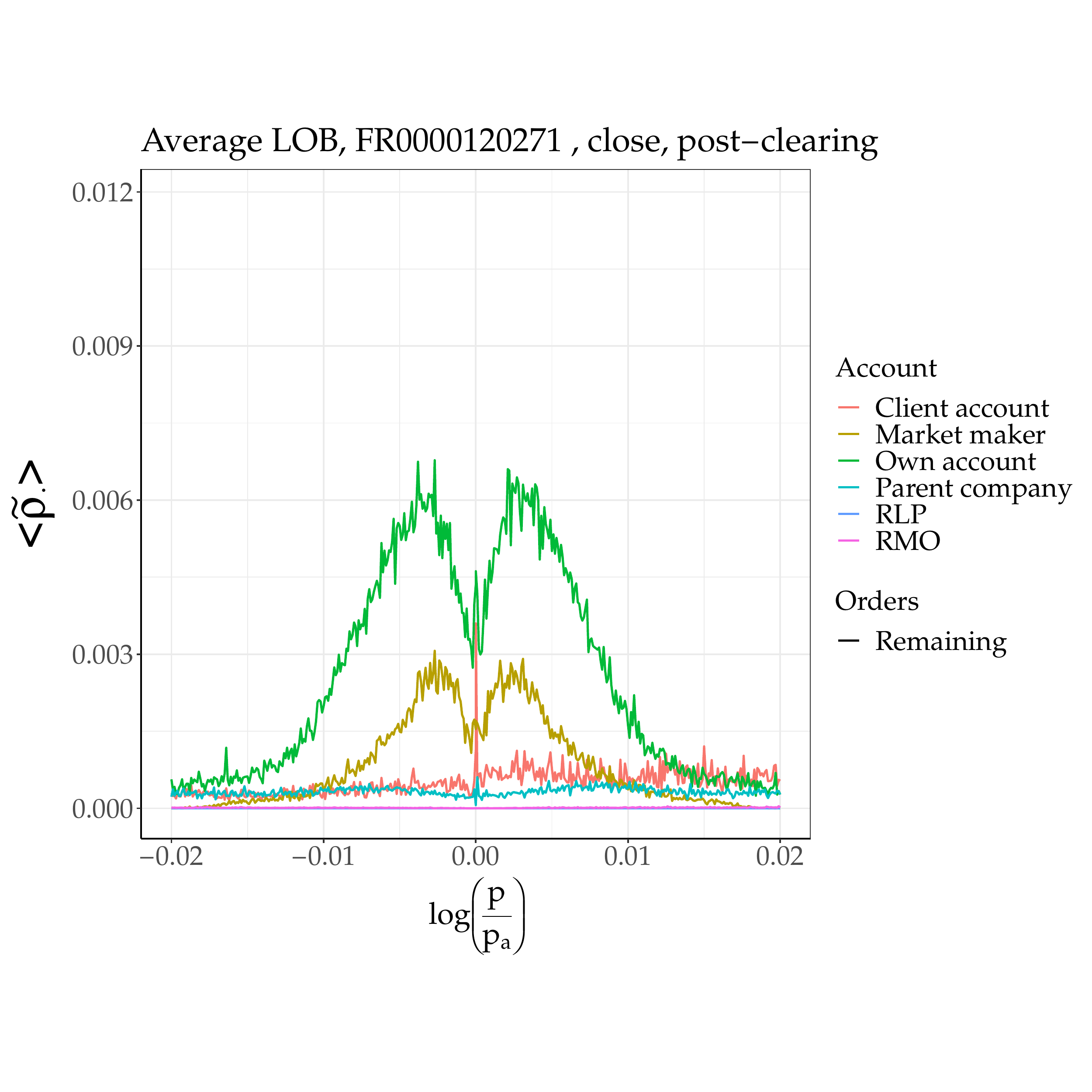}
    \includegraphics[scale=0.35]{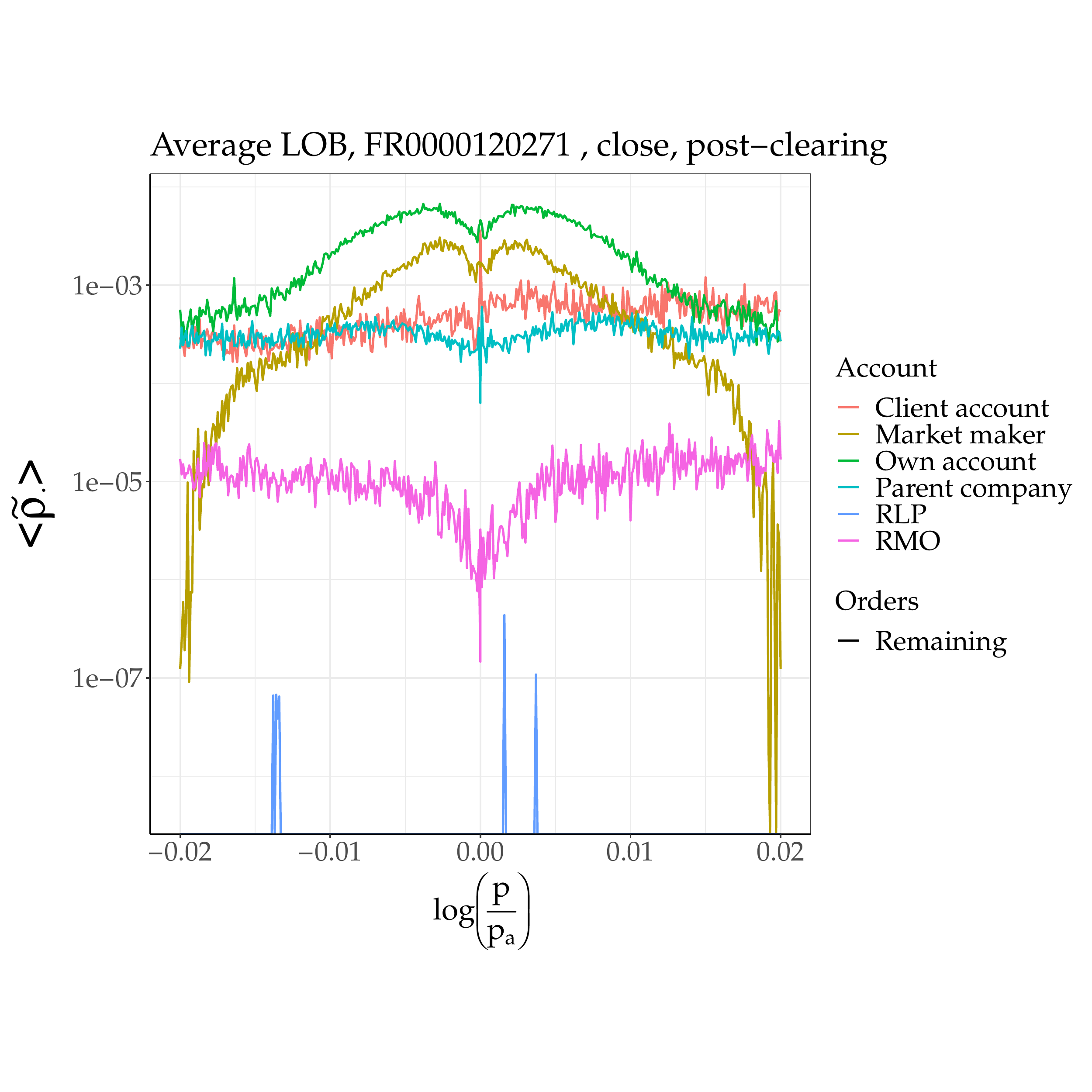}
    \caption{Average density of the limit order book $\left<\tilde \rho_{\bullet}\right>$ as a function of the log difference from the auction price $p_a$: breakdown of the average LOB during the closing auction by user account type just before the clearing (left), right after the clearing (right and bottom), with Y-axis in a log-scale (bottom) for TTE.PA between 2013 and 2017. Colors represent orders executed on the behalf of: a client account, a market maker, an own account, a parent company account, a retail market organization (RMO), and a retail liquidity provider (RLP).}
    \label{fig:byaccount}
\end{figure}

Figure \ref{fig:byaccount} shows a breakdown of the average empirical densities $\left<\tilde{\rho}_{\bullet}\right>$ at the closing auction by the account type. This particular flag tells on whose behalf an order was sent:  client account,  market marker,  own account, parent company account, retail market organization (RMO), and retail liquidity provider (RLP).
We notice that traders operating on behalf of their own account, which includes a significant fraction of investment bank activities, and market markers provide most of the liquidity in the vicinity of the auction price. In addition, the density of orders sent on behalf of clients and slow traders have the same shape (see Fig. \ref{fig:bylatency}). This decomposition will be valuable in designing realistic agent-based models in addition to incorporating multi-time scale liquidity\footnote{There are only 126 authorized participants on the cash market (that includes equities) of Euronext Paris. See: https://live.euronext.com/en/resources/members-list.}. 

\section{Price impact}\label{Price impact}

This section investigates a set of statistical regularities of price impact in equity auctions focusing on closing auctions. In the first part, we study price impact at the auction time, which was fixed at 17:35:00 before the $28^{\text{th}}$ of September 2015, and then randomly between 17:35:00 and 17:35:30: we assume that a trader wishes to know by how much the auction price would have moved if she had sent a market order right before the clearing, supposing that she could know the clearing time in advance. In the second part, we study the behavior of price impact before the auction time. To this end, we examine the evolution of the virtual/instantaneous price impact throughout the accumulation period. Then, we relate the price impact at auction time with that at 17:35:00. Finally, we compute the average impact of actual submissions/cancellations during auctions and discuss why it is markedly different from that of open markets.

\subsection{At the auction time}

In this first part, we investigate the impact of a market order submitted (or canceled) to the exchange just before the clearing. We explicitly assume that the trader would have been able to insert or cancel her order just before the clearing process. In this setting, we highlight the existence of a significant zero impact volume below which the auction price would not have changed and explain why this zero impact is purely mechanical. We then show that any additional volume has a linear price impact over a volume range that we determine, not only on average but for most stocks and days. We also derive a simple formula for the impact slope that we validate empirically using a simplified optimization routine. Finally, we examine the influence of derivative expiry days on closing auctions.

\subsubsection{Zero impact: $\omega<\omega^{(0)}_{\bullet}$}

\begin{figure}
    \centering
    \includegraphics[scale=0.4]{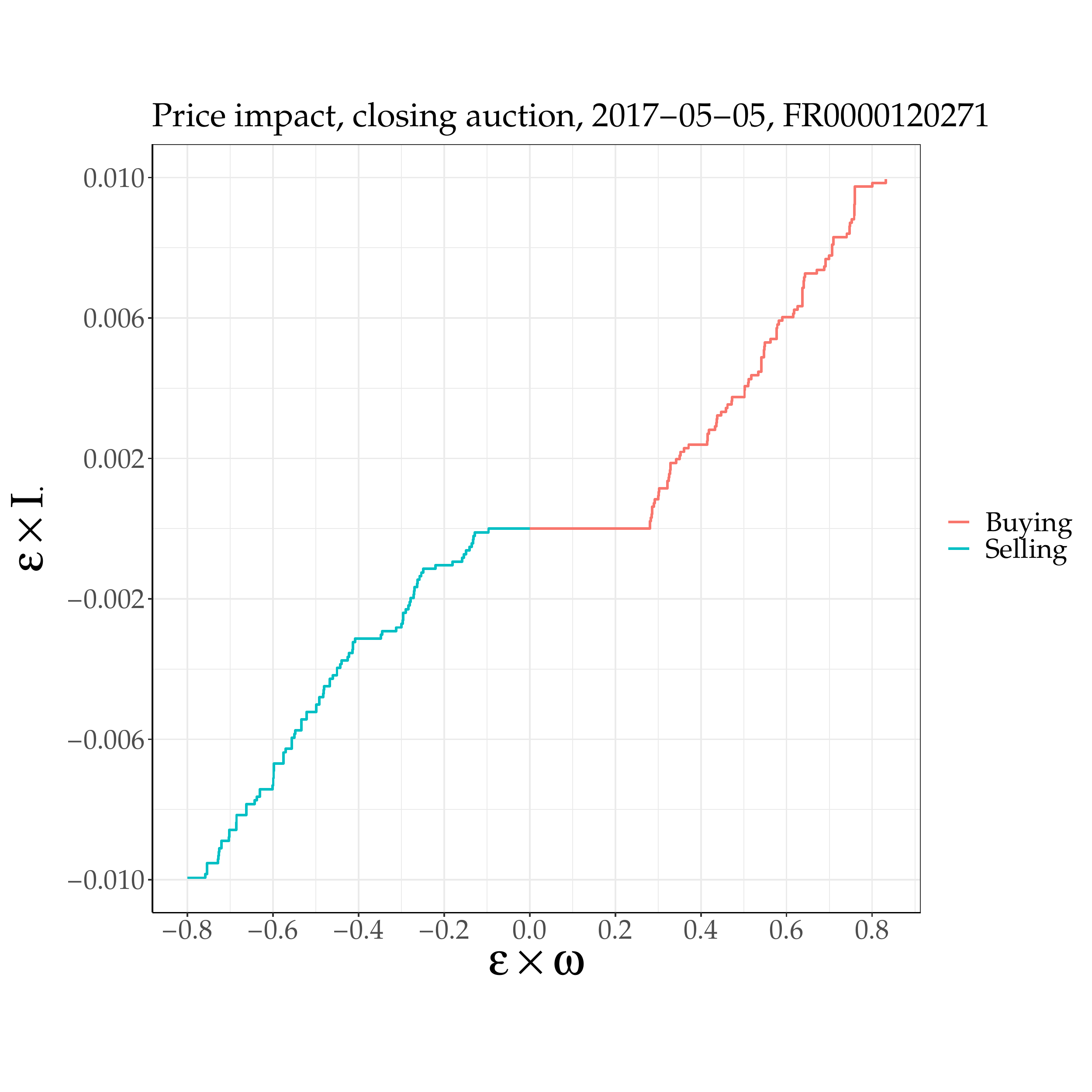}
    \caption{Virtual price impact $\varepsilon \cdot I$ as a function of the (scaled) added signed volume $\varepsilon \cdot \omega$ at the closing auction of TTE.PA on 2017-05-05. $\varepsilon = +1$ for a buy market order and $\varepsilon = -1$ for sell market order.}
    \label{fig:Price impact example}
\end{figure}

When inspecting the price impact function over several days and auctions, we observe that the minimal volume necessary to change the auction price ($Q_a \times \omega_{\bullet}^{(0)}$ using the notations of Proposition \ref{prop:T}), can be much larger than the typical volumes needed to impact the price further ($Q_a \times \delta \omega_{\bullet}^{(i)}$ , $i \geq 1$).
A compelling example is given by Figure \ref{fig:Price impact example}, which shows the price impact function for TTE.PA at the closing auction of May 5, 2017, with the following quantities: $p_a = 48.00$\euro, $Q_a = 2,246,617$, $\omega_B^{(0)} = 27.45\% $, and $\omega_S^{(0)} = 9.61\%$.
Hence, if sent just before $T_a$, a buy order of a cash volume lower than $Q_a \times \omega_B^{(0)} \times p_a = 29.6$ million\euro \text{ } would not have resulted in an auction price change. Similarly a sell order of a cash volume lower than $Q_a \times \omega_S^{(0)} \times p_a = 10.3$ million\euro \text{ } would have had zero impact.

 In our sample, zero price impact is present in more than 98\% of the total processed days and sides. This means that in more than 98\% of the time, sending one share, either on the buy or the sell side, will not change the auction price. In addition, and maybe more surprisingly, zero impact on both sides simultaneously is by far the most common situation. This comes from the fact that the prices are discrete and thus the cumulative buy and sell volumes $D(p)$ and $S(p)$ are step functions. At the auction price, these steps only overlap partially. To change the auction price, one needs to shift vertically either $D$ or $S$ in such a way that the overlap at the auction price disappears (see Figure \ref{fig:remV} for an illustration). Thus, zero price impact only disappears when both $V_B(p_a)$ and $V_S(p_a)$ only have one share at most at $p_a$.

Price impact can be zero for relatively large orders because of the peak of volume at $p_a$: recall that the (scaled) zero-impact volume $\omega^{(0)}_{\bullet}$ is the minimal volume needed to change the auction price; by the definition of $\omega^{(0)}_{\bullet}$ in Proposition \ref{prop:T}, having large matched buy and sell volumes at the auction price leads to large zero impact volumes on both sides (see Figure \ref{fig:remV}). This is confirmed empirically:  we report in Table \ref{tab:correls} the probability $\mathbb{P}_{1\%}$ to send a market order of size $q = 1\% \times Q_a$ just before the clearing without moving the closing price. For the stocks in our sample, this probability ranges from 46\% to 74\%. The randomization of the clearing time prevents fast agents from using their low latency to size their trades so as to have zero impact.

\begin{figure}
    \centering
    \includegraphics[scale=0.33]{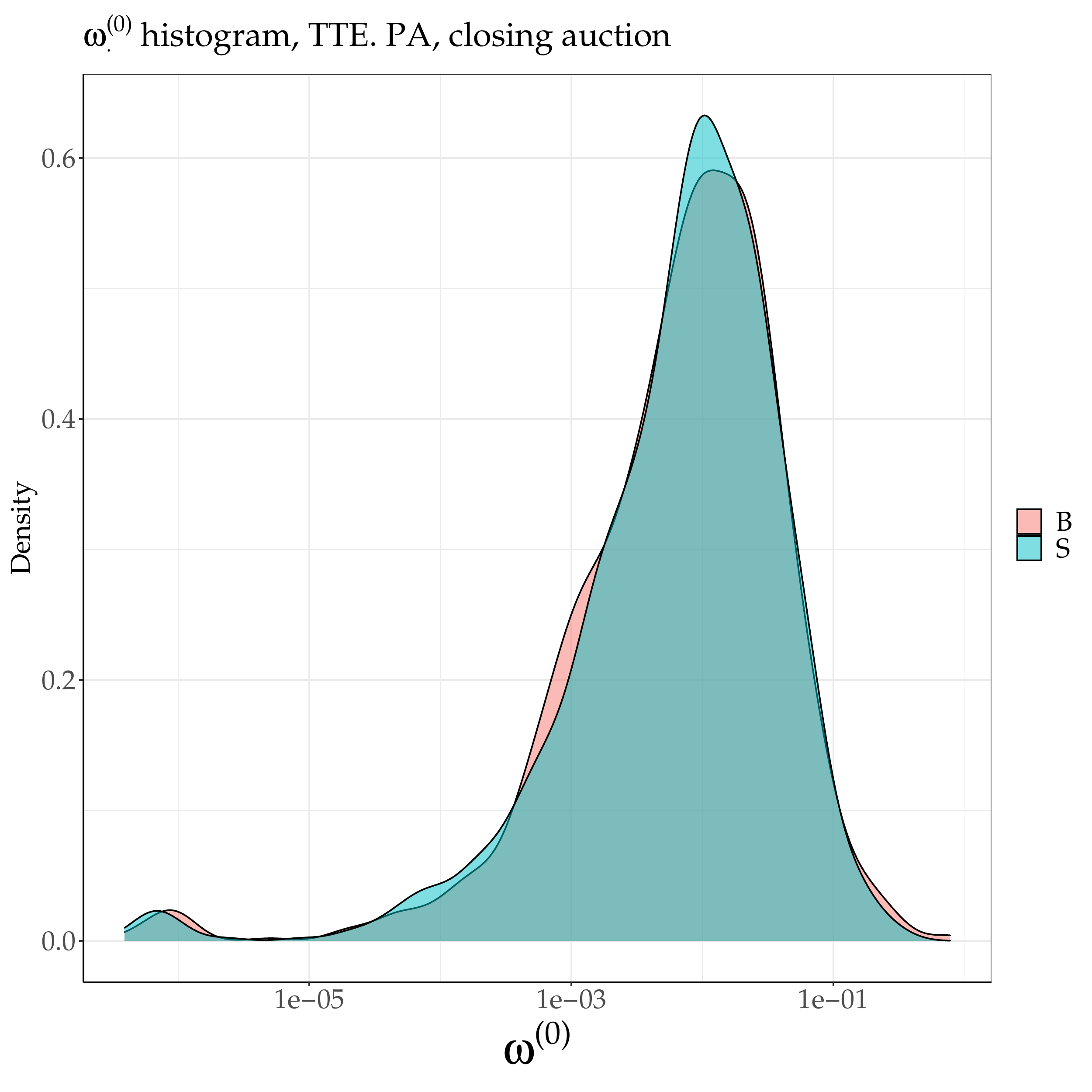}
    \includegraphics[scale=0.33]{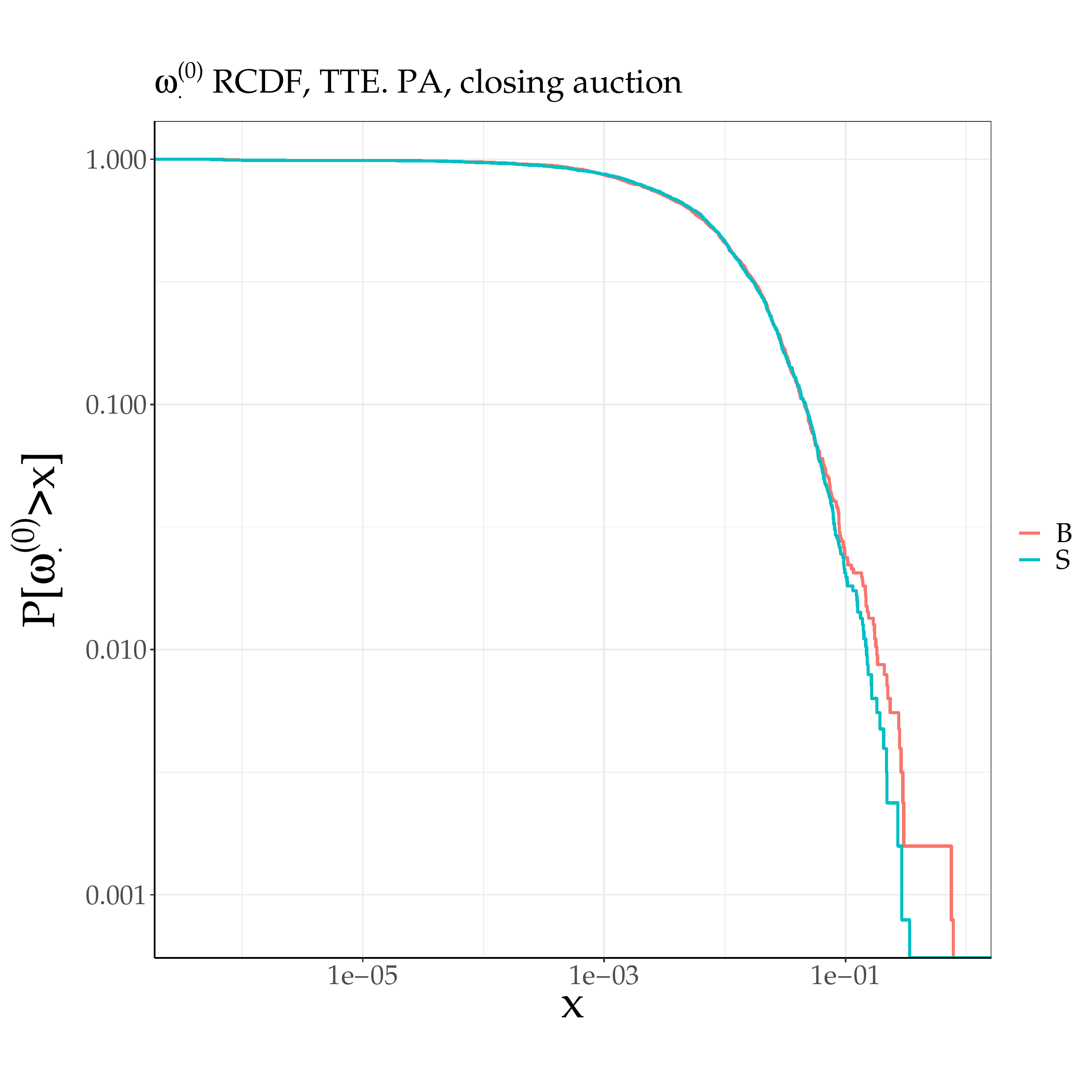}
    \caption{Left panel: smoothed histograms of zero impact volumes (buy and sell) $\omega_{\bullet}^{(0)}$; right panel: empirical reverse cumulative distribution function (RCDF) of zero impact volumes $\omega_{\bullet}^{(0)}$.}
    \label{fig:omega0_distrubtion}
\end{figure}

\begin{table}
\caption{Spearman correlation and Kolmogorov-Smirnov test statistics for $\omega_B^{(0)}$ and $\omega_S^{(0)}$, as well as the probability $\mathbb{P}_{1\%}$ to send a market order of size $q = 1\% \times Q_a$ without impacting the auction price just before the clearing across the stocks in our sample.}
\label{tab:correls}
\begin{ruledtabular}
\begin{tabular}{lcccr}
\multicolumn{1}{c}{ISIN}&\multicolumn{1}{c}{$\text{cor}(\omega_B^{(0)},\omega_S^{(0)})$}&\multicolumn{1}{c}{KS statistic$(\omega_B^{(0)},\omega_S^{(0)})$}&\multicolumn{1}{c}{$\mathbb{P}_{1\%}$}&\multicolumn{1}{c}{Observations}\tabularnewline
\hline
CH0012214059&-0.559***&0.088*&64\%& 510\tabularnewline
FR0000031122&-0.267***&0.056&67\%&1009\tabularnewline
FR0000045072&-0.321***&0.035&65\%&1014\tabularnewline
FR0000073272&-0.102**&0.049&66\%&1015\tabularnewline
FR0000120073&-0.210***&0.025&64\%&1014\tabularnewline
FR0000120172&-0.156***&0.045&66\%&1268\tabularnewline
FR0000120271&-0.048&0.022&46\%&1266\tabularnewline
FR0000120354&-0.152***&0.088***&70\%&1014\tabularnewline
FR0000120404&-0.089**&0.052&69\%&1015\tabularnewline
FR0000120537&-0.004&0.044&70\%& 504\tabularnewline
FR0000120578&-0.139***&0.054*&48\%&1268\tabularnewline
FR0000120628&-0.263***&0.043&64\%&1261\tabularnewline
FR0000120644&-0.166***&0.027&60\%&1012\tabularnewline
FR0000120685&-0.209***&0.044&68\%&1014\tabularnewline
FR0000121014&-0.272***&0.021&68\%&1014\tabularnewline
FR0000121147&-0.015&0.027&73\%&1013\tabularnewline
FR0000121261&-0.225***&0.027&65\%&1013\tabularnewline
FR0000121501&-0.311***&0.053&68\%&1014\tabularnewline
FR0000121667&-0.338***&0.021&69\%&1012\tabularnewline
FR0000121972&-0.095***&0.035&58\%&1264\tabularnewline
FR0000124141&-0.288***&0.026&73\%&1012\tabularnewline
FR0000125007&-0.14***&0.036&57\%&1265\tabularnewline
FR0000125338&-0.172***&0.042&68\%&1012\tabularnewline
FR0000125486&-0.132***&0.038&59\%&1013\tabularnewline
FR0000127771&-0.248***&0.04&68\%&1015\tabularnewline
FR0000130338&-0.098*&0.033&74\%& 613\tabularnewline
FR0000130809&-0.061*&0.032&56\%&1259\tabularnewline
FR0000131104&-0.128***&0.049&53\%&1264\tabularnewline
FR0000131708&-0.118**&0.042&68\%& 771\tabularnewline
FR0000131906&-0.075**&0.032&62\%&1269\tabularnewline
FR0000133308&-0.242***&0.043&67\%&1012\tabularnewline
FR0010208488&-0.271***&0.025&68\%&1010\tabularnewline
FR0013176526&-0.349***&0.065&57\%& 401\tabularnewline
NL0000235190&-0.096***&0.035&61\%&1264\tabularnewline
\end{tabular}
\end{ruledtabular}
  \begin{tablenotes}
   \item The symbols
***,**, and * indicate significance at the 0.1\%, 1\%, and 5\% level, respectively.
  \end{tablenotes}
\end{table}
We also report several statistical observations on $\omega_B^{(0)}$ and $\omega_S^{(0)}$. First, their statistical distribution can not be distinguished (as shown in Figure \ref{fig:omega0_distrubtion}). 
This is confirmed by a  Kolmogorov-Smirnov test reported in Table \ref{tab:correls}, which also reports the empirical Spearman correlation between these two quantities: quite surprisingly, given the observation above, the correlation between $\omega_B^{(0)}$ and $\omega_S^{(0)}$ is rather weak, $-0.15$ on average, and is non-significant for some very liquid stocks (e.g., TTE.PA the most traded stock in our dataset). This confirms that zero-impact is mostly a mechanical effect, not a strategic one. 

\begin{figure}
    \centering
    \includegraphics[scale=0.33]{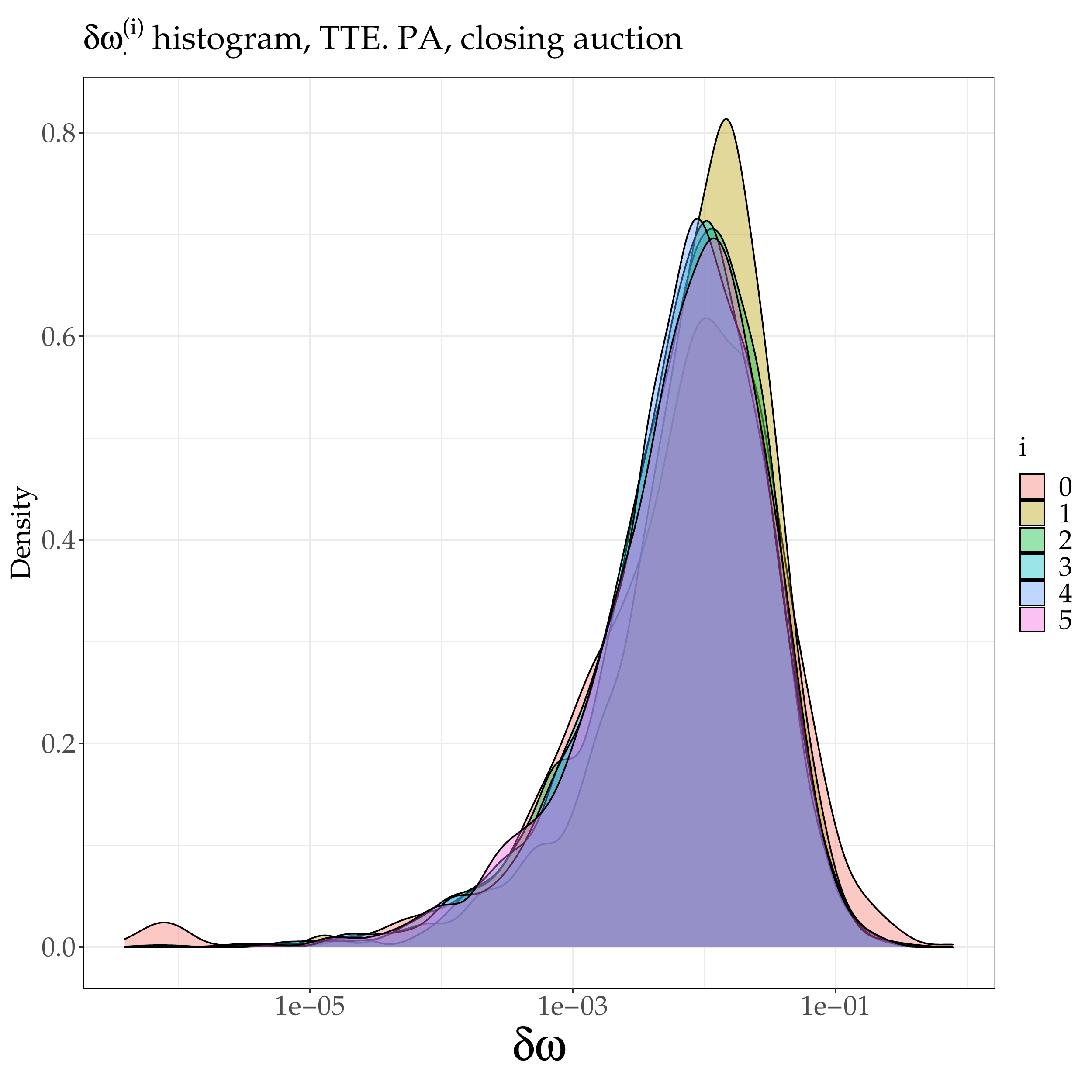}
    \includegraphics[scale=0.33]{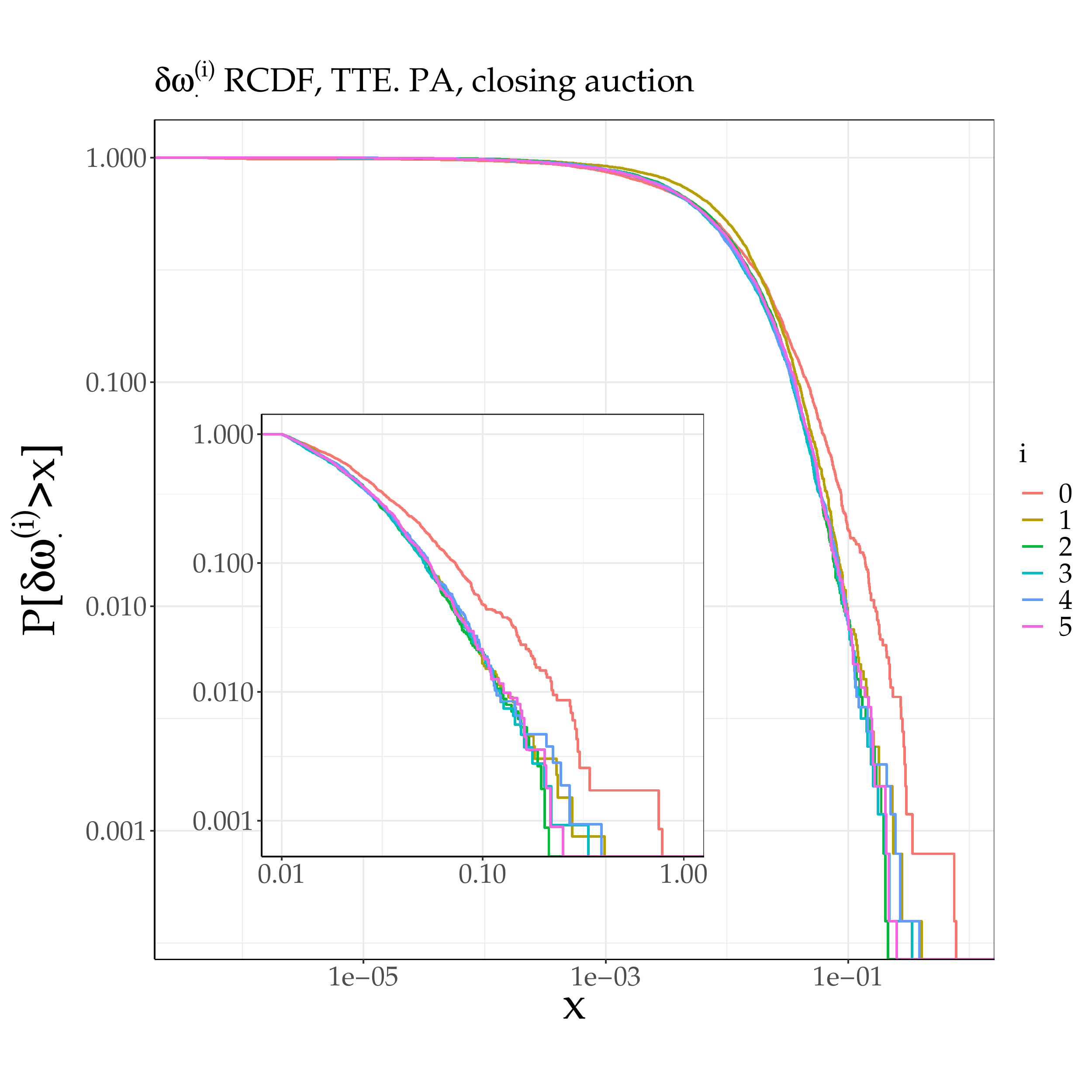}
    \caption{Left panel: smoothed histograms of scaled incremental volumes $\delta \omega_{\bullet}^{(i)}$ for $i = 0 \cdots 5$;  right panel: empirical reverse cumulative distribution function (RCDF) of scaled incremental volumes $\delta \omega_{\bullet}^{(i)}$ for $i = 0 \cdots 5$.}
    \label{fig:domegas_distrubtion}
\end{figure}

Let us finally compare $\delta\omega_{\bullet}^{(0)}=\omega_{\bullet}^{(0)}$, the minimal scaled volume needed to move the auction price, to $\delta \omega_{\bullet}^{(i)} =  \omega^{(i)}_{\bullet}  - \omega^{(i-1)}_{\bullet}$, $i\geq 1$, the minimal scaled volumes needed to take the price from $p_{\bullet}^{(i)}$ to $p_{\bullet}^{(i+1)}$ (see Proposition \ref{prop:T}). Table \ref{tab:domegas_etests} presents results for the stock TTE.PA of pairwise Kolmogorov-Smirnov tests on the empirical distribution functions of $\delta\omega_{\bullet}^{(i)}$ and $\delta\omega_{\bullet}^{(j)}$. For the sake of brevity, results are presented for $i,j \leq 10$, but the statistical testing has actually been conducted up to $i,j=40$.
\begin{table}
\caption{Kolmogorov-Smirnov statistics for pairs of rescaled incremental volumes $\delta \omega_{\bullet}^{(i)}$ and $\delta \omega_{\bullet}^{(j)}$ for the stock TTE.PA.}
\label{tab:domegas_etests}
\begin{ruledtabular}
\begin{tabular}{c|ccccccccccc}
\multicolumn{1}{c}{}&\multicolumn{1}{c}{$\delta\omega^{(0)}$}&\multicolumn{1}{c}{$\delta\omega^{(1)}$}&\multicolumn{1}{c}{$\delta\omega^{(2)}$}&\multicolumn{1}{c}{$\delta\omega^{(3)}$}&\multicolumn{1}{c}{$\delta\omega^{(4)}$}&\multicolumn{1}{c}{$\delta\omega^{(5)}$}&\multicolumn{1}{c}{$\delta\omega^{(6)}$}&\multicolumn{1}{c}{$\delta\omega^{(7)}$}&\multicolumn{1}{c}{$\delta\omega^{(8)}$}&\multicolumn{1}{c}{$\delta\omega^{(9)}$}&\multicolumn{1}{c}{$\delta\omega^{(10)}$}\tabularnewline
\hline
$\delta\omega^{(0)}$&&&&&&&&&&&\tabularnewline
$\delta\omega^{(1)}$&0.091***&&&&&&&&&&\tabularnewline
$\delta\omega^{(2)}$&0.047**&0.08***&&&&&&&&&\tabularnewline
$\delta\omega^{(3)}$&0.063***&0.103***&0.034&&&&&&&&\tabularnewline
$\delta\omega^{(4)}$&0.057***&0.105***&0.033&0.023&&&&&&&\tabularnewline
$\delta\omega^{(5)}$&0.053**&0.089***&0.02&0.026&0.028&&&&&&\tabularnewline
$\delta\omega^{(6)}$&0.055**&0.112***&0.04*&0.021&0.028&0.032&&&&&\tabularnewline
$\delta\omega^{(7)}$&0.068***&0.117***&0.044*&0.019&0.026&0.031&0.024&&&&\tabularnewline
$\delta\omega^{(8)}$&0.043*&0.086***&0.02&0.03&0.03&0.018&0.037.&0.037&&&\tabularnewline
$\delta\omega^{(9)}$&0.047**&0.091***&0.019&0.025&0.021&0.02&0.032&0.036&0.016&&\tabularnewline
$\delta\omega^{(10)}$&0.042*&0.081***&0.022&0.04*&0.037&0.022&0.043*&0.041*&0.016&0.023&\tabularnewline
\end{tabular}
\end{ruledtabular}
  \begin{tablenotes}
   \item The symbols ***, **, and * indicate significance at the 0.1\%, 1\%, and 5\% level, respectively.
  \end{tablenotes}
\end{table}
We clearly observe that $\delta\omega_{\bullet}^{(0)}$ and $\delta\omega_{\bullet}^{(1)}$ have specific statistical properties, while the distributions of the incremental volumes $\delta\omega_{\bullet}^{(i)}$ for $2 \leq i \leq 32$ could hardly be distinguished as the null hypothesis could not be rejected at the 1\% significance level. Figure \ref{fig:domegas_distrubtion} shows smoothed histograms and empirical reverse cumulative distribution function for $\delta\omega_{\bullet}^{(i)}$, $0 \leq i \leq 5$.
This observation is not easily generalized to all stocks since additional factors come into play: small tick vs. large tick stocks and the randomization of the clearing time. These factors have a non-negligible influence on the distribution of $\delta\omega_{\bullet}^{(i)}$s across different stocks and over the years.

\subsubsection{Linear impact: $\omega^{(0)}_{\bullet}<\omega<\omega^{(\max)}_{\bullet}$}
\label{sec:linear_impact}
According to \cite{donier2016walras}, in a Walrasian auction with continuous prices, average volumes around the auction price are non-null, which leads to a linear impact (in a first-order expansion), while in a continuous double auction, average volumes vanish around the current price and lead to a square root impact. 

It is useful to first assume that price is continuous in order to derive a simple condition for the price impact to be strictly linear. If we send a buy market order of size $\omega \times Q_a$ before the auction clearing, and assuming we work in a log-price frame of reference $x=\log(p/p_a)$ in a continuous price setting, we have
\begin{equation}
    \begin{cases}
    S(0) = D(0),
    \\
    S\left(I_B(\omega)\right) = D\left(I_B(\omega)\right) + \omega Q_a,
    \end{cases}
\end{equation}
hence,
\begin{equation}
    \label{eq:per_eq}
    S\left(I_B(\omega)\right) - S(0) = D\left(I_B(\omega)\right) - D(0) + \omega Q_a.
\end{equation}
\citet{donier2016walras} perform a first-order expansion to write
\begin{equation}
    \partial_x S(0) \times (I_B(\omega)-0) = \partial_x D(0) (I_B(\omega)-0) + \omega Q_a,
\end{equation}
and approximate
\begin{equation}
     I_B(\omega) = \frac{1}{\tilde{\rho}_S(0)+\tilde{\rho}_B(0)} \times \omega.
\end{equation}
However, instead, we use equation (\ref{eq:per_eq}) to find exactly
\begin{equation}
        \int_{0}^{I_B(\omega)} (\tilde{\rho}_S+\tilde{\rho}_B) (x)\mathrm{d}x = \omega,
\end{equation}
thus
\begin{equation}
\begin{aligned}
I_B(\omega) &= F^{-1}(\omega), \\
F(x)&= \int_0^x (\tilde{\rho}_S+\tilde{\rho}_B)(u) \mathrm{d}u.
\end{aligned}
\end{equation}

Having a linear impact requires that $F^{-1}$ and $F$ are linear functions, therefore that $x \mapsto (\tilde{\rho}_S+\tilde{\rho}_B)(x)$ is constant.

\begin{figure}
    \centering
    \includegraphics[scale=0.35]{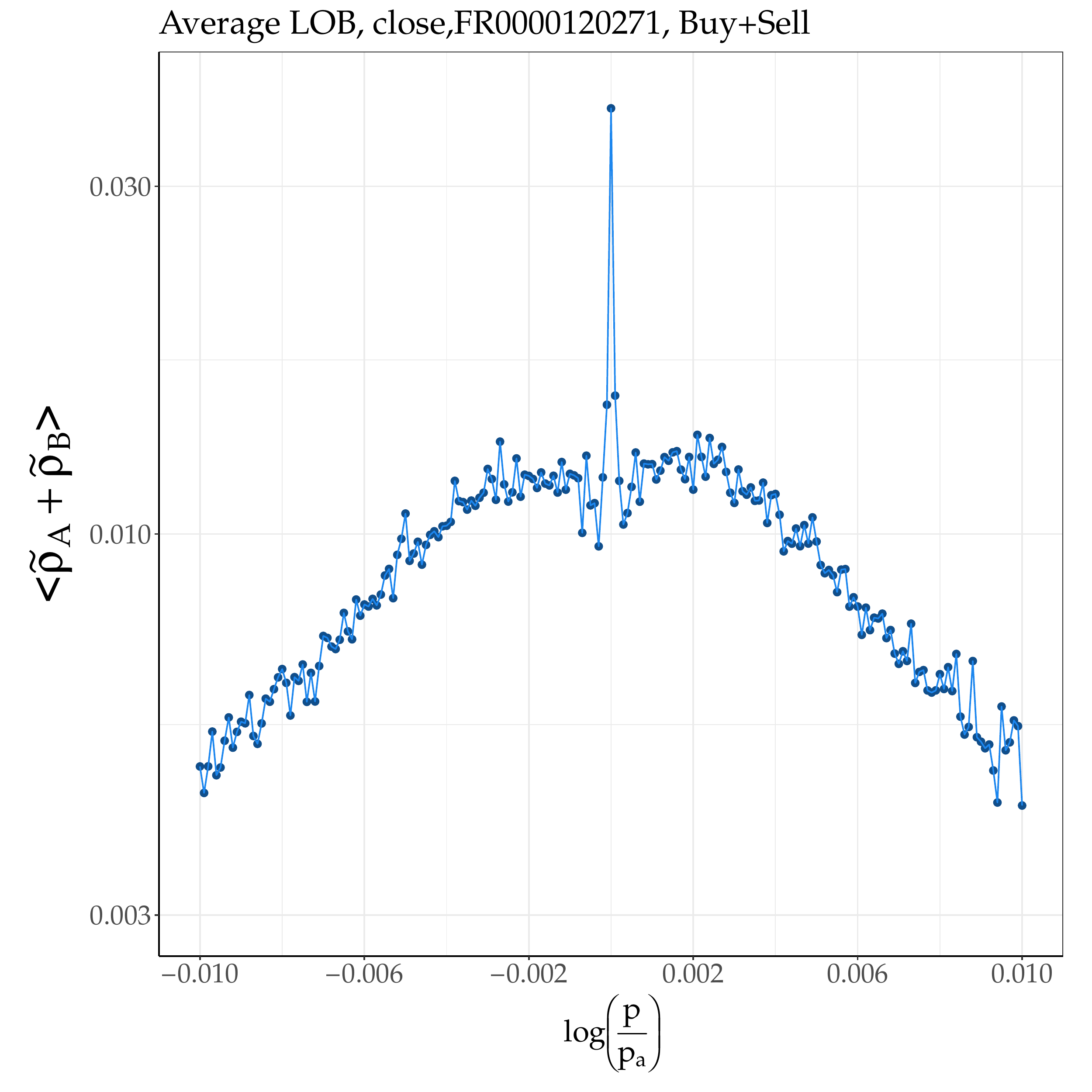}
    \caption{Average empirical density of total (buy + sell) volumes for TTE.PA at the closing auction.} 
    \label{fig:av_sum_BS}
\end{figure}

Figure \ref{fig:av_sum_BS} shows the average empirical density $\left< \tilde{\rho}_S+\tilde{\rho}_B \right>_d$ of the sum of buy and sell volumes for the most liquid stock in the sample.
It strongly suggests the existence of a price interval, on each side of the auction price, in which the sum of buy and sell volumes can be well approximated by a constant.

We now include this observation in the discrete-price theoretical framework introduced Section \ref{framework} and we prove in Proposition \ref{prop:lin imp} that if buy and sell densities sum up to a constant around the auction price $p_a$ (removing the zero-impact part),  price impact is linear.
\begin{proposition}
\label{prop:lin imp}
If $x \mapsto (\tilde{\rho}_S+\tilde{\rho}_B)(x)$ is constant on some intervals $]-\Delta_S, 0[$ and $]0, \Delta_B[$, then the price impact $I_{\bullet}$ is linear. More precisely, if $\tilde{\rho}_S(x)+\tilde{\rho}_B(x)= \widetilde{\mathcal{L}}_B$ positive constant for all $x\in]0, \Delta_B[$ and $\tilde{\rho}_S(x)+\tilde{\rho}_B(x)= \widetilde{\mathcal{L}}_S$  positive constant for all $x\in]-\Delta_S, 0[$, then for all $i$ such that $I(\omega^{(i)}) < \Delta$, we have
\begin{equation}
I(\omega^{(i)}) - I(\omega^{(0)}) =\frac{1}{p^{(1)}\widetilde{\mathcal{L}}} \left(\omega^{(i)} - \omega^{(0)}\right),
\label{eq:slopeifp}
\end{equation}
where we omitted the $\bullet \in \{B,S\}$ notation from $I$, $\omega$, $p^{(1)}$ and $\widetilde{\mathcal{L}}$. Recall that $p_{\bullet}^{(1)}$ is the first non empty price tick after (resp. before) the auction price when $\bullet = B$ (resp. when $\bullet = S$), as in Proposition \ref{prop:T}. 
\end{proposition}
The proof of Proposition \ref{prop:lin imp} is given in Appendix \ref{sec:proofs2}. Notice that $\widetilde{\mathcal{L}}$ represents a constant scaled liquidity around $p_a$. Also, since $\widetilde{\mathcal{L}}$ and $\omega$ are both scaled by $Q_a$, the price impact as written in the right-hand side of equation \eqref{eq:slopeifp} does not depend on the auction volume $Q_a$. For large-tick stocks, if $V_B(p) + V_S(p)=V_c$ constant around $p_a$, then the scaled liquidity is given by $\widetilde{\mathcal{L}}= V_c/(Q_a\theta)$, where $\theta$ is the tick size. For small-tick stocks, one can obtain an approximation by substituting $\theta$ with a fraction of the average spread.

\begin{figure}
    \centering
    \includegraphics[scale=0.33]{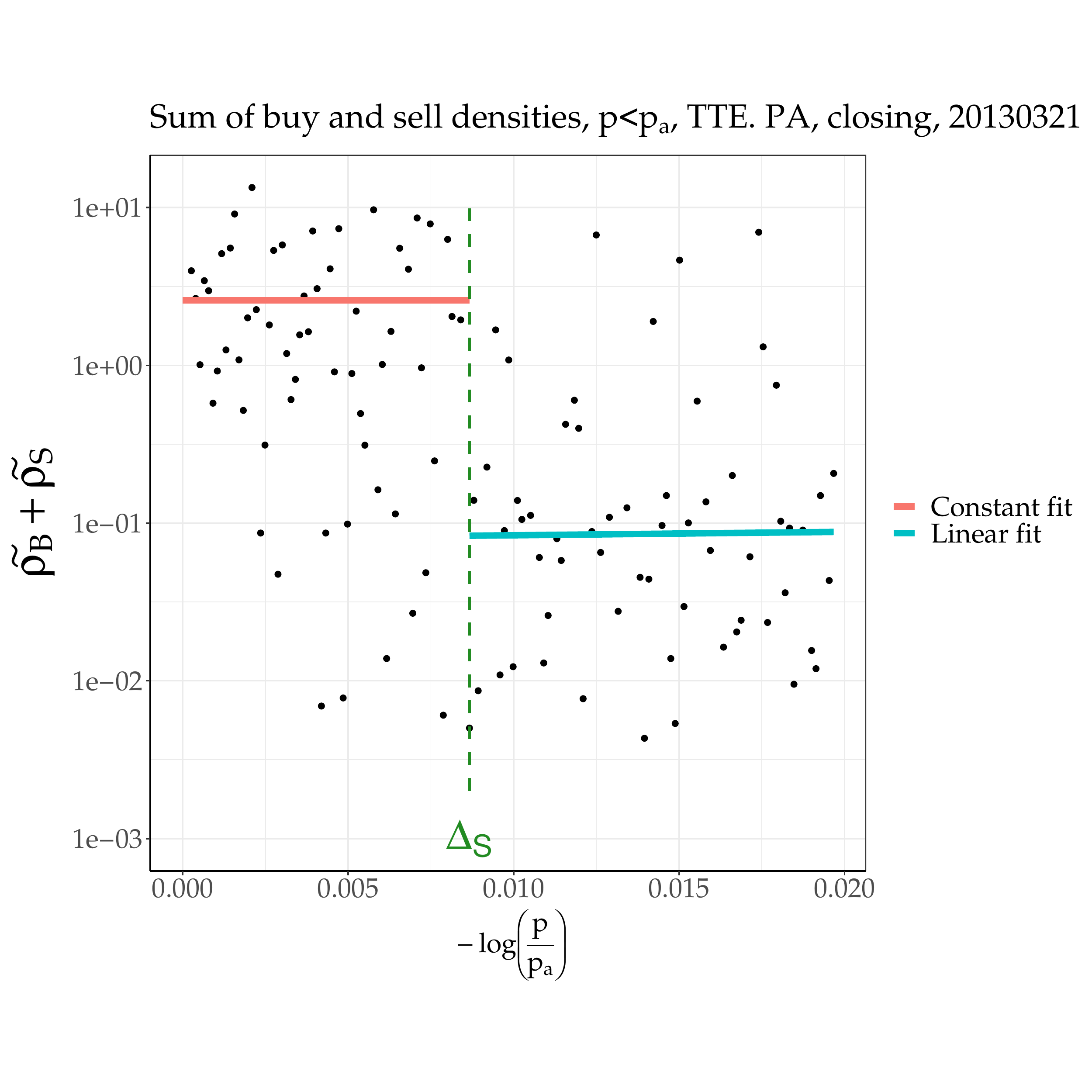}
    \includegraphics[scale=0.33]{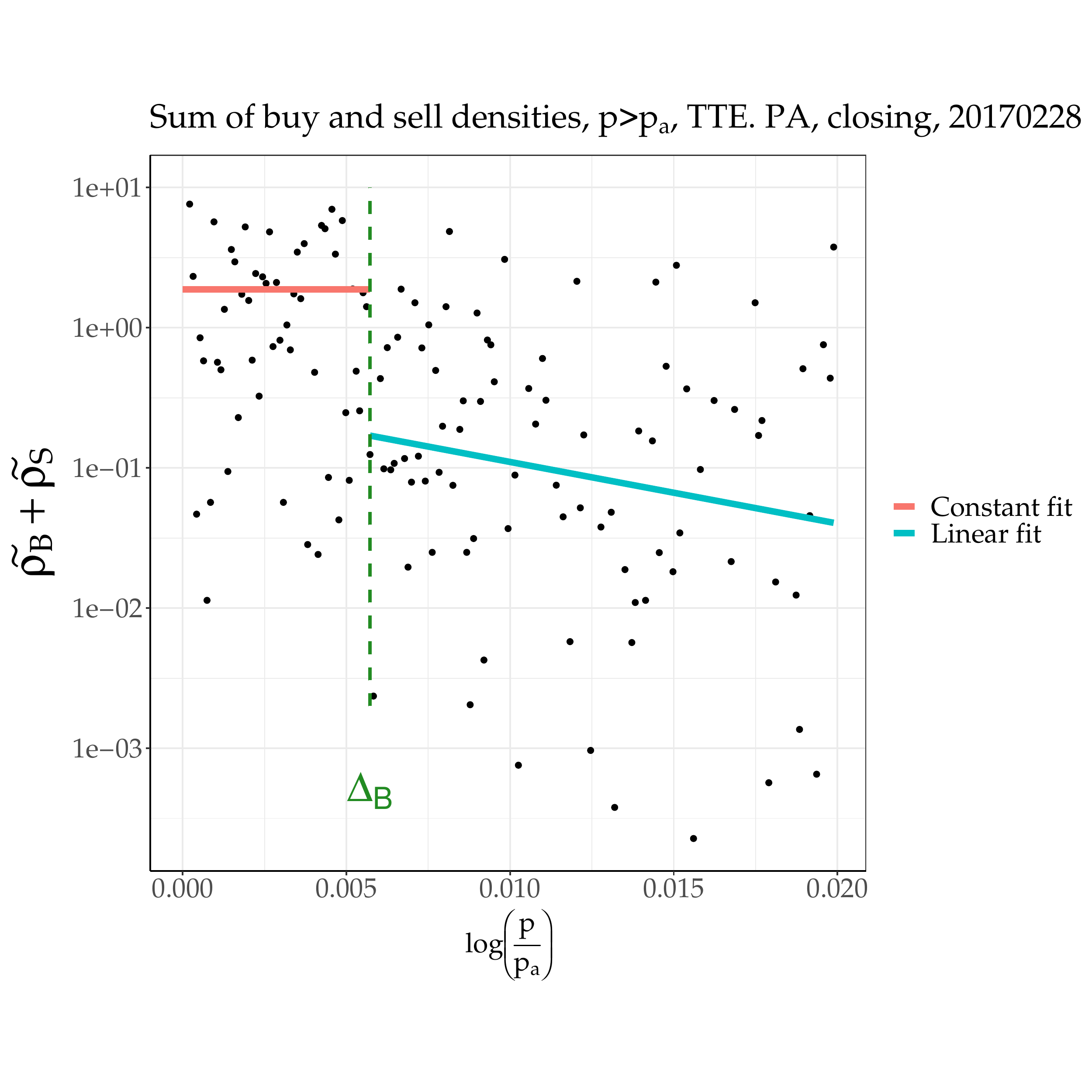}
    \includegraphics[scale=0.33]{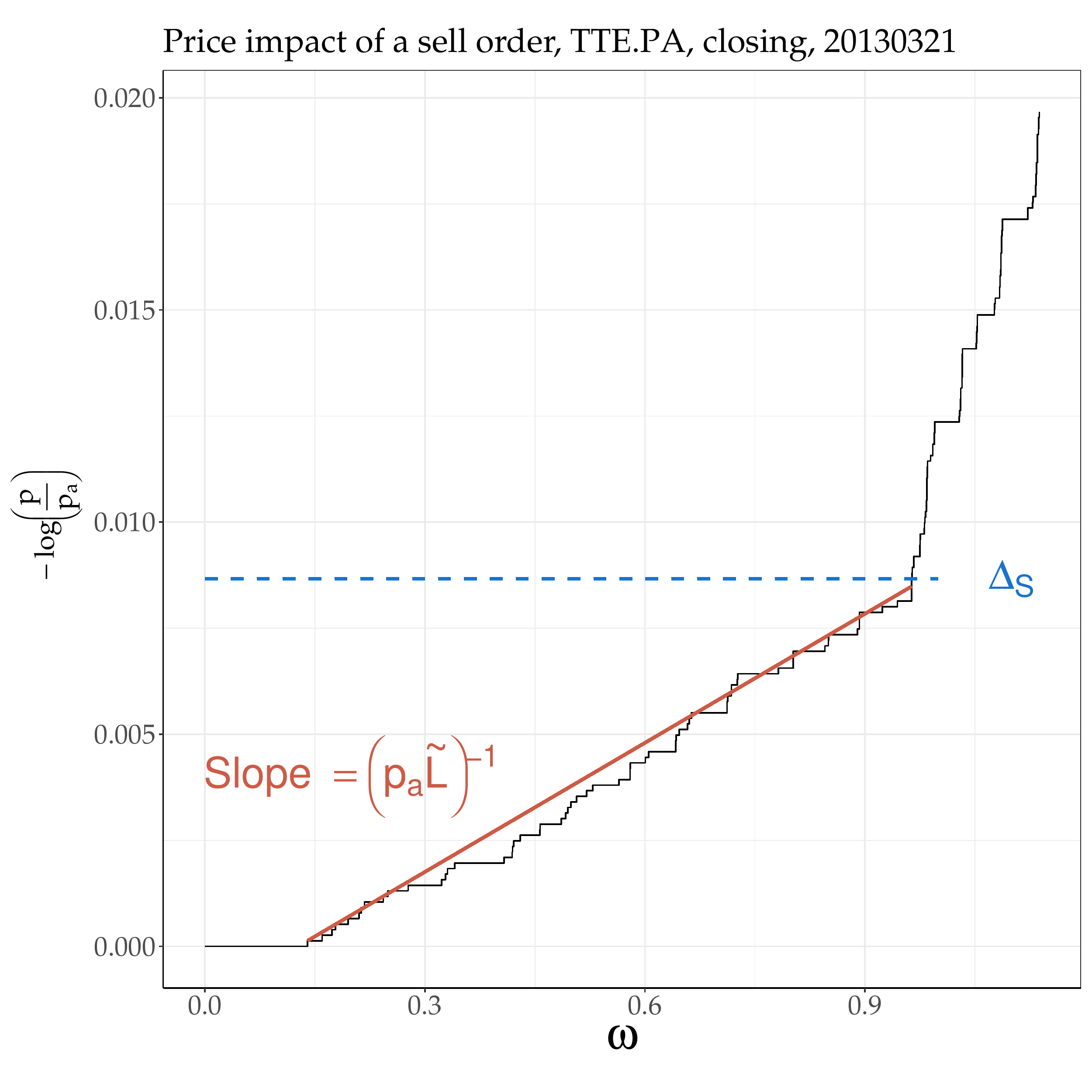}
    \includegraphics[scale=0.33]{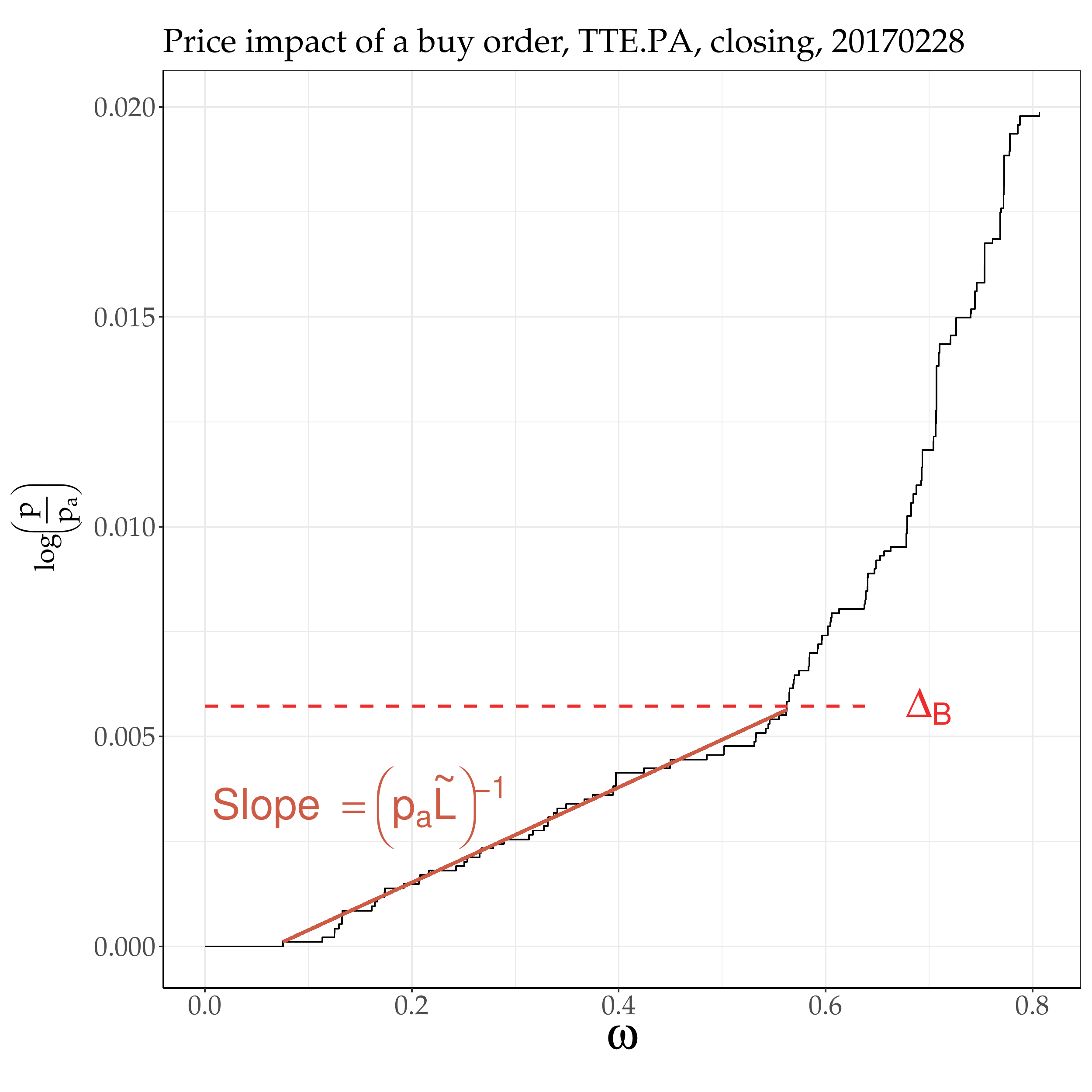}
    \caption{Simplified change point detection algorithm applied on the buy side of the closing auction of TTE.PA at 2013-03-21 (left) and the sell side of the closing auction of TTE.PA at 2017-02-28 (right). The upper plots show the sum of the buy and sell empirical densities and estimated cut-off $\Delta$ with a green dashed line. Lower plots show the fit of the estimated impact slope $(p^{(1)} \widetilde{\mathcal{L}})^{-1} \approx (p_a \widetilde{\mathcal{L}})^{-1} $ on the corresponding impact functions.}
    \label{fig:Examples}
\end{figure}

\begin{figure}
    \centering
    \includegraphics[scale=0.33]{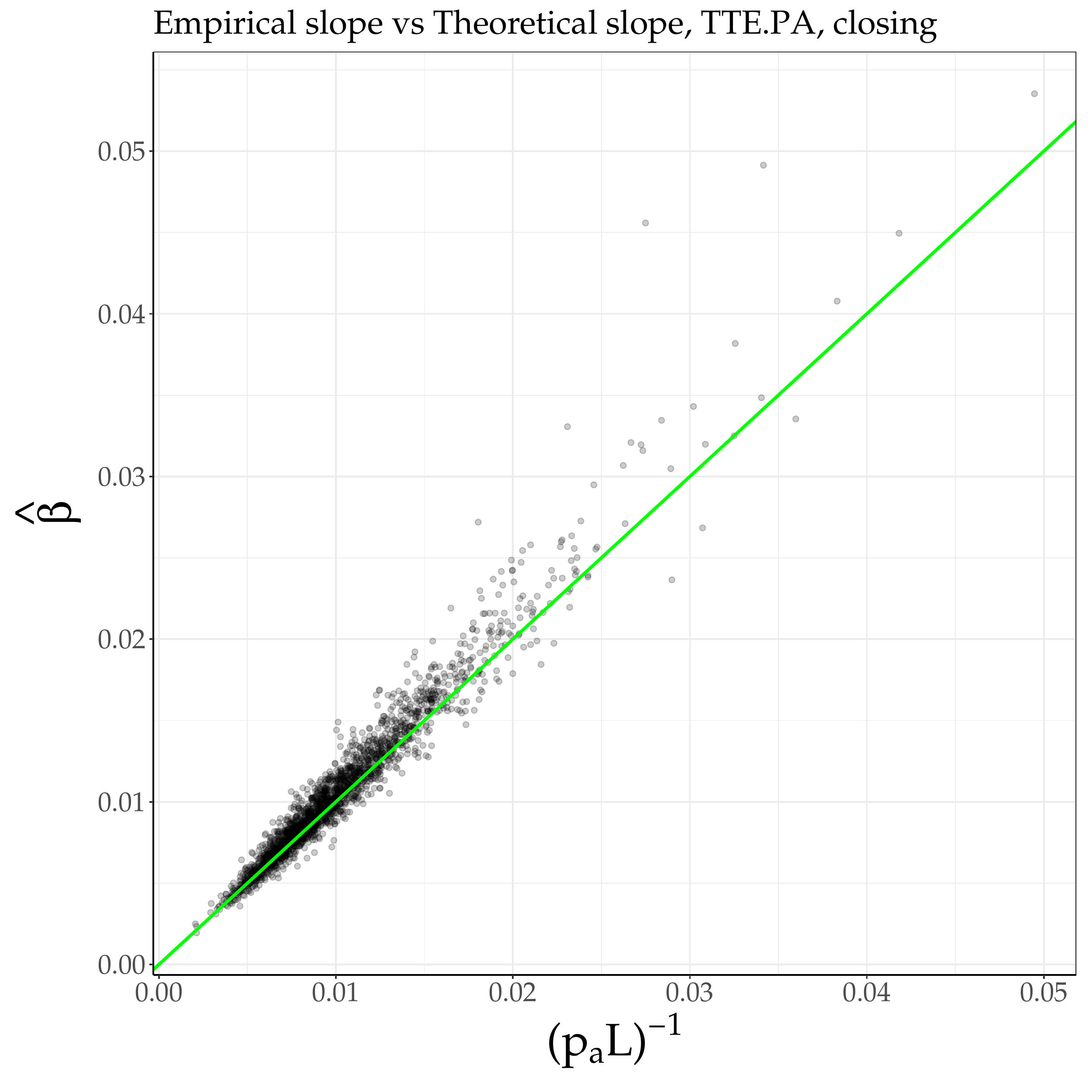}
    \includegraphics[scale=0.33]{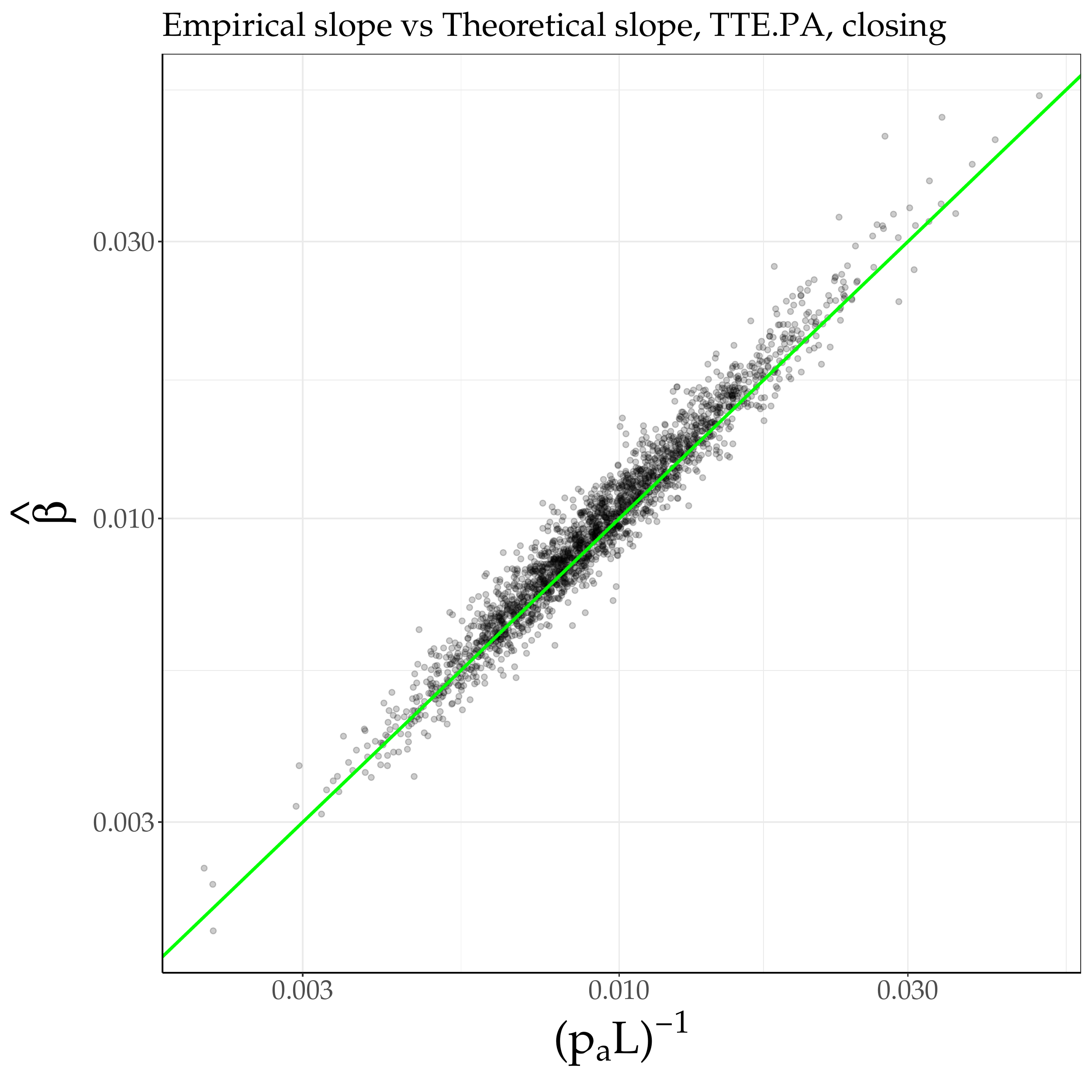}
    \caption{Empirical slope $\hat{\beta}$ vs Theoretical slope $(p^{(1)} \widetilde{\mathcal{L}})^{-1} \approx (p_a \widetilde{\mathcal{L}})^{-1} $ given by equation \eqref{eq:slopeifp} at the closing auction for TTE.PA. Left panel: normal scale. Right panel: log-log scale. A straight line with unit slope and null intercept is plotted in green for visual guidance.}
    \label{fig:Fslope_vs_Tslope} 
\end{figure}

Following Proposition \ref{prop:lin imp}, we want to characterize the intervals in which $\tilde{\rho}_S+\tilde{\rho}_B$ can be considered constant. Therefore we need to find $\widetilde{\mathcal{L}}_{\bullet}$ and $\Delta_{\bullet}$ for $\bullet \in \{B,S\}$, such that
\begin{equation}
\label{eq:cst}
\begin{aligned}
\tilde{\rho}_S(x)+\tilde{\rho}_B(x)= \widetilde{\mathcal{L}}_B & \text{ for all } x\in]0, \Delta_B[ ; \\
\tilde{\rho}_S(x)+\tilde{\rho}_B(x)= \widetilde{\mathcal{L}}_S & \text{ for all } x\in]-\Delta_S, 0[.
\end{aligned}
\end{equation}
For symmetry reasons, we focus on $\Delta_B$ and $\widetilde{\mathcal{L}}_B$:  the problem is to find $\Delta_B$ and $\widetilde{\mathcal{L}}_B$ for a given day $d$ by resorting to a simple change point detection algorithm. This method minimizes the residual sum of squared errors between $\log \left(\tilde{\rho}_S(x)+\tilde{\rho}_B(x) \right)$ and its mean $\eta(y)$ for $x \in ]0 , y]$ plus the residual sum of errors of a linear fit of $\log \left(\tilde{\rho}_S(x)+\tilde{\rho}_B(x) \right)$ for $x>y$. We choose to work with logarithms, since errors are multiplicative. The resulting cost function is
\begin{equation}
\begin{aligned}
    f(y) &= \sum_{0< x \leq y} \left| \log \left(\tilde{\rho}_S(x)+\tilde{\rho}_B(x) \right) - \eta(y)\right|^2 + \sum_{x>y} \left|\log \left(\tilde{\rho}_S(x)+\tilde{\rho}_B(x) \right) - \hat{\beta}(y) x - \hat{\alpha}(y) \right|^2; \\
    \eta(y) &= \frac{1}{N_y} \sum_{0< x \leq y} \log \left(\tilde{\rho}_S(x)+\tilde{\rho}_B(x) \right),
\end{aligned}
\label{eq:cost func cpt}
\end{equation}
where $(\hat{\alpha}(y),\hat{\beta}(y))$ is the linear regression estimate of $\log \left(\tilde{\rho}_S(x)+\tilde{\rho}_B(x) \right)$ over $x$ for $x>y$, and $N_y$ is the number of non-null observations $\left(x, \log \left(\tilde{\rho}_S(x)+\tilde{\rho}_B(x) \right) \right) $ for  $x \in ]0 , y]$.
We then define
\begin{equation}
    \Delta_B =\underset{y}{\arg\min} f(y).
\end{equation}
This definition means that for $x \leq \Delta_B$, the sum of the logarithm of the sum of scaled empirical buy and sell densities is better approximated by its mean than by a non-constant (linear) fit, whereas for $x > \Delta_B$, the opposite holds. Then, we calculate $\widetilde{\mathcal{L}}_B$ as the mean of $(\tilde{\rho}_S+\tilde{\rho}_B)(x)$ for $0<x\leq \Delta_B$ in order to avoid an underestimation due to the convexity of the exponential function. Finally, we define $\omega_{\bullet}^{(\max)}$ as the maximum scaled volume of a market order that would result in a null or linear impact, i.e.,
\begin{equation}
    \omega_{\bullet}^{(\max)} = \omega_{\bullet}^{(0)} + \sum_{0 < |x| \leq  \Delta_{\bullet} } \frac{V_B(x) + V_S(x)}{Q_a}.
\end{equation}
Figure \ref{fig:Examples} shows examples for $\Delta$ detection using the previous optimisation for two different days at the closing auction, and plots in each case the theoretical impact given by Proposition \ref{prop:lin imp} with respect to the actually observed impact function.
One  sees that the estimated cut-off $\Delta$, as well as the slope estimate $(p^{(1)} \widetilde{\mathcal{L}})^{-1} \approx (p_a \widetilde{\mathcal{L}})^{-1} $, fit very well the actual slope and domain of the linear price impact. This is actually the case of most days, as shown by Figure \ref{fig:Fslope_vs_Tslope}, where we plot the observed slope against the theoretical slope.

\begin{figure}
    \centering
    \includegraphics[scale=0.33]{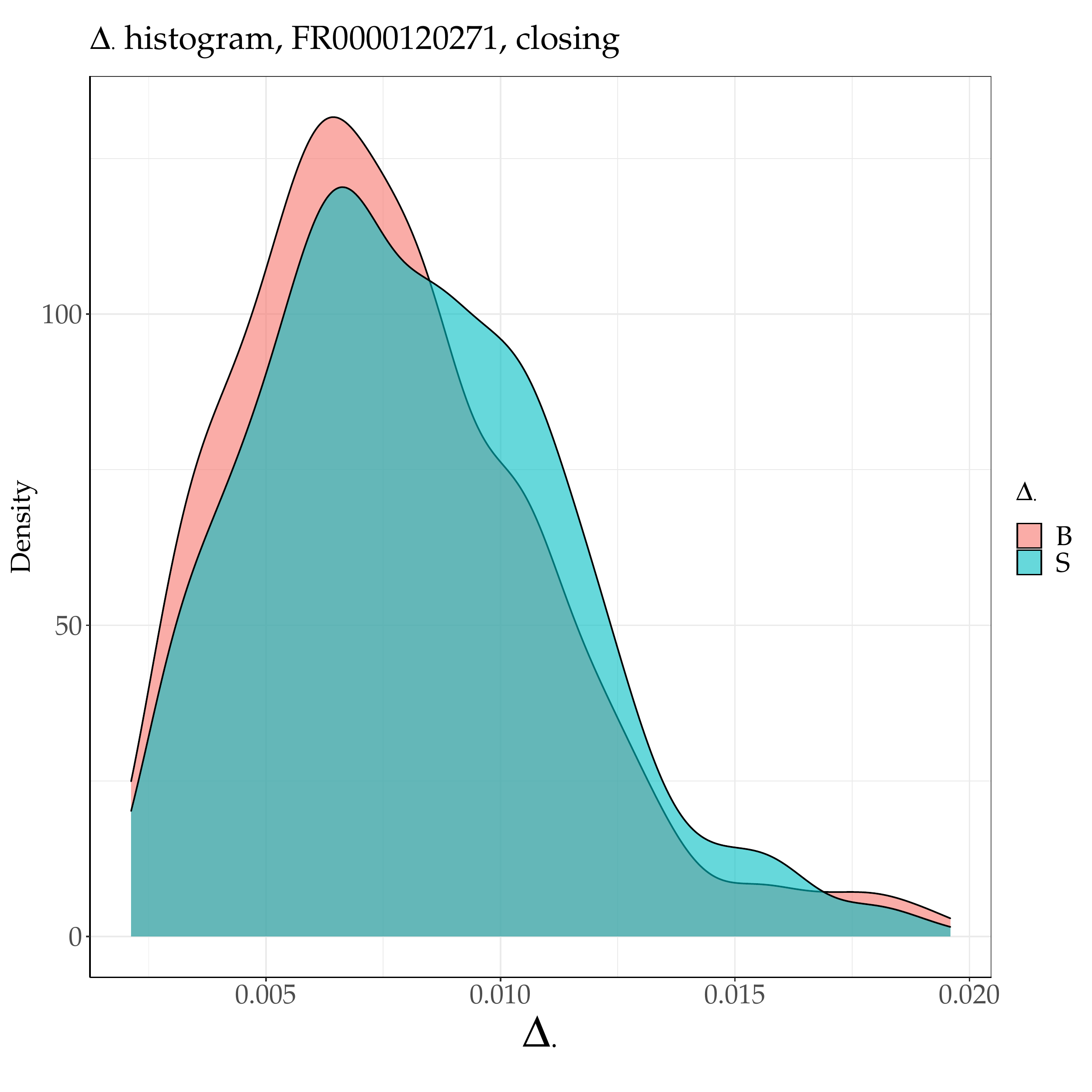}
    \includegraphics[scale=0.33]{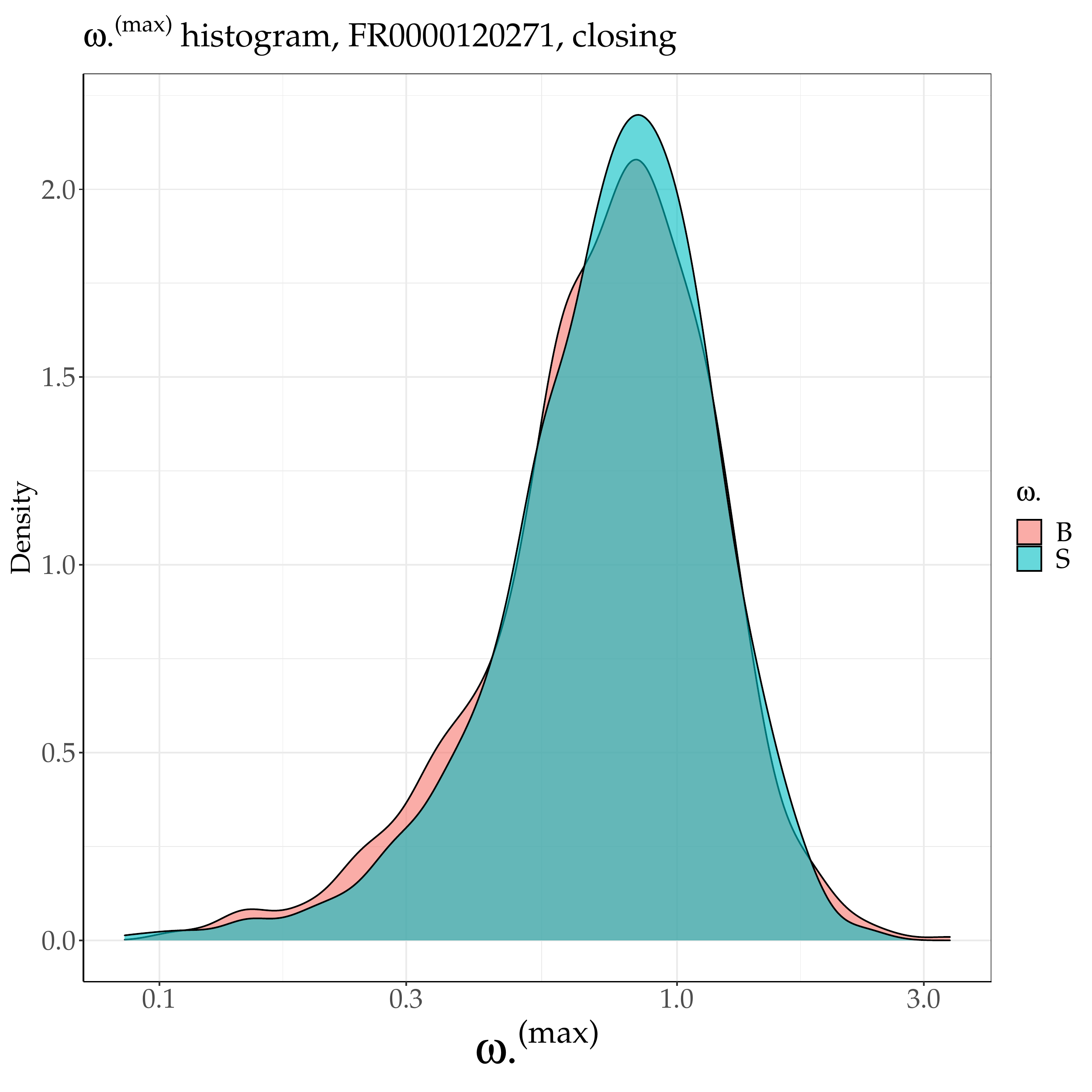}
    \caption{Smoothed histograms of the maximum log-distance $\Delta$ over which the sum of the buy and sell densities can be considered constant (left) and the maximum scaled volume $\omega^{(\max)}$  that results in a null or linear impact. These are outputs of the optimization of equation (\ref{eq:cost func cpt}) applied to closing auctions of TTE.PA between 2013 and 2017.}
    \label{fig:Delta and Vmax}
\end{figure}

We also plot the smoothed histograms of $\Delta$ and $\omega^{(\max)}$ issued by our detection algorithm for the stock TTE.PA between 2013 and 2017 ($1266$ stock-days and two sides (buy and sell)) (see Fig. \ref{fig:Delta and Vmax}) . 
Note that we truncated the closing auction snapshots at a maximum log-price distance $x \leq 2\%$, which is twice the average impact of a market order of a size equal to the auction volume $Q_a$. In addition, only fits with a number of points $\geq20$ are kept, which happens in about $\approx 90\%$ of the days and sides: this shows that the price impact is linear for most of the days with an average value of $\Delta$ above 50 basis points. Finally, $\mathbb{P}\left[\omega^{(\max)} > 0.5\right] = 0.73$: this means that a trader has 73\% chance to execute 50\% of the total auction volume just before the close clearing and still result in zero or linear impact.

Appendix \ref{sec:empirical_slopes} reports empirical properties of the impact slope at auction time computed for every asset, which may be of some use in transaction cost analysis.

\subsubsection{Influence of derivatives expiry dates}
 
\begin{figure}
    \centering
    \includegraphics[scale=0.33]{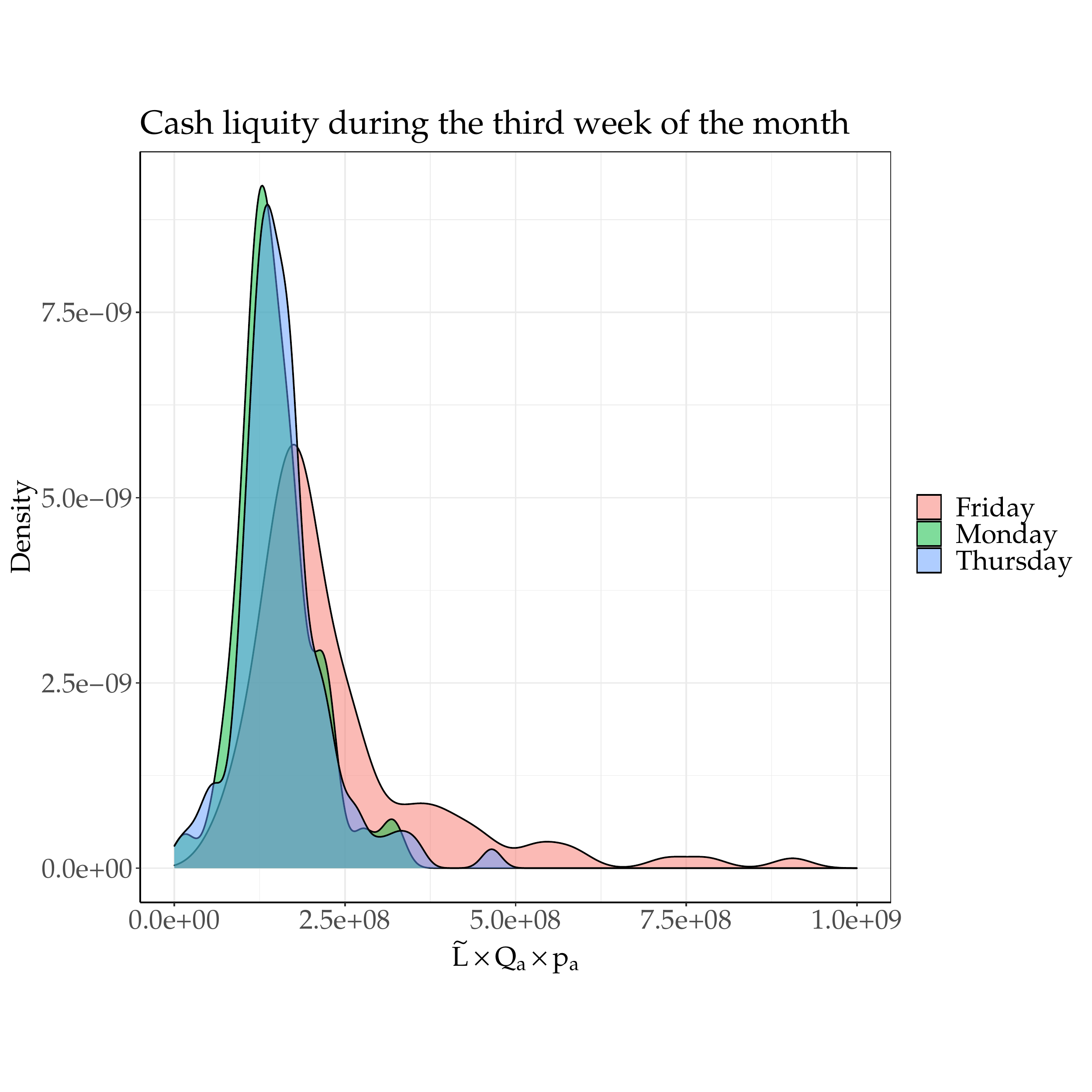}
    \includegraphics[scale=0.33]{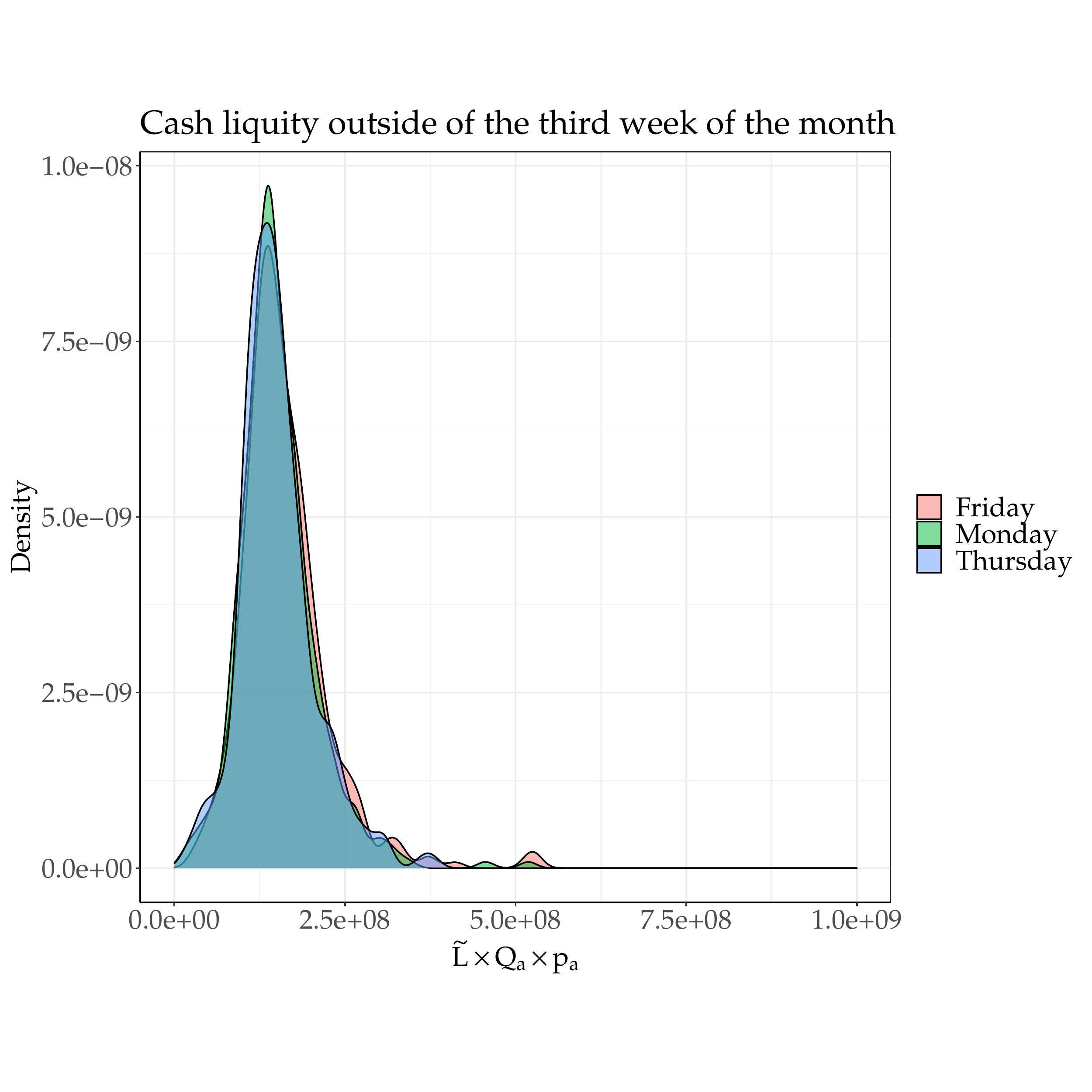}
    \caption{Smoothed histograms of the closing auction estimated cash liquidity $L^{\$} := \widetilde{\mathcal{L}} \times Q_a \times p_a$ during the third week of the month (left) and outside of the third week of the month (right).} 
    \label{fig:L_Va_ThirdWeek} 
\end{figure}

When there is no derivatives expiry, the liquidity in currency units defined by $L^{\$} := p_a \times Q_a \times \widetilde{\mathcal{L}}$ whether on Friday or other days of the week (Fig. \ref{fig:L_Va_ThirdWeek}, right panel) seem to be drawn from the same distribution, as we could not reject the null hypothesis associated with Kolmogorov-Smirnov tests for any pair of weekdays outside the third week of the month.
However, on expiry days (third Fridays of the month), liquidity in currency units is typically larger than for other weekdays during the same week and seems to be drawn from a different shifted distribution to the right (Fig. \ref{fig:L_Va_ThirdWeek}, left panel). This finding is confirmed by one tailed Kolmogorov-Smirnov tests for Friday and any other weekday during third weeks of a month. Therefore, the impact slope is typically smaller during expiry days and the final auction order book is more resistant to price changes.

\FloatBarrier

\subsection{Before the auction time}

In this second part, we study price impact before the auction time. First, we examine the evolution of virtual price impact throughout the accumulation period by looking at the evolution of liquidity as well as the maximum volume resulting in a linear impact. Second, we assume that traders have means to infer the impact slope at 17:35:00, which is the latest time that ensures not missing the clearing with certainty. We then relate zero impacts and the impact slope at 17:35:00 with those at the auction time. Finally, we study the average impact on the indicative price of actual submissions/cancellations between 17:30:30 and the auction time by means of response functions.

\subsubsection{Price impact evolution}

We investigate how the virtual price impact behaves throughout the accumulation period. We construct successive snapshots at 5-second intervals for TTE.PA at the closing auction. Then, we compute the (virtual/ instantaneous) price impact for $t \leq T_a$ with $p_a \leftarrow p_t^{\text{ind}}$ and $Q_a \leftarrow Q_t^{\text{ind}}$. We define the absolute liquidity $\mathcal{L}_t$ as the (constant) sum of buy and sell empirical densities at time $t$: $\mathcal{L}_t = (V_B+V_S)(t)/\delta p$ (Recall that the buy and sell densities sum up to a constant around the current indicative price). Similarly, we define $Q^{(\max)}_t$ as the maximum (absolute) volume that results in a null or linear impact time $t$.
\begin{figure}
    \begin{center}
        \includegraphics[scale=0.33]{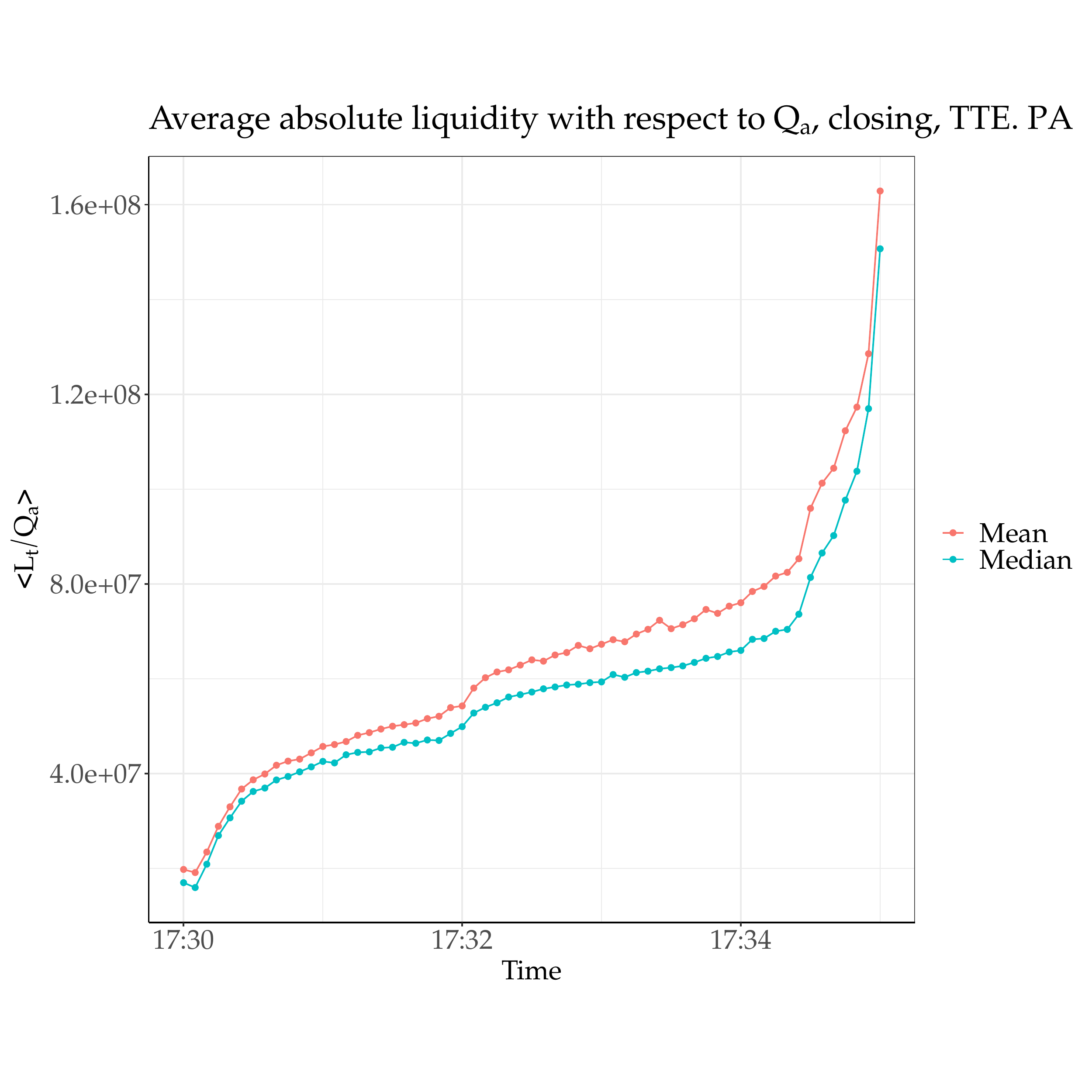}
        \includegraphics[scale=0.33]{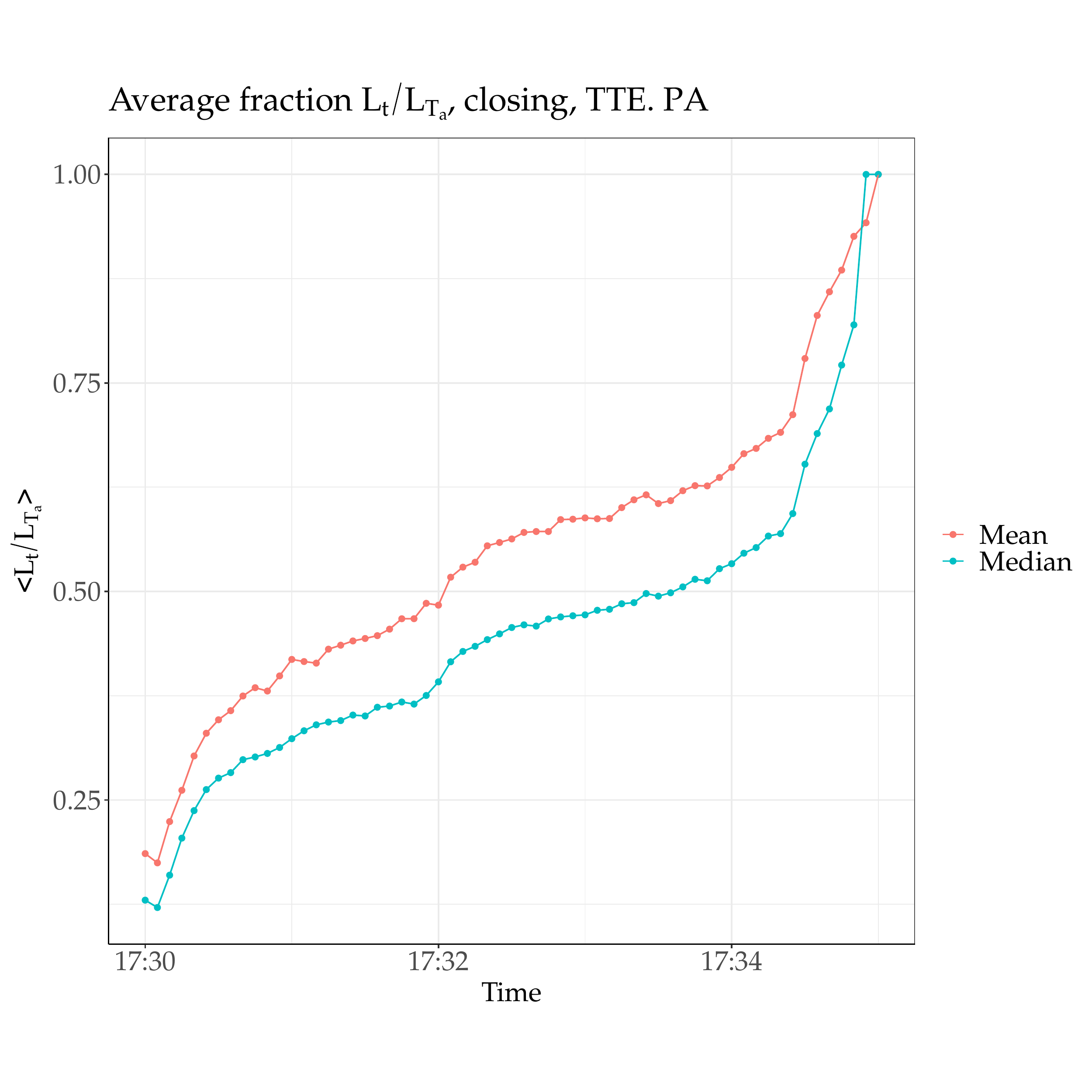}
        \includegraphics[scale=0.33]{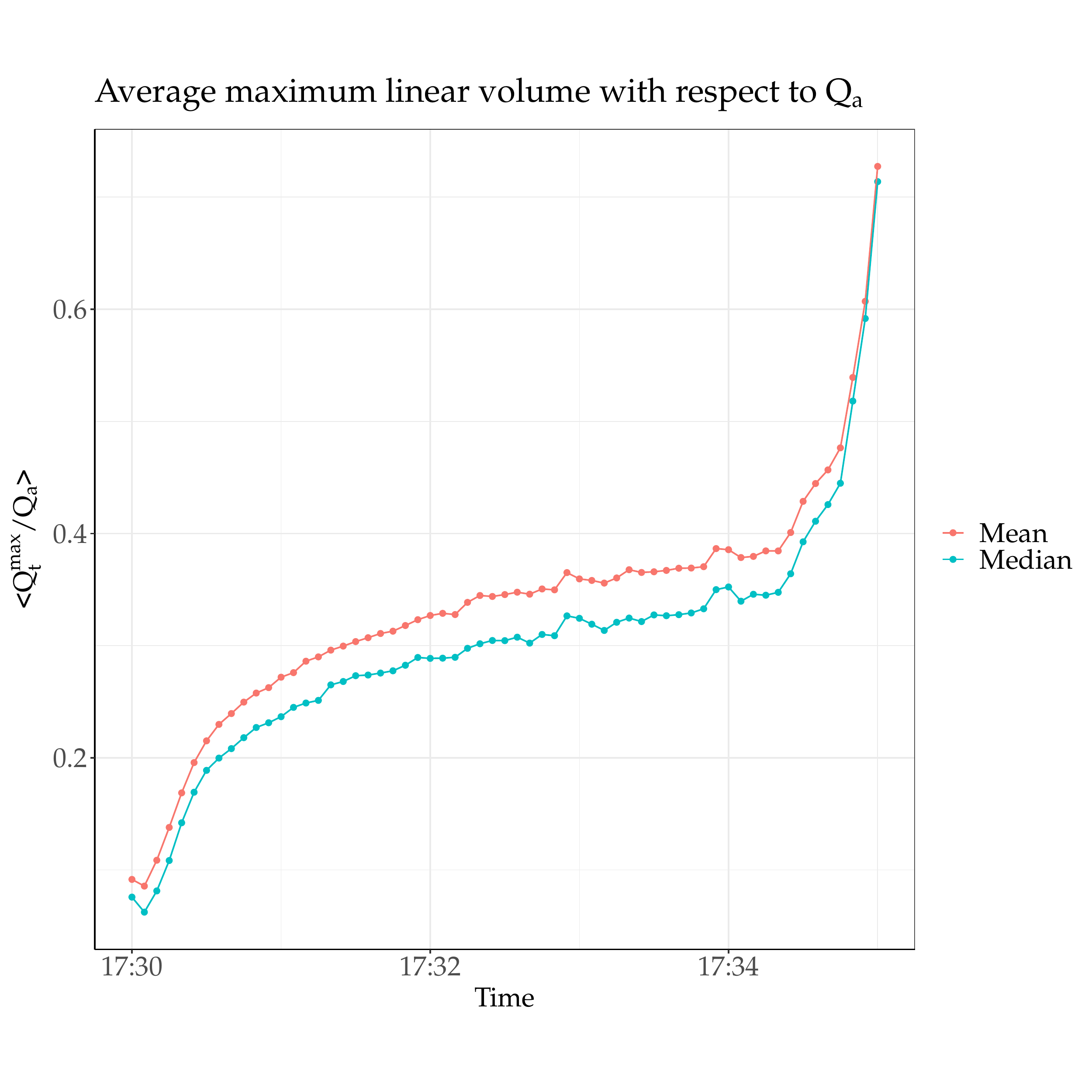}
        \includegraphics[scale=0.33]{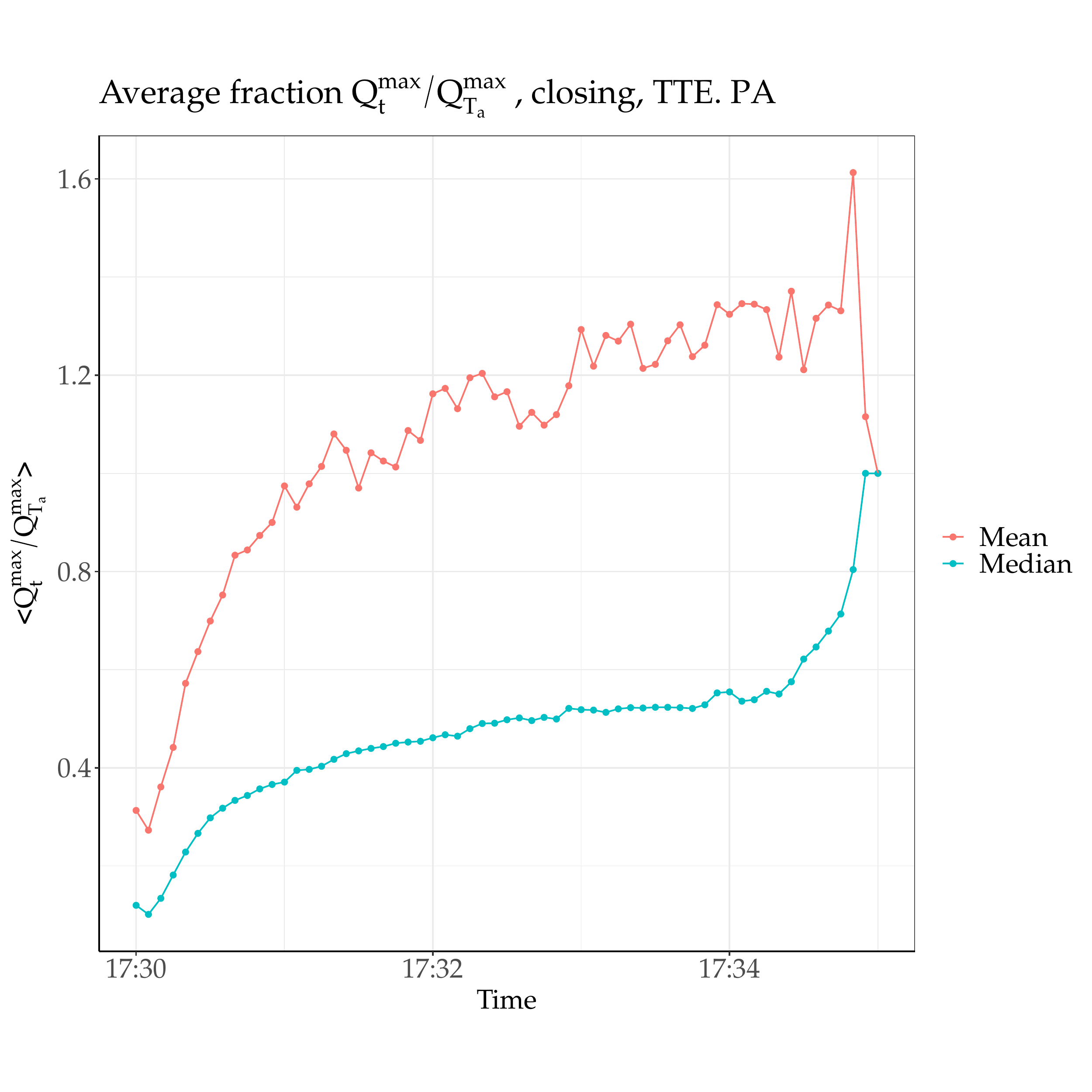}
    \end{center}
    \caption{Average absolute liquidity with respect to $Q_a$ during the closing auction (upper left). Average fraction $\left< \mathcal{L}_t/\mathcal{L}_{T_a} \right>$ of the absolute liquidity at time t $\mathcal{L}_t$ by final -at the clearing- liquidity $\mathcal{L}_{T_a}$ (upper right). Average (absolute) maximum linear-impact volume  with respect to $Q_a$ during the closing auction $\left<Q^{(\max)}_t/Q_a \right>$ (lower left). Average fraction $\left< Q^{(\max)}_t/Q^{(\max)}_{T_a} \right>$ of the absolute maximum linear-impact volume at time t $ Q^{(\max)}_t$ by final -at the clearing- absolute maximum volume $Q^{(\max)}_{T_a}$ (lower right).}
    \label{fig:Omega_mo Slope t}
\end{figure}
Figure \ref{fig:Omega_mo Slope t} shows that averages of both the absolute liquidity (w.r.t. to $Q_a$) $\mathcal{L}_t / Q_a$ and the fraction of the final liquidity $\mathcal{L}_t / \mathcal{L}_{T_a}$ follow the same pattern, i.e., a strong concave monotonicity at the start of the accumulation period followed by strong convex evolution as the clearing nears. Likewise, the average of the maximum linear volume with respect to the final volume $Q^{(\max)}_t/Q_a$  has the same shape. Nonetheless, the average mean of $Q^{(\max)}_t/Q^{(\max)}_{T_a}$ has a more complex pattern and suggests a strong effect of cancellations. 

\subsubsection{Impact at auction time vs. 17:35:00}

Let us now relate the virtual market impact at 17:35:00 and at auction time after the introduction of the randomized clearing time. Because the limit order book is not disseminated, traders have no direct way to estimate its shape, hence their virtual impact, at either time. However, sending a large market order and gradually cancelling it is a way around, and is observed at times. 

\begin{figure}
    \centering    \includegraphics[scale=0.33]{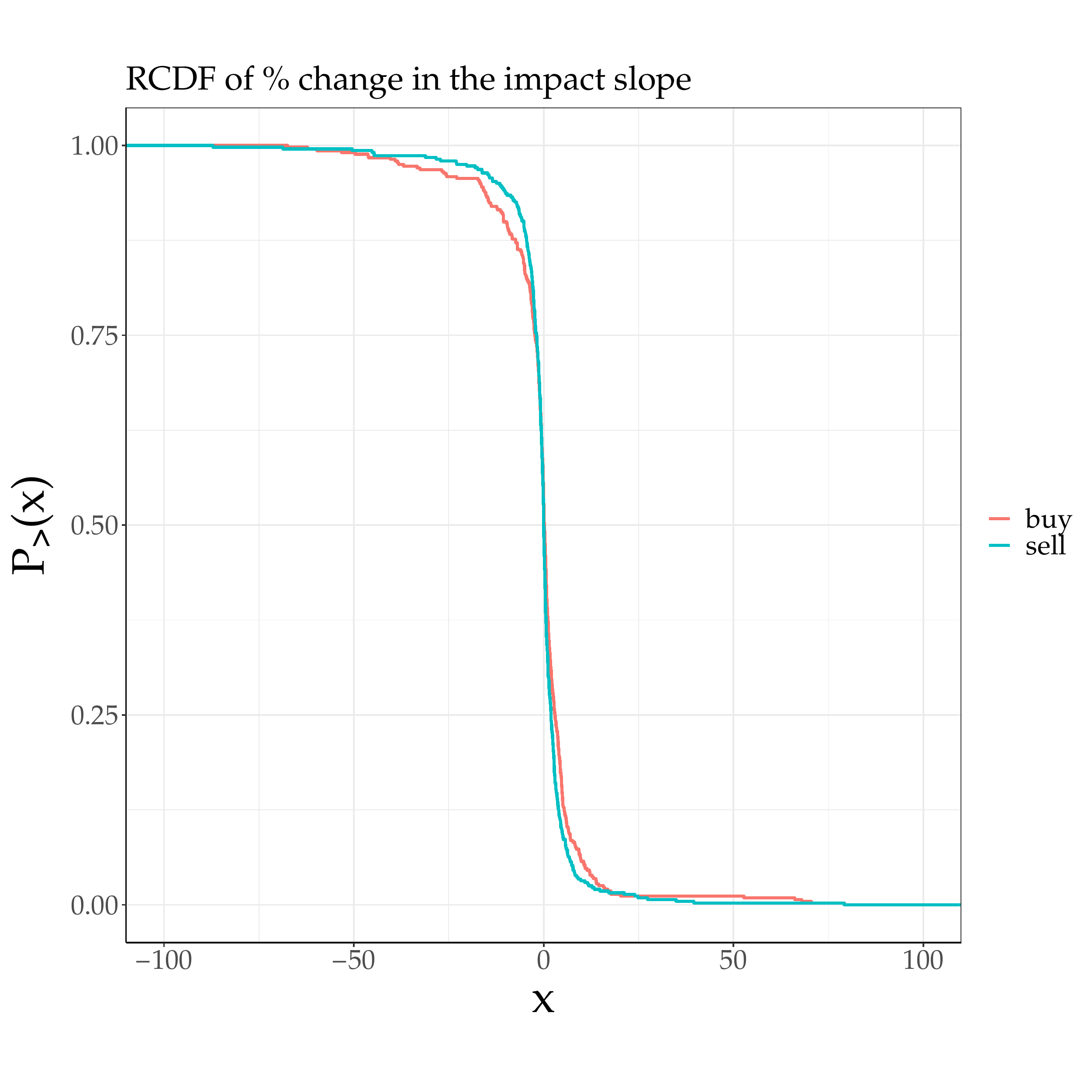}
    \includegraphics[scale=0.33]{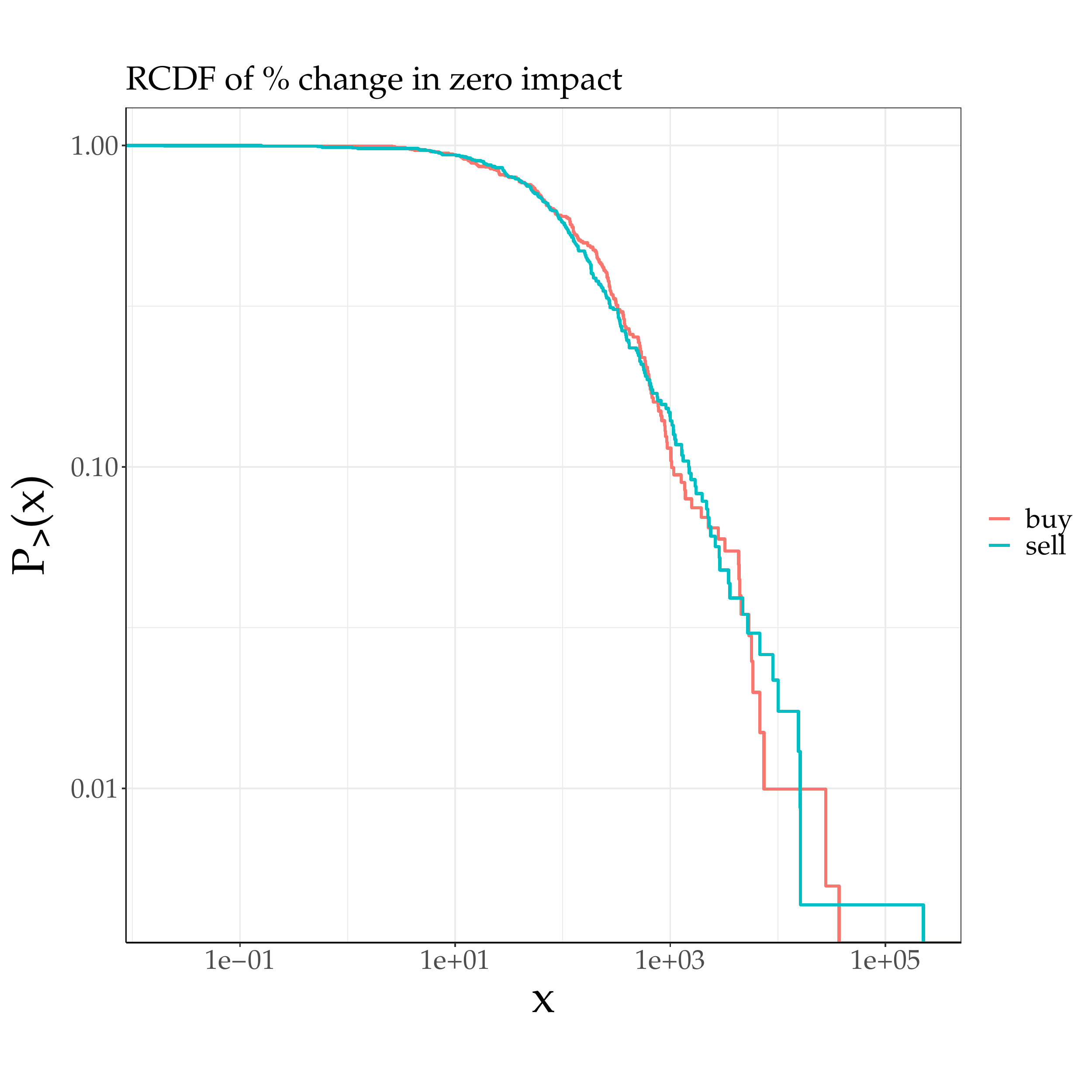}
    \caption{Empirical reverse cumulative distribution of the \% change in the linear impact slope (left panel), and the positive \% changes in zero impact (right panel) between 17:35:00 and the clearing after the introduction of a random clearing window between 17:35:00 and 17:35:30 at closing auctions. Negative changes in \% of zero impact are broadly distributed between 0\% and -100\%. Shown results are for TotalEnergies stock.}
    \label{fig:pct_change}
\end{figure}

The relationship between the two parts of price impact  (zero, then linear) at both times is markedly different. The relative change of zero impact volumes is distributed over several orders of magnitude  (see Fig. \ref{fig:pct_change}); agents do have an incentive to send zero-impact orders between 17:35:00 and the auction time. On the contrary, the slopes of the linear impact part are closely related: in 90\% of the days, the relative change in the impact slope is smaller than 12\% in absolute value (see Fig. \ref{fig:pct_change}). This means that the auction book stabilizes after 17:35:00 as one can expect since the clearing can occur at any time after 17:35:00. For TotalEnergies stock, the average absolute price change between 17:34:55 and 17:35:00 is 7 basis points. It is only 1.6 basis points between 17:35:00 and the auction time. 

\subsubsection{The linear impact of market order submission/cancellation before the auction}

Finally, we evaluate the average impact of actual submissions/cancellations during the accumulation period. To this end, we compute the one lag response function $R^1$ for marketable orders (market orders and limit orders with an aggressive limit price) conditional on the order (scaled) size $\omega$
\begin{equation}
    R^1(\omega) = \left<\varepsilon_t \cdot (p_{t+1}-p_t) | \omega\right>,
\end{equation}
where, $p_t$ is the indicative price just before the arrival of $t^{\text{th}}$ marketable order submission, $\varepsilon = +1$ for a buy, $-1$ for a sell, and the time is incremented at each marketable order submission. Additionally, we compute the one lag mechanical response function $R^M$ conditional on the order (scaled) size $\omega$
\begin{equation}
    R^M(\omega) = \left<\varepsilon_t \cdot (p_t^+ - p_t) | \omega\right>,
\end{equation}
where $p_t^+$ is the indicative price just after the marketable order arrival. 

\begin{figure}
    \centering
    \includegraphics[scale=0.33]{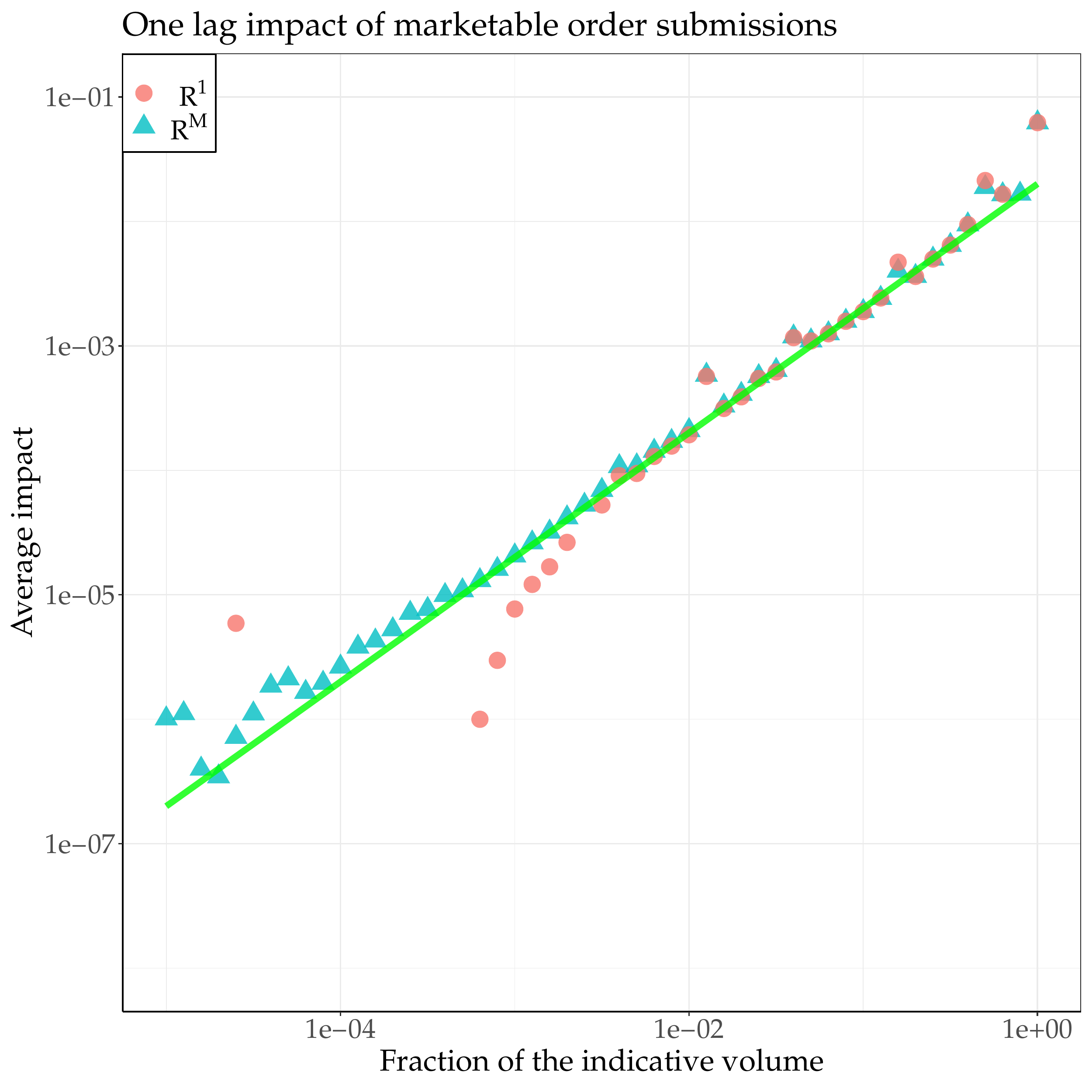}
    \includegraphics[scale=0.33]{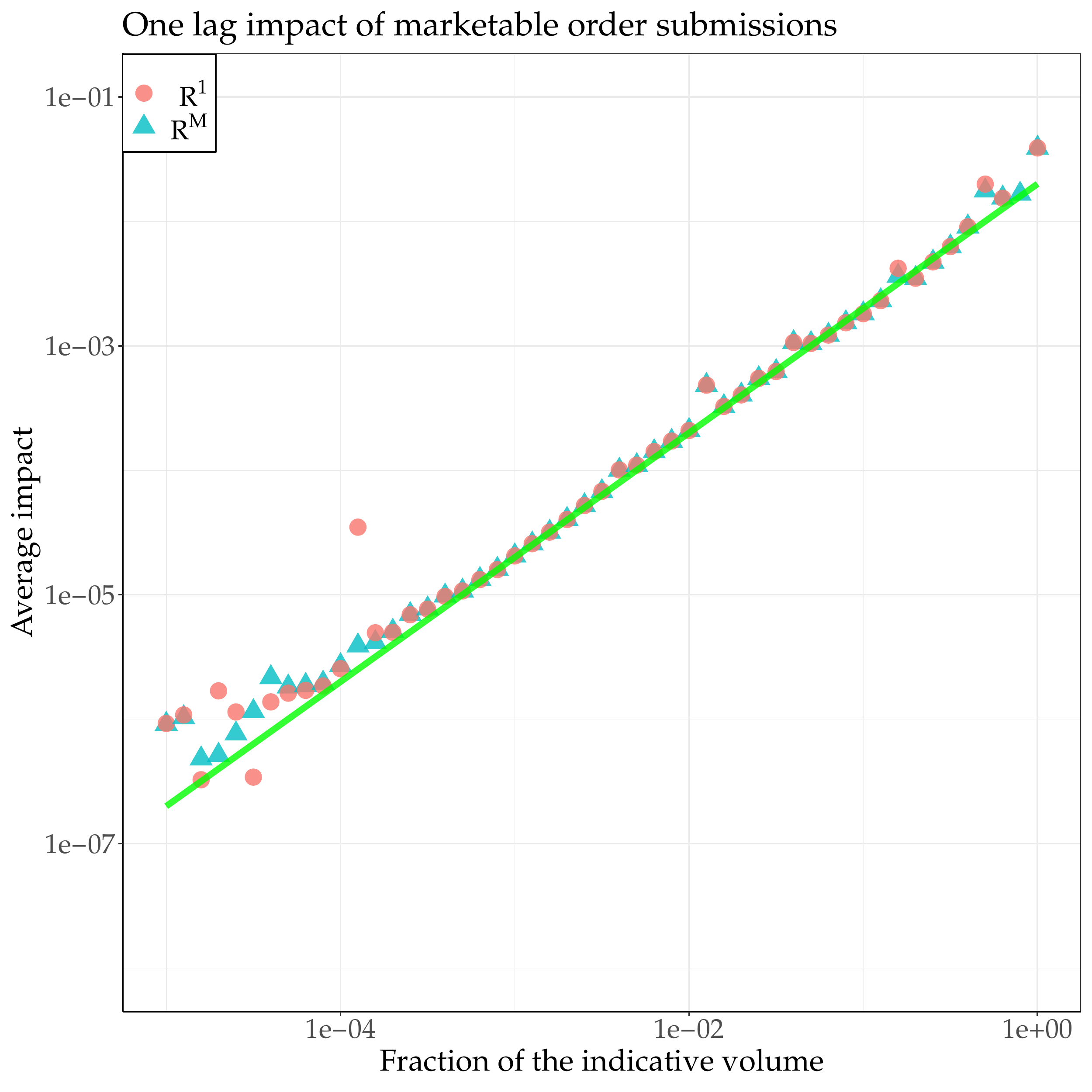}
    \caption{Left panel: the average one lag response function $R^1$ and the mechanical impact $R^M$ for marketable order submissions as a function of the scaled order size $\omega = V/Q^{\text{ind}}_t$; right panel: $R^1$ and $R^M$ for submissions and cancellations. We used tick-by-tick closing auction data from BEDOFIH for TotalEnergies stock between 2013 and 2017. We discarded the first 30 seconds of each auction as it contains abnormal submissions related to the activation of VFA/VFC orders (Valid For Auction/Valid For Closing). The green line is the curve of $y =0.02 \times x$. }
    \label{fig:av_imp}
\end{figure}

In contrast to open markets where $R^1$ is sub-linearly dependent on the volume \cite{lillo2003master,potters2003more, bouchaud2018trades}, we observe in Fig. \ref{fig:av_imp} that $R^1$ scales linearly with $\omega$ for marketable orders larger than a certain threshold $\omega^* \approx 3\cdot10^{-3}$. For $\omega<\omega^*$, values of $R^1$ can be negative indicating a strong mean reversion of the price, with values smaller than a tenth of a basis point in absolute value. For $\omega>\omega^*$, we have essentially $R^1 \approx R^M$ indicating that the price impact of individual orders is mostly mechanical and linear in $\omega$. As agents can not access the full order book during auctions, there is no selective liquidity taking: this is confirmed as $R^1$ and $R^M$ scale linearly with $\omega$. Incorporating marketable order cancellations with $\varepsilon = +1$ for sell cancellations and $-1$ for buy cancellations yields  $\omega^* \longrightarrow 0$, as we account for almost all price-changing events (right panel of Fig. \ref{fig:av_imp}).
These results imply that the nature of price impact is the same during the accumulation time and at the auction time, and contrasts with results for open markets, where selective liquidity taking causes very different shapes between the average virtual impact (using the instantaneous shape of the book) and market impact of actual trades \cite{weber2005order, bouchaud2009markets}.

\section{Conclusion}

The discrete nature of prices in limit order books mechanically causes the price impact at auction time to be zero at first, sometimes for quite a substantial fraction of the total exchanged volume. Surprisingly, zero price impact happens most of the time simultaneously on both sides of the auction book, for additional sell and buy market orders or equivalently for cancellations of buy or sell market orders. For volumes larger than zero-impact ones, price impact at auction time is linear in a limited price range around the auction price not only on average but for more than 90\% of days. The theoretical work of \citet{donier2016walras} shows the linearity of the auction impact locally around the auction price using a first-order expansion and under strong regularity assumptions of supply and demand in a continuous price setting. Here, we showed that the linearity of auction impact is due instead to the fact that the sum of buy and sell volumes around the auction price is constant.

While this work mainly describes the final result of the order accumulation process and characterizes the limit order book at the auction time, a more microscopic description of the dynamics of order submission, cancellation, and perhaps diffusion (price update) is needed. Even though market orders submitted during the accumulation period do not play a significant role in shaping the price response of the final limit order book, the action-reaction game between market orders and limit orders throughout the auction \cite{raillon2020growing, euronext2021} is probably a major driver of its dynamics. Similarly, the interplay between the various categories of agents (HFTs, market markers, agents trading on their behalf, or agents trading on behalf of their clients, \dots) is clearly of great interest. 
 For example, \citet{boussetta2017role} show that HFTs submit their orders in a markedly different way than slow traders. A good starting point would be a substantial modification of the model of \citet{donier2016walras} in the spirit of the work done by \citet{lemhadri2019price}.

\section*{Acknowledgements}
This publication stems from a partnership between CentraleSup\'elec and BNP Paribas.

The authors would like to acknowledge the support of the``Equipex PLADIFES ANR-21-ESRE-0036 (France 2030)'' responsible for the EUROFIDAI BEDOFIH's database. 
\nocite{*}
\bibliographystyle{plainnat}
\bibliography{apssamp}

\appendix

\section{Proof of Proposition \ref{prop:T}}
\label{sec:proofs}
\begin{proof}
We only prove the proposition for an additional buy market order resulting in a price impact denoted by $I_B$. The case of a sell market order resulting in an impact $I_S$ is symmetric. By definition, $\omega\mapsto p_\omega$ is a non-decreasing right-continuous step function; the same holds for $\omega\mapsto I(\omega)$. Obviously $I(0)=0$ and $I(\omega)=0$ if and only if $p_w=p_a$. Since $\omega^{(0)}$ denotes the first point of discontinuity of $I$,  by monotonicity, the condition $p_w=p_a$ is equivalent to $\omega<\omega^{(0)}$. In the original auction $\mathcal A$ with auction price $p_a$ and auction volume $Q_a$, we have
\begin{equation}
\begin{cases}
    \begin{aligned}
        S(p_a) -V_S^R(p_a) &= D(p_a) - V_B^R(p_a) = Q_a, \\ 
        V_S^R(p_a) \times V_B^R(p_a) &= 0.
    \end{aligned}
\end{cases}
    \label{eq:o2}
\end{equation}
All these quantities are fixed by the original auction setting. If we add a buy market order of size $q=\omega \times Q_a$ in this setting, the new auction price $p_{\omega}$ satisfies
\begin{equation}
\begin{cases}
\begin{aligned}
 S(p_{\omega}) - V_S^R(\omega) &= D(p_{\omega}) - V_B^R(\omega) + q, \\
 V_S^R(\omega) \times V_B^R(\omega) &= 0,
\end{aligned}
\end{cases}
\label{eq:o1}
\end{equation}
where $S$ and $D$ are the original supply and demand functions, and $V_S^R(\omega)$ (resp. $V_B^R(\omega)$) is the remaining sell quantity (resp. buy quantity) \emph{at price $p_\omega$} in the new setting. These volumes depend clearly on $\omega$.

Let us now determine the first point of discontinuity $\omega_B^{(0)}$. It is clear that the first price change due to the addition of a market order of size $q=\omega_B^{(0)} Q_a $ occurs when $V_S^R(\omega) = V_S(p_{\omega})$, $V_B^R(\omega)=0$, and the new auction price $p_{\omega} = p_B^{(1)}$ is the first non empty price tick after $p_a$ in the sense of $V_S+V_B$, i.e., the first tick price strictly greater than the auction price which contains buy or sell shares. (see Figure \ref{fig:remV} to build an intuition). Equation (\ref{eq:o1}) yields
\begin{equation}
\begin{aligned}
S(p_B^{(1)}) -V_S(p_B^{(1)}) = D(p_B^{(1)}) + q .
\end{aligned}
\end{equation}
Using the fact that $S(p_B^{(1)}) = S(p_a) + V_S(p_B^{(1)})$ and $D(p_B^{(1)}) = D(p_a) -V_B(p_a)$ we obtain
\begin{equation}
\begin{aligned}
S(p_a) = D(p_a) - V_B(p_a) + q,
\end{aligned}
\end{equation}
hence, using equation (\ref{eq:o2}), one finds
\begin{equation}
\begin{aligned}
V_S^R(p_a) = V_B^R(p_a) - V_B(p_a) + q.
\end{aligned}
\end{equation}
Using $V_B(p_a) = V_B^R(p_a) + V_B^M(p_a)$, we obtain
\begin{equation}
\begin{aligned}
q = V_S^R(p_a) + V_B^M(p_a), 
\end{aligned}
\end{equation}
which yields
\begin{equation}
\begin{aligned}
\omega_B^{(0)} = \frac{1}{Q_a} \left( V_S^R(p_a) + V_B^M(p_a) \right)
\end{aligned}
\end{equation}

Let us now determine $\omega_B^{(i)}$, $i\geq 1$: which is the $(i+1)^{\text{th}}$ point of discontinuity of $I_B$. We proceed similarly: the $(i+1)^{\text{th}}$ price change due to the injection of a market order occurs when $V_S^R(\omega) = V_S(p_{\omega})$, $V_B^R(\omega)=0$, and $p_{\omega}=p_B^{(i+1)}$ is the $(i+1)^{\text{th}}$ non empty price tick greater than $p_a$ (in the sense of $V_S+V_B$). Equation (\ref{eq:o1}) yields
\begin{equation}
S(p_B^{(i+1)}) -V_S(p_B^{(i+1)}) = D(p_B^{(i+1)}) + q,
\end{equation}
\begin{equation}
\sum_{p' < p_B^{(i+1)}} V_S(p') = \sum_{p' \geq p_B^{(i+1)}} V_B(p') + q,
\end{equation}
\begin{equation}
S(p_a) + \sum_{p_a < p' < p_B^{(i+1)}} V_S(p') = D(p_a) - \sum_{p_a \leq p' < p_B^{(i+1)}} V_B(p') + q.
\end{equation}
Using equation (\ref{eq:o2}), we obtain
\begin{equation}
 Q_a +V_S^R(p_a)  + \sum_{p_a < p' < p_B^{(i+1)}} V_S(p') =  Q_a +V_B^R(p_a)  -(V_B(p_a) +  \sum_{p_a < p' < p_B^{(i+1)}} V_B(p')) + q.
\end{equation}
Finally,
\begin{equation}
\sum_{p_a < p' < p_B^{(i+1)}} (V_S+V_B)(p') = q - (V_S^R(p_a) + V_B^M(p_a)).
\end{equation}
Thus,
\begin{equation}
\omega^{(i)}_B  = \omega^{(0)}_B + \frac{1}{Q_a}\sum_{p_a<p'<p_B^{(i+1)}}\left(V_S + V_B\right) (p') \quad , \quad i\geq 1,
\end{equation}
which leads to
\begin{equation}
\omega^{(i)}_B  = \omega^{(i-1)}_B +  \frac{V_S(p_B^{(i)})+V_B(p_B^{(i)})}{Q_a} \quad , \quad i \geq 1.
\end{equation}

\end{proof}

\section{Proof of Proposition \ref{prop:lin imp}}\label{sec:proofs2}

\begin{proof}

Using Proposition \ref{prop:T} we have
\begin{equation}
\label{eq:n}
\begin{aligned}
\omega_B^{(i)} - \omega_B^{(0)} &=  \frac{1}{Q_a} \sum_{p_a < p' < p_B^{(i+1)}}V_S(p')+ V_B(p')  \\
 & =  \frac{1}{Q_a} \sum_{k=1}^{i} (V_S+ V_B)(p_B^{(k)}) \\
 & =  \sum_{k=1}^{i} (p_B^{(k+1)}-p_B^{(k)}) \left(\tilde{\rho}_S+ \tilde{\rho}_B\right)(p_B^{(k)})  \\
 & = \widetilde{\mathcal{L}}_B \sum_{k=1}^{i} (p_B^{(k+1)}-p_B^{(k)}) \\
 & =\widetilde{\mathcal{L}}_B  (p_B^{(i+1)} - p_B^{(1)}) \\
 & \approx\widetilde{\mathcal{L}}_B p_B^{(1)} \left[I_B\left(\omega_B^{(i)}\right) - I_B\left(\omega_B^{(0)}\right) \right],
\end{aligned}
\end{equation}
where we used the approximation $I_B\left(\omega_B^{(i)}\right) - I_B\left(\omega_B^{(0)}\right) = \log(p_B^{(i+1)}/p_B^{(1)}) \approx p_B^{(i+1)}/p_B^{(1)} - 1$.

\end{proof}

\section{Empirical properties of impact slopes at auction time}
\label{sec:empirical_slopes}

In this appendix, we report empirical observations on the impact slope at auction time on day $d$ defined as
\begin{equation}
    \widetilde{S}_d = (p^{(1)}\widetilde{\mathcal{L}})^{-1}.
\end{equation}

\begin{figure}
    \centering
    \includegraphics[scale=0.49]{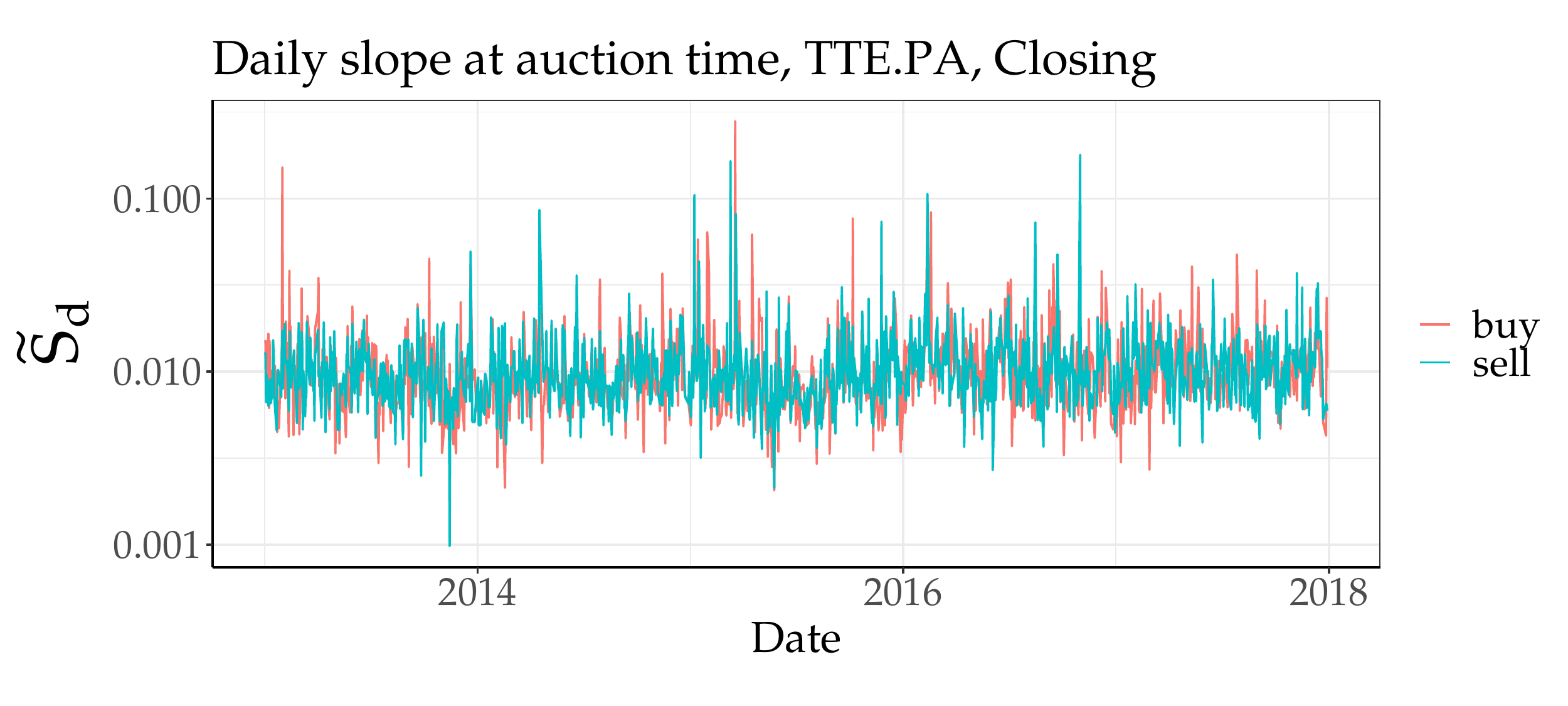}
    \caption{Daily impact slope at the closing auction time for TotalEnergies stock.}
    \label{fig:daily_slope}
\end{figure}

\begin{figure}
    \centering
    \includegraphics[scale=0.56]{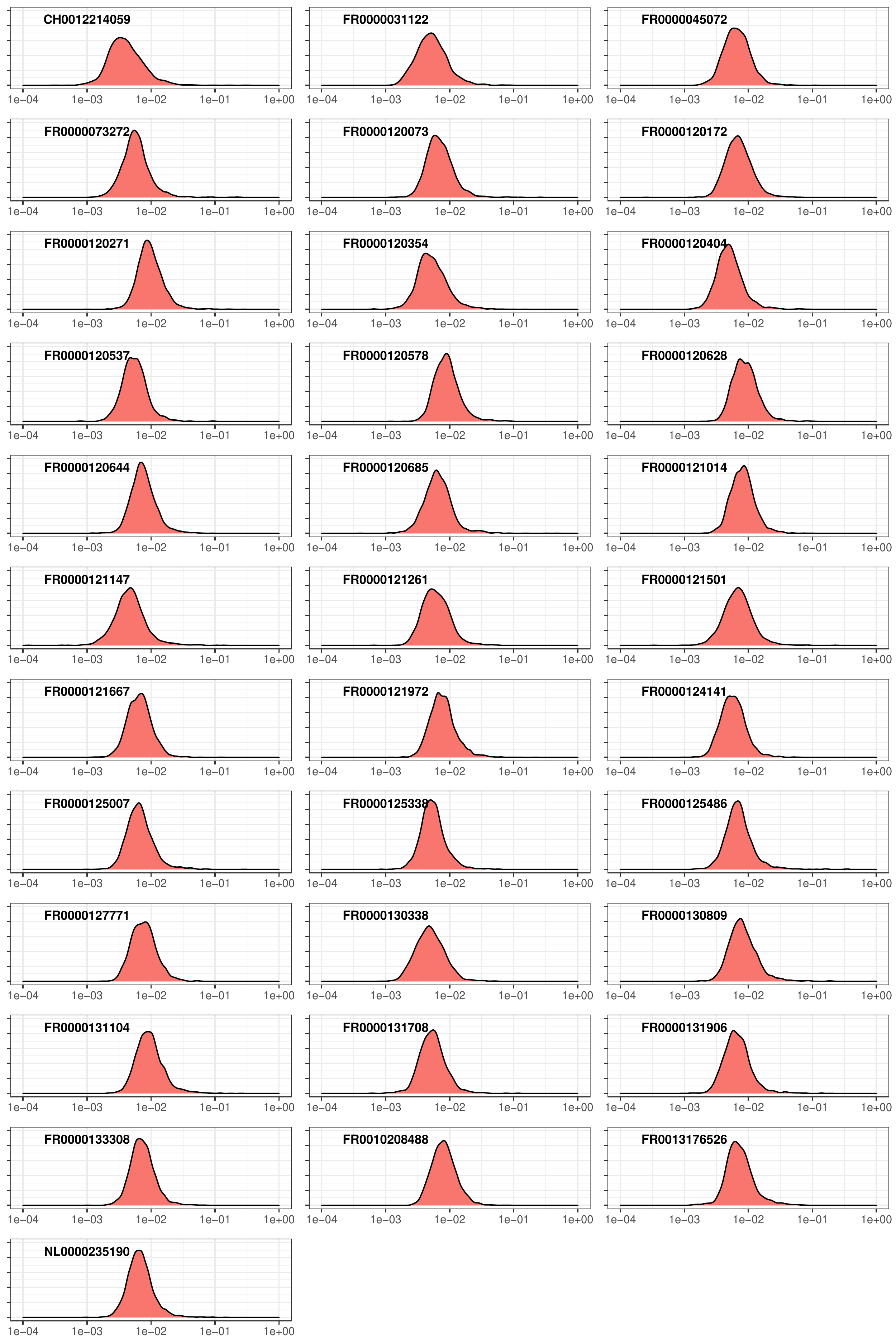}
    \caption{Kernel density of the impact slope at the closing auction time for the 34 studied assets.}
    \label{fig:hist_slopes}
\end{figure}

Figure \ref{fig:daily_slope} plots $\widetilde{S}_d$ for TotalEnergies as a function of time. It oscillates around a typical value and has a positive autocorrelation over a few days. The distribution of $\widetilde{S}_d$ for the 34 stocks is reported in Fig.\ \ref{fig:hist_slopes}: while its shape is similar for all the assets, its parameters depend on each stock.

\begin{figure}
    \centering
    \includegraphics[scale=0.6]{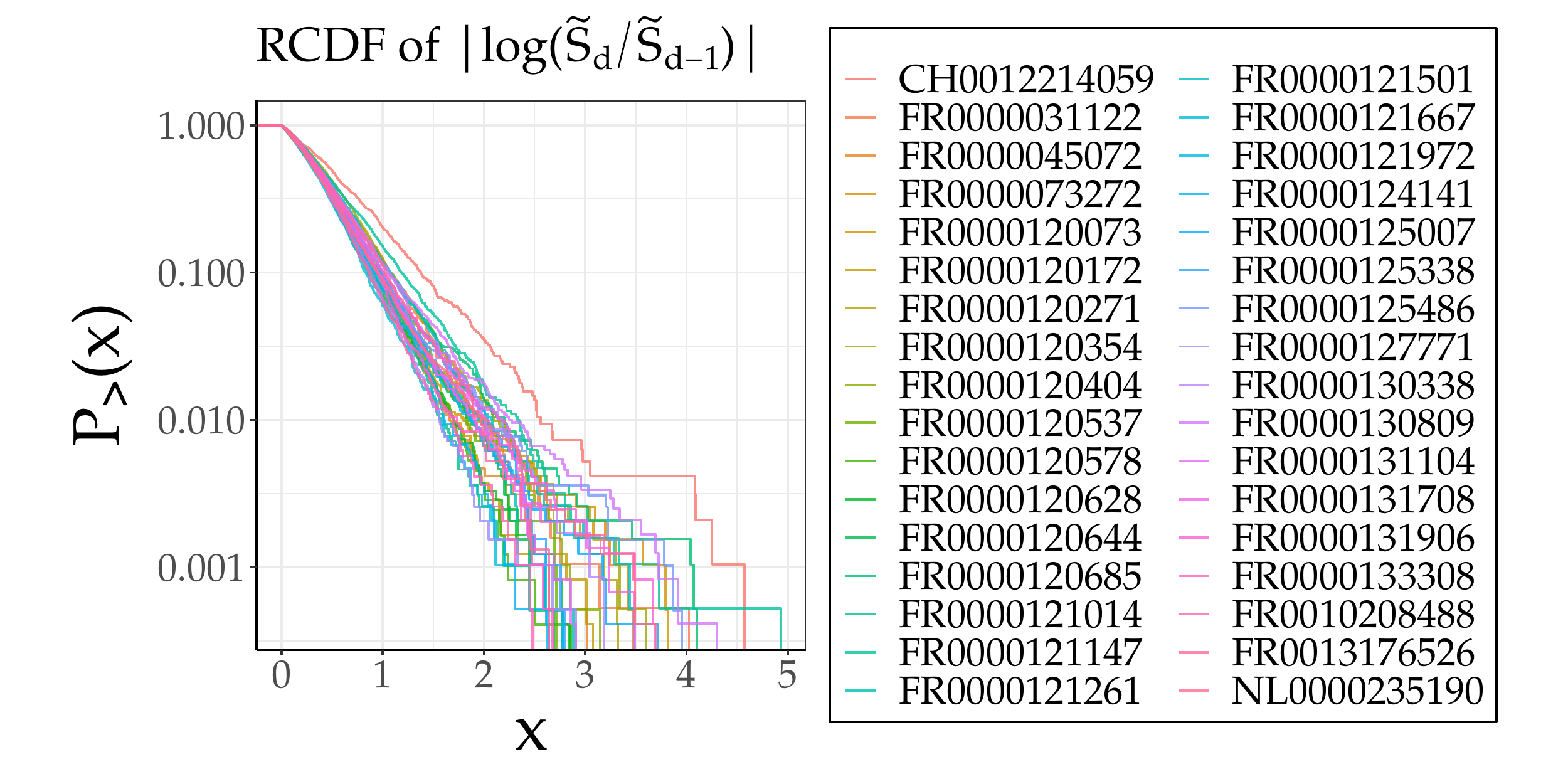}
    \caption{Reverse cumulative distribution function of the absolute change in the logarithm of the slope.}
    \label{fig:RCDF_logret_slopes}
\end{figure}

We also report the distribution of the absolute value log-changes of the slopes in Fig. \ref{fig:RCDF_logret_slopes}, which clearly appear to be exponentially distributed. Its one-step autocorrelation is negative.

\end{document}